\documentclass[prb,twocolumn,superscriptaddress,a4paper,twoside]{revtex4-2}

\usepackage{amsmath}
\usepackage{amssymb}
\usepackage{todonotes}
\usepackage{nicefrac}
\usepackage{graphics}
\usepackage{color}
\usepackage{graphicx}
\usepackage{verbatim}
\usepackage{float}

\makeatletter
\newcommand*{\rom}
[1]{\expandafter\@slowromancap\romannumeral #1@}
\makeatother

\begin{document}

\title{Density-Functional Green Function Theory:\\ Dynamical exchange-correlation
field in lieu of self-energy}

\author{F. Aryasetiawan}
\affiliation{
Department of Physics, Division of Mathematical Physics, 
Lund University, Professorsgatan 1, 223 63, Lund, Sweden}
\affiliation{
LINXS Institute of advanced Neutron and X-ray Science,
IDEON Building: Delta 5, Scheelevägen 19, 223 70 Lund, Sweden}

\begin{abstract}
The one-particle Green function of a many-electron system is traditionally
formulated within the self-energy picture.
A different formalism was recently proposed,
in which the self-energy is replaced by a dynamical exchange-correlation field,
which acts on the Green function locally in both space and time.
It was found that there exists a fundamental quantity,
referred to as the dynamical exchange-correlation hole, which can be
interpreted as effective density fluctuations induced in a many-electron system
when a hole or an electron is introduced into the system, as in photoemission
and inverse photoemission experiments.
The dynamical exchange-correlation potential is simply the Coulomb potential
of this exchange-correlation hole, which
fulfils a sum rule and an exact constraint, identical to those satisfied
by the static exchange-correlation hole in density-functional theory.
The proposed formalism has been applied to a number
of model systems such as the half-filled one-dimensional Hubbard model,
the one-dimensional antiferromagnetic Heisenberg model, and the single-impurity
Anderson model. The dynamical exchange-correlation hole and field of the
homogeneous electron gas have also been studied with the view of constructing
a density-functional approximation such as the local-density approximation.
The availability of simple but accurate
approximations for the exchange-correlation potential
would circumvent costly computations of the traditional self-energy.
The formalism may also provide new perspectives and insights into the many-body
problem.
\end{abstract}

\maketitle


\section{Introduction}

The arrival of density-functional theory (DFT) in 1964-1965 
\cite{kohn1964,kohn1965} witnessed one of the
most important developments in condensed-matter theory and quantum chemistry
\cite{jones1989,becke2014,jones2015}.
The fundamental theorem of Hohenberg and Kohn \cite{kohn1964}
states that the ground-state density
of a many-electron system uniquely determines the external potential. This
seemingly simple theorem has far-reaching consequences: since the
external potential uniquely determines the Hamiltonian it implies that
\emph{all} electronic properties of the system, 
ground-state as well as excited-state, are determined by the
ground-state density, i.e., they are functionals of the ground-state density.

By itself, the Hohenberg-Kohn theorem has no immediate use, since
the theorem provides no prescription on how to calculate electronic properties
of the many-electron system. However, the ground-state energy
possesses a special property in that it is minimised
by the ground-state density.
Were the ground-state energy as a functional of the ground-state density known,
this variational property would yield,
in principle, the Euler-Lagrange equation. An early
example of this, long before the Hohenberg-Kohn theorem was known,
is the Thomas-Fermi functional \cite{thomas1927,fermi1928},
where the kinetic energy
is approximated by that of a homogeneous electron gas in the spirit of the
local-density approximation (LDA) \cite{kohn1965}. However,
the Thomas-Fermi energy functional is not
sufficiently accurate for applications to real materials. It
is too crude to reproduce, for example, the shell structure in
the ground-state density of atoms.

The practical development of DFT
was made possible by the Kohn-Sham scheme \cite{kohn1965},
which introduces the concept
of an auxiliary noninteracting system with the same density as that of 
the interacting system. The total
ground-state energy is reshuffled into a sum of the noninteracting
kinetic energy of the auxiliary system and the rest, which includes the
difference between the interacting and the noninteracting kinetic energies,
lumped into the exchange-correlation (xc) energy.
The variational property of the total energy
leads to the famous Kohn-Sham equation, which is 
a one-particle Schr\"odinger equation. Since the kinetic energy is calculated
directly for the noninteracting system, the need for an explicit
kinetic energy functional is avoided. 
There remains the problem of
finding a good approximation for the xc functional,
which fortunately can be approximated by explicit expressions of the density
sufficiently accurate for making theoretical predictions.
Starting with the LDA,
more accurate functionals have been continuously developed over
the years \cite{becke1993,perdew1996a,perdew1996,heyd1996}
(for reviews, see, for example, \cite{burke1997,becke2014}), 
and the recent use of machine learning \cite{nagai2020,fiedler2022}
may accelerate further development.
The Kohn-Sham scheme is rather revolutionary, since the complicated
many-electron problem of calculating the ground-state energy
is reduced to solving the
single-particle Kohn-Sham equation self-consistently. 

Although the Kohn-Sham scheme has been very successful in calculating
ground-state properties, the extension of DFT to calculations
of excitation spectra remains an open challenge. 
The challenge of developing a density-functional 
formalism for excited states arose almost immediately
after the birth of DFT.
Almost 60 years ago,
Sham and Kohn \cite{sham1966} addressed this challenge by considering
the Green function approach with the associated self-energy.
The self-energy of the homogeneous electron gas, which can be calculated
within the random-phase approximation (RPA) \cite{pines1952}
or the $GW$ approximation \cite{hedin1965,hedin1969,aryasetiawan1998,onida2002},
is used as a reference. The LDA for the self-energy can then
be applied. This approach was implemented
by Wang and Pickett \cite{wang1983} and later
improved by Godby \emph{et al.} \cite{godby1987} to account for nonlocality
of the self-energy. The applicability of this approach appears to be
rather limited and has not found widespread use. 

A promising approach to extend DFT to calculations of excited-state
energies is via ensemble DFT \cite{gross1988,oliveira1988,gould2020,scott2024}.
By generalising DFT to mixtures of ground
and excited states, it is possible to determine
specifically chosen excitation energies at low cost.
However, the method
may not be suitable for calculating excitation spectra which are continuous
such as those of solids. 

An extension of DFT to time-dependent DFT (TDDFT)
\cite{stott1980,zangwill1980,runge1984,ullrich,marques2006,maitra2016}
provides a route for calculating
excitation energies.
TDDFT offers a rigorous formalism for describing
the time evolution of an electron density under
a time-varying external field \cite{runge1984}.
This allows one to obtain the linear density-response function from which
excitation energies can be extracted.
However, the excitation energies correspond
to the $N$-electron system, i.e., electron-number-conserving excitations.
Excitations involving removal and addition of an electron
resulting in $N\pm 1$ electrons corresponding to photoemission and inverse
photoemission are difficult to simulate within TDDFT. 
To overcome this difficulty, a steady-state DFT was recently proposed
\cite{jacob2018,kurth2018}.

This article describes a recently
proposed density-functional approach to determine the Green function
\cite{aryasetiawan2022}
from which excitation spectra measured in angle-resolved
photoemission and inverse photoemission can be extracted.
This formalism replaces the self-energy by a dynamical xc field or potential,
hereafter referred to as Vxc,
which acts locally in both space and time on the Green function.
Vxc is shown to be the Coulomb potential of a dynamical xc hole,
which as the xc hole in DFT
fulfils a sum rule and an exact constraint, providing a natural
link to the ground-state density. This is analogous to
the Slater exchange potential \cite{slater1951}, which can in fact be seen as
a special case of Vxc when correlations are neglected.

The dynamical xc hole can be thought of as density fluctuations induced when a hole or
an electron is introduced into a many-electron system, as in photoemission and
inverse photoemission experiments. The concept of xc hole offers a fundamentally
different
perspective of viewing the many-electron problem from the traditional self-energy.
On the one hand, the xc hole connects the Vxc formalism to DFT
and on the other hand opens a path to dynamical correlations
not accessible in Kohn-Sham DFT.
This duality allows for harnessing a wealth of DFT-inspired approximations
while at the same time taking advantage of well-established approximations
from many-body theory such as the RPA to incorporate dynamical correlations.

Since the formalism is exact, there is no restriction in applying it
to systems with strong electron correlations. This is illustrated in
Sec. \ref{sec:Applications} in which applications to the one-dimensional
Hubbard and Heisenberg models as well as the single-impurity Anderson model
are described. For weakly to moderately correlated systems, LDA-based approximations
can be expected to work well, as in traditional DFT. A first step towards this direction
is to calculate Vxc for the homogeneous electron gas elaborated in Sec. \ref{sec:HEG}
with the aim of parametrising the resulting Vxc as a function of density.

A potentially important advantage of the Vxc formalism is in dealing with
nonequilibrium systems. Due to the local character of Vxc in both space and time,
the equation of motion of the Green function can be solved piecewise in time, in
contrast to the traditional self-energy formalism, which requires knowledge of
the self-energy for all time.

Although the present article is focused on fermions, the formalism can be
extended to bosons and spin systems. An example of the latter is given in
Sec. \ref{sec:HeisenbergModel}, which describes an application to the
one-dimensional antiferromagnetic Heisenberg model.


\section{Slater exchange hole and potential}
\label{sec:Slaterxhole}

In his attempt in the early 1950's \cite{slater1951}
to simplify the Hartree-Fock equation by replacing
the nonlocal exchange potential by a local one Slater discovered
the exchange hole, a quantity of great importance in DFT, both from the
conceptual and practical points of view.
Slater's motivation was numerical; considering the state of computers at that
time, it was not feasible to solve the Hartree-Fock equation for
molecules and solids with a nonlocal exchange potential.

The nonlocal Fock exchange is given by
\begin{align}
    \Sigma_\mathrm{x}(r,r')=-v(r-r')\sum_i^\mathrm{occ} 
    \phi_i(r) \phi_i^*(r'),
\end{align}
where the sum is over the lowest occupied (occ) orbitals and
$v(r-r')=1/|\mathbf{r}-\mathbf{r}'|$.
The nonlocal exchange potential appears in the Hartree-Fock equation as
\begin{align}
    \int dr'\, \Sigma_\mathrm{x}(r,r') \phi_k(r').
\end{align}
Slater defined an orbital-dependent local potential
\begin{align}
    V_k(r)= \frac{1}{\phi_k(r)} \int dr'\, \Sigma_\mathrm{x}(r,r') 
    \phi_k(r')
\end{align}
so that
\begin{align}
     \int dr'\, \Sigma_\mathrm{x}(r,r') \phi_k(r')
     = V_k(r) \phi_k(r).
\end{align}
Thus, the nonlocal Fock exchange when acting on a given orbital
has been converted into a local potential, albeit orbital-dependent.

The local but orbital-dependent potential is in essence a rephrasing of the Fock
exchange since orbital dependence implies nonlocality.
Slater proceeded further
and introduced an orbital-averaged potential:
\begin{align}
   & V^\mathrm{S}(r) =\sum_k^\mathrm{occ} \frac{|\phi_k(r)|^2}{\rho(r)} V_k(r)
    \nonumber\\
    &= - \int dr'\, v(r-r') \frac{1}{\rho(r)}
    \sum_{ik}^\mathrm{occ}
\phi_k^*(r)\phi_i(r) \phi_i^*(r') \phi_k(r'),
\label{eq:SlaterPot}
\end{align}
where $\rho(r)$ is the ground-state density,
\begin{align}
    \rho(r) =\sum_k^\mathrm{occ} |\phi_k(r)|^2.
\end{align}
The local orbital-independent
potential $V^\mathrm{S}(r)$ is known as the Slater $X\alpha$
exchange potential, which replaces the nonlocal Fock exchange potential:
\begin{align}
    \left[ -\frac{1}{2}\nabla^2 +V_\mathrm{ext}(r)+V^\mathrm{H}(r)
    +V^\mathrm{S}(r)
    \right] \phi_k(r) = \varepsilon_k \phi_k(r).
\end{align}
This equation is much more amenable to a numerical solution compared to
the original Hartree-Fock equation.

Slater made an important observation:
In any electronic system, the quantity
\begin{align}
    \rho_\mathrm{x}(r,r')=
    -\frac{1}{\rho(r)}
    \sum_{ik}^\mathrm{occ} 
    \phi_k^*(r)\phi_i(r) \phi_i^*(r') \phi_k(r')
\label{eq:Slaterx-hole}
\end{align}
fulfils a sum rule and an exact constraint for \emph{any} $r$:
\begin{align}
    \int d^3r'\, \rho_\mathrm{x}(r,r')= -\delta_{\sigma\sigma'},\qquad \rho_\mathrm{x}(r,r)=-\rho(r),
\label{eq:SlaterConstraint}
\end{align}
which can be readily verified.
Note that the notation $r=(\mathbf{r},\sigma)$ is used.
Since the sum rule is independent of $r$ one may write
for $\sigma=\sigma'$,
\begin{align}
 \rho_\mathrm{x}(\mathbf{r},\mathbf{r}')=\rho_\mathrm{x}(\mathbf{r},\mathbf{r+R}),
 \quad \mathbf{R}=\mathbf{r}'-\mathbf{r},
\end{align}
so that
\begin{align}
    \int d^3R\, \rho_\mathrm{x}(\mathbf{r},\mathbf{r}+\mathbf{R})= -1.
\end{align}

The above sum rule and
the exact constraint motivate the interpretation of
$\rho_\mathrm{x}(r,r')$ as an exchange hole.
For an electron at $\mathbf{r}$, the exchange-hole density at
the position of the electron is the negative
of the electron density.
The exchange interaction (Pauli principle)
pushes away electrons of the same spin,
creating a hole around the given electron
at $\mathbf{r}$, which
as a function of $\mathbf{R}$
integrates to a negative unit charge, \emph{independent} of $\mathbf{r}$.

From Eq. (\ref{eq:SlaterPot}), it is evident that
the Slater potential can be regarded as
the Coulomb potential generated by the exchange hole:
\begin{align}
    V^\mathrm{S}(r)&=\int dr' v(r-r') \rho_\mathrm{x}(r,r').
\end{align}
Because the exchange hole integrates to a negative unit charge,
it is to be
expected that the Slater potential is negative.
The sum rule and the exact constraint have a very important consequence.
As explained by Slater \cite{slater1968}, if the exchange
hole is approximated by a sphere with radius $R_0$ and assumed to have a
constant density $\rho(r)$ as a function
of $r'$ one finds
\begin{equation}
\frac{4\pi}{3}R_0^{3}\rho(r)=1\rightarrow R_0=\left[  \frac{3}{4\pi\rho
(r)}\right]  ^{1/3}.%
\end{equation}
The Coulomb potential due to this exchange hole at $\mathbf{r}$,
the centre of the sphere, is given by
\begin{align}
    \rho(r) \int d^3 r' \frac{\theta(R_0-|\mathbf{r}'-\mathbf{r}|)}
                             {|\mathbf{r}'-\mathbf{r}|}
    = 2\pi\rho(r) R_0^2      
\end{align}
which is proportional to $\rho^{1/3}$, the well-known local-density
dependence of the exchange potential.
Note that
this $\rho^{1/3}$-dependence does not depend on the homogeneous electron gas.
If the exchange hole of the homogeneous electron gas is used as a model,
the Slater potential is given by
\begin{align}\label{eq:SlaterVx}
    V^\mathrm{S}(r)=-3\left[ \frac{3}{4\pi}\rho(r)
    \right]^{1/3}.
\end{align}
The finding that the exchange potential can be approximated by the
local density and varies as $\rho^{1/3}$ can be regarded as the precursor
of the LDA in DFT.
The exchange potential in the LDA of DFT
obtained by taking the functional derivative
of the exchange energy of the homogeneous
electron gas with respect to the density
is equal to $\frac{2}{3}V^\mathrm{S}(r)$.

The Slater local potential was meant to be an approximation to the
non-local Fock exchange, but it was discovered that the bond energies of molecules
obtained
from the Slater potential were curiously superior to those calculated from the
Hartree-Fock equation \cite{becke2014}. 
The bond energy of a molecule is the difference between
the total energies of the molecule and the constituent atoms, which is
a very important quantity in chemistry and physics. 

Another indication of the significance of the Slater local potential
is provided by prediction of
the quasiparticle dispersion of the homogeneous electron gas.
In this case, the Slater potential is simply constant, implying that
the quasiparticle dispersion up to a constant shift
is identical to that of the free-electron
gas, namely, $\varepsilon(k)=\frac{1}{2}k^2$. 
This is in stark contrast to the dispersion
calculated within the Hartree-Fock approximation,
which widens the occupied band width by approximately a
factor of two for $r_s\approx 3 \sim 4$ \cite{ashcroft1976},
representative of metallic systems.
The logarithmic
singularity of the derivative of the Hartree-Fock 
dispersion at the Fermi wave vector leads to
a zero density of states at the Fermi level, which is nonphysical
\cite{ashcroft1976}.
The exact dispersion is not known, but one can expect with confidence
that the dispersion calculated within the
$GW$ approximation should be close to the exact one.
The $GW$ dispersion essentially resembles that of the free-electron gas
with $\approx 10\%$ narrowing of the occupied band \cite{holm1998}.
This implies that the band widening resulting from a nonlocal exchange is
cancelled to a great extent by correlations.

One can conclude that the Slater potential is actually
a rather poor approximation to the nonlocal exchange
since it yields a superior electron gas dispersion
and bonding energy in molecules. 
This implies that the Slater potential
contains to a large extent, in the sense of many-body perturbation theory,
the effects of correlations.
In systems where the valence electrons can be modelled by the
electron gas, the combined effects of
exchange and correlations may be better represented by a local potential.

There is a subtle difference between the Slater potential
and the exchange potential in Kohn-Sham DFT.
This difference can be understood by
considering the exact expression for the exchange energy:
\begin{align}
    E_\mathrm{x}&=\frac{1}{2}\int dr' dr''\rho(r') v(r'-r'')
\rho_\mathrm{x}(r',r'')
    \nonumber\\
    &=\frac{1}{2}\int dr' dr''\rho(r') v(r'-r'')
    [g_\mathrm{x}(r',r'')-1]\rho(r''),
\end{align}
where 
\begin{align}
   \rho_\mathrm{x}(r,r')= [g_\mathrm{x}(r,r')-1]\rho(r').
\end{align}
$g_\mathrm{x}(r,r')$ is the exchange-only pair distribution function.
The exchange potential in DFT is given by
\begin{align}
    &V_\mathrm{x}(r)= \frac{\delta E_\mathrm{x}}{\delta \rho(r)}
    \nonumber\\
    &=V^\mathrm{S}(r) 
    +\frac{1}{2}\int dr' dr''\rho(r') v(r'-r'')
    \frac{\delta g_\mathrm{x}(r',r'')}{\delta \rho(r)}
    \rho(r'').
\label{eq:VxVS}
\end{align}

From the above equation it is clear that
the Kohn-Sham exchange potential is not the Coulomb potential of the
exchange hole.
In addition to the Coulomb potential,
there is a complicated term involving a three-point vertex function
$\delta g_\mathrm{x}(r',r'')/\delta\rho(r)$, which may be the origin of
the pathology that is sometimes observed in the Kohn-Sham potential.
In fact, this pathology was convincingly demonstrated for the beryllium atom
by explicitly calculating the vertex function.
The resulting vertex contribution exhibits a step structure,
which is constant within the atomic shells and changes abruptly
at the shell boundaries \cite{vanLeeuwen1995}.

Slater's work anticipates much of
the practical aspect of DFT.
However, the idea that the local exchange potential is the Coulomb potential
of the exchange hole has not been further developed.
Development in DFT has followed the
Kohn-Sham scheme of having an auxiliary noninteracting system.
For example, an extension to nonequilibrium systems in the presence of
a time-dependent potential also employs a Kohn-Sham noninteracting auxiliary
system \cite{runge1984,ullrich}.

\section{Traditional self-energy formalism}

The time-ordered Green function is defined as
\cite{fetter-walecka,negele-orland,aryasetiawan2025a}
\begin{align}
    iG(rt,r't') = \langle T [\hat{\psi}(rt) \hat{\psi}^\dagger(r't')] \rangle ,
\end{align}
where the expectation value is taken with respect to the 
interacting many-electron ground state.
${T}$ is the time-ordering symbol and
$\hat{\psi}(rt)$ is the field operator in the Heisenberg picture,
\begin{align}
    \hat{\psi}(rt) =\hat{U}(0,t)\hat{\psi}(r)\hat{U}(t,0),
\end{align}
where the time-evolution operator is given by
\begin{align}
    \hat{U}(t,0) = {T}e^{-i\int_0^t dt' \hat{H}(t')}.
\end{align}

For systems in equilibrium, the time-evolution operator simplifies to
$\hat{U}(t,0)= \exp(-i\hat{H}t)$, in which
the many-electron Hamiltonian is given by
\begin{align}
    \hat{H} &=\int dr\, \hat{\psi}^\dagger(r) h_0(r) \hat{\psi}(r)
    \nonumber\\
    &+\frac{1}{2} \int drdr' \hat{\psi}^\dagger(r)\hat{\psi}^\dagger(r')
    v(r-r') \hat{\psi}(r')\hat{\psi}(r),
\end{align}
where
\begin{align}
    h_0(r) = -\frac{1}{2} \nabla^2 + V_\mathrm{ext}(r).
\label{eq:h0}
\end{align}
In this case, the Green function depends only on the time
difference $t-t'$ so that $t'$ can be set to zero. 
For a Hamiltonian without a spin-flip term such as the one above,
the Green function is diagonal in spin space, that is, $\sigma=\sigma'$.
The equation of motion of the Green function
is obtained from the Heisenberg equation of motion of the field operator:
\begin{align}
    i\partial_t \hat{\psi}(rt)&= \left[\hat{\psi}(rt),\hat{H}  \right]
    \nonumber\\
    &= \left( h_0(r) + \hat{V}_\mathrm{H}(rt)\right) \hat{\psi}(rt),
\label{eq:EOMfieldop}
\end{align}
where
\begin{align}
    \hat{V}_\mathrm{H}(rt) =\int dr' v(r-r')\hat{\rho}(r't).
\end{align}
Here, $\hat{\rho}(rt)=\hat{\psi}^\dagger(rt)\hat{\psi}(rt)$ is the density
operator. 
Constructing two equations by multiplying Eq. (\ref{eq:EOMfieldop}) from the left
and right by $\hat{\psi}^\dagger(r')$ ($t'=0$) leads to
the equation of motion of the Green function,
\begin{align}
    &\left[ i\partial_t -h_0(r)
    \right] G(r,r';t) 
    \nonumber\\
    & \qquad +i\int dr'' v(r-r'') \langle {T} 
    [ \hat{\rho}(r''t) \hat{\psi}(rt)\hat{\psi}^\dagger(r')] \rangle
\nonumber\\    &\qquad\qquad
    =\delta(r-r')\delta(t).
\label{eq:EOMG3}
\end{align}
The interaction term with the Coulomb potential
contains the following,
\begin{align}
    iG^{(2)}(r,r',r'';t)= \langle T 
    [ \hat{\rho}(r''t) \hat{\psi}(rt)\hat{\psi}^\dagger(r')] \rangle ,
\end{align}
which is a special case of the two-particle Green function.
(Note that the definition of $G^{(2)}$ in this article differs from
the one in Ref. \cite{aryasetiawan2022a} by a factor $-i$.)
The equation of motion of the two-particle Green function in turn contains the
three-particle Green function, and so forth. 
Traditionally, this hierarchy of equations is truncated
by introducing the self-energy $\Sigma$ defined as \cite{hedin1969}
\begin{align}
    &\int dr'' dt'' \,\Sigma(r,r'';t-t'') G(r'',r';t'')
    \nonumber\\
    &=\int dr''\, v(r-r'') G^{(2)}(r,r',r'';t)- V_\mathrm{H}(r) G(r,r';t),
\label{eq:SigmaTrad}
\end{align}
where the Hartree mean-field $V_\mathrm{H}$ 
has been subtracted from $G^{(2)}$. 
The equation of motion becomes
\begin{align}
    &\left[ i\partial_t -h_0(r) - V_\mathrm{H}(r)
    \right] G(r,r';t) 
    \nonumber\\
    &-\int dr'' dt'' \,\Sigma(r,r'';t-t'') G(r'',r';t'')
    =\delta(r-r')\delta(t).
\label{eq:EOMGSigma}
\end{align}
The self-energy then
accounts for the effects of exchange and correlations
of the many-electron system. 
It is also a natural quantity that emerges when the Green function is
calculated using
the many-body perturbation theory based on Wick's theorem
\cite{fetter-walecka,negele-orland}. 

\section{Dynamical xc hole}

In this section, it is shown that there is a universal quantity,
referred to as the dynamical xc hole,
which fulfils a sum rule and a strict constraint
independent of the particular systems. It will be shown later
in Sec. \ref{sec:RelnSlaterxhole}
that its static ($t=0$) and equal-space
($r'=r)$ limit in the absence of correlations
is identical with the Slater exchange hole.
The Coulomb potential of this dynamical xc hole generates Vxc,
formally replacing the nonlocal
self-energy. Thus, it may be seen that the Slater exchange potential
is a special case of static Vxc without correlations.

In the proposed framework,
a correlation function $g$, which correlates the electron density
and the Green function, is introduced \cite{aryasetiawan2022}:
\begin{align}
    G^{(2)}(r,r',r'';t)&= -i\langle T 
    [ \hat{\rho}(r''t) \hat{\psi}(rt)\hat{\psi}^\dagger(r')] \rangle
    \nonumber\\
     &=G(r,r';t) g(r,r',r'';t) \rho(r'').
\label{eq:defG2}
\end{align}
$G^{(2)}$ can be rewritten as
\begin{align}
    G^{(2)}(r,r',r'';t)= [\rho(r'')+\rho_\mathrm{xc}(r,r',r'';t)] G(r,r';t),
\label{eq:G2}
\end{align}
where
\begin{align}
    \rho_\mathrm{xc}(r,r',r'';t)= [g(r,r',r'';t)-1]\rho(r'')
\label{eq:rhoxc}
\end{align}
can be interpreted as an xc hole since
it integrates to $-1$ for $t<0$ as will be shown below.

To show the sum rule,
consider integrating $G^{(2)}$ over the space variable $\mathbf{r}''$.
For $t<0$ one finds
\begin{align}
i\int d^3r''\, &G^{(2)}(r,r',r'';t<0) 
\nonumber\\
&=-\int d^3r'' \langle 
     \hat{\psi}^\dagger(r')\hat{\rho}(r''t) \hat{\psi}(rt) \rangle 
\nonumber\\
     &=-\langle 
     \hat{\psi}^\dagger(r')\hat{N}_{\sigma''}(t) \hat{\psi}(rt) \rangle,
\end{align}
where
\begin{align}
    \hat{N}_\sigma = \int d^3r\, \hat{\rho}(r)
\end{align}
is the number operator that counts the number of electrons with
spin $\sigma$ in the system.
%
%
It then follows that
\begin{align}
 \int d^3r''\, &G^{(2)}(r,r',r'';t<0)
 \nonumber\\
     &=i(N_{\sigma''}-\delta_{\sigma\sigma''}) \langle 
     \hat{\psi}^\dagger(r')\hat{\psi}(rt) \rangle
     \nonumber\\
     &=(N_{\sigma''}-\delta_{\sigma''\sigma}) G(r,r';t<0).
\label{eq:intG2}
\end{align}
From Eq. (\ref{eq:G2}),
\begin{align}
   &\int d^3r'' G^{(2)}(r,r',r'';t)
   \nonumber\\
   &=\left[ 
   N_{\sigma''} +\int d^3r''\,\rho_\mathrm{xc}(r,r',r'';t)
   \right]G(r,r';t),
\end{align}
which when compared with Eq. (\ref{eq:intG2})
yields the sum rule,
\begin{align}
    \int d^3r''\,\rho_\mathrm{xc}(r,r',r'';t<0) = -\delta_{\sigma\sigma''}
\end{align}
for \emph{any} $r$, $r'$, and $t<0$.
This justifies the interpretation of $\rho_\mathrm{xc}$ 
as an xc hole.

A similar derivation can be carried out for $t>0$ and one finds
\begin{align}
  \int d^3r''\, G^{(2)}(r,r',r'';t>0)
     &=N_{\sigma''} G(r,r';t>0).
\end{align}
The sum rule for $t>0$ is therefore
\begin{align}
    \int d^3r''\,\rho_\mathrm{xc}(r,r',r'';t>0) =0. 
\end{align}
The sum rule for any $r$, $r'$, and $t$ can be summarised as
\begin{align}
    \int d^3r''\,\rho_\mathrm{xc}(r,r',r'';t) =-\delta_{\sigma\sigma''}
    \theta(-t).
\label{eq:SumRule}
\end{align}

It can also be seen from the definition of $G^{(2)}$ in Eq. (\ref{eq:defG2})
that
\begin{align}
    iG^{(2)}(r,r',r;t) &= 
    \langle T 
    [ \hat{\rho}(rt) \hat{\psi}(rt)\hat{\psi}^\dagger(r')] \rangle
    \nonumber\\
    &=\langle T 
    [ \hat{\psi}^\dagger(rt) \hat{\psi}(rt)\hat{\psi}(rt)\hat{\psi}^\dagger(r')] \rangle
    =0
\end{align}
since $\hat{\psi}(rt)\hat{\psi}(rt)=0$
so that
\begin{align}
    g(r,r',r;t)=0.
\end{align}
It follows from Eq. (\ref{eq:rhoxc}) that the xc hole 
fulfils a strict constraint,
\begin{align}
    \rho_\mathrm{xc}(r,r',r;t) =-\rho(r),
\label{eq:ExactReln}
\end{align}
for \emph{any} $r$, $r'$, and $t$.

\section{Vxc and the equation of motion of the Green function}

Using the expression for $G^{(2)}$ in Eq. (\ref{eq:G2}), the equation of motion
of the Green function in Eq. (\ref{eq:EOMG3}) becomes
\begin{align}
    &\left[ i\partial_t -h(r) - V_\mathrm{xc}(r,r';t)
    \right] G(r,r';t) 
    =\delta(r-r')\delta(t),
\label{eq:EOMGVxc}
\end{align}
where
\begin{align}
    h(r) = h_0(r) + V_\mathrm{H}(r),
\end{align}
and
\begin{align}
    V_\mathrm{xc}(r,r';t)=\int dr'' \,v(r-r'') \rho_\mathrm{xc}(r,r',r'';t),
\label{eq:phixc}
\end{align}
which is simply the classical Coulomb potential of the xc hole.
Mathematically it can also be viewed as the Lagrange parameter which ensures
that the Green function is correctly reproduced.
This potential may in general be complex so that the Hamiltonian is nonhermitian.
The form of the equation of motion of the Green function
is identical to the time-dependent Schr\"odinger equation.
Treating $r'$ as a parameter, 
the Green function is analogous to the wave function but it
can decay due to the potential $V_\mathrm{xc}$ being complex.
This property is elaborated further in Sec. \ref{sec:QPWF}.

The xc hole may be interpreted as
effective density fluctuations induced
in a many-electron system when a hole ($t<0$) or an electron ($t>0$) is added.
The added hole or electron experiences the Coulomb potential arising
from the density
fluctuations (xc hole) as it propagates through the system.
However, these density fluctuations are in general complex and therefore not
experimentally observable. 

An important feature that distinguishes Vxc from
the self-energy is its local character in time. Due to this feature,
the Green function can be propagated point-wise in time since its value at time
$t$ can be used to calculate the value at an infinitesimal time later
without knowing its previous values at earlier time than $t$. 
In this sense, as with the time-dependent Schr\"odinger equation,
the equation of motion has no memory.
This is in contrast to the self-energy approach
which requires
knowledge of the Green function at all time due to the time
convolution between the self-energy and the Green function.

Vxc and the self-energy
differ in a fundamental way in that the former is local
whereas the latter is nonlocal in space.
The nonlocality in space renders expressing the self-energy
as a functional of the density difficult, which is illustrated by
the nonlocal exchange self-energy. The form of the nonlocal exchange
in terms of products of orbitals permits no simple approximation
in terms of the density.
The problem is akin to the kinetic energy, which is also nonlocal.
The kinetic energy expressed as a density functional such as
the Thomas-Fermi functional \cite{thomas1927,fermi1928}
and its improvements has not been particularly
successful. This problem is avoided in the Kohn-Sham DFT \cite{kohn1965}
by directly calculating the kinetic energy using orbitals.  
In contrast, the
Slater local exchange potential can be naturally approximated
in terms of the density, as described in Sec. \ref{sec:Slaterxhole}.
However, the local
exchange potential is not a good approximation to the
nonlocal one, as illustrated by the electron-gas dispersion, for which the
local exchange potential and the nonlocal Fock exchange
give very different results.
The Slater local exchange potential performs much better than the nonlocal
Fock exchange.

\subsection{Equation of motion in orbital basis}

Expressed in a set of base orbitals $\{\varphi_{i}\}$, 
the Green function can be written as
\begin{align}
    G(r,r';t)=\sum_{ij}\varphi_i(r)G_{ij}(t)\varphi_j(r').
\end{align}
In the orbital basis, the equation
of motion in Eq. (\ref{eq:EOMGVxc}) takes the form
\begin{equation}
i\frac{\partial}{\partial t}G_{ij}(t)-\sum_{k}h_{ik}G_{kj}(t)-\sum
_{kl}V_{ik,lj}^\mathrm{xc}(t)G_{kl}(t)=\delta_{ij}\delta(t),
\end{equation}
where $G_{ij}$ and $h_{ik}$ are the matrix elements of $G$ and $h$ in the
orbitals and
\begin{equation}\label{eq:Vxc_ijkl}
V_{ik,lj}^\mathrm{xc}(t)=\int d^{3}rd^{3}r^{\prime}\text{ }\varphi_{i}^{\ast
}(r)\varphi_{k}(r)V_\mathrm{xc}(r,r^{\prime};t)\varphi_{l}^{\ast}(r^{\prime}%
)\varphi_{j}(r^{\prime}).
\end{equation}
The presence of four orbital indices is
indicative of the two-particle bosonic character of $V_\mathrm{xc}$, in
contrast to the self-energy which is fermionic.

\subsection{Spectral representation of $V_\mathrm{xc}$}

Although Vxc is naturally expressed in the time domain,
its spectral properties may provide useful information when constructing
approximations.
To analyse the spectral representation of $V_\mathrm{xc}$, it is
necessary to investigate the spectral representation of $G^{(2)}$. 
This is achieved by inserting in the definition of $G^{(2)}$
a complete set of eigenstates of the $N\pm 1$-system,
\begin{align}
    \hat{H} |\Psi^{N\pm 1}_m\rangle =E^{N\pm 1}_m |\Psi^{N\pm 1}_m\rangle.
\end{align}
For $t>0$
\begin{align}
&iG^{(2)}=\langle \Psi_0 | \hat{\rho}(r''t) \hat{\psi}(r,t)\hat{\psi}^\dagger(r') |\Psi_0\rangle
\nonumber\\
   &= \langle \Psi_0 | e^{i\hat{H}t}\hat{\rho}(r'') \hat{\psi}(r)e^{-i\hat{H}t} |\Psi^{N+1}_m \rangle 
    \langle \Psi^{N+1}_m|\hat{\psi}^\dagger(r')|\Psi_0\rangle
    \nonumber\\
    &=\sum_m\langle \Psi_0 | \hat{\rho}(r'')
     \hat{\psi}(r) |\Psi^{N+1}_m \rangle 
    \langle \Psi^{N+1}_m| \hat{\psi}^\dagger(r')|\Psi_0\rangle 
    \nonumber\\
    &\qquad\qquad \times e^{-i(E^{N+1}_m-E_0)t}.
\end{align}
For $t<0$
\begin{align}
iG^{(2)}&=-\langle \Psi_0 | \hat{\psi}^\dagger(r') \hat{\rho}(r''t) 
\hat{\psi}(r,t)|\Psi_0\rangle
\nonumber\\
&=-\sum_m
\langle \Psi_0 | \hat{\psi}^\dagger(r') |\Psi_m^{N-1}\rangle \langle \Psi_m^{N-1}|
\hat{\rho}(r'') \hat{\psi}(r)|\Psi_0\rangle 
\nonumber\\
&\qquad\qquad\times e^{i(E_m^{N-1}-E_0)t}.
\end{align}
This spectral decomposition shows that $G^{(2)}$ shares the same spectral structure
as $G$ in terms of the excitation energies of the $N\pm 1$ system but with
different spectral weights. Since $G$ is fermionic, it follows that
$G^{(2)}$ is also fermionic.

From the relation
\begin{align}
    V_\mathrm{xc}(r,r';t)G(r,r';t)& = \int dr''\,v(r-r'') G^{(2)}(r,r',r'';t) 
    \nonumber\\
&\qquad    -V_\mathrm{H}(r)G(r,r';t)
\end{align}
it is clear that the spectral property of the left-hand side must be fermionic
since the Coulomb potential $v$ and the Hartree potential $V_\mathrm{H}$ are
static and do not alter the spectral property of $G^{(2)}$ and $G$, which
are both fermionic.

To find out the spectral property of Vxc, it is useful to make comparison
with the self-energy in the $GW$ approximation,
\begin{align}
    \Sigma(r,r';t)= iG(r,r';t)W(r,r';t).
\end{align}
Here, the screened interaction $W$ is bosonic whereas $\Sigma$ is fermionic.
Since $V_\mathrm{xc} G$ is fermionic and $G$ itself
is fermionic, it can be conjectured that analogous to $W$,
$V_\mathrm{xc}$ should be bosonic.
$V_\mathrm{xc}$ consists of static and dynamic terms:
\begin{align}
    V_\mathrm{xc}(r,r';t)= V_\mathrm{xc}^\mathrm{S}(r,r')
    + V_\mathrm{xc}^\mathrm{D}(r,r';t).
\end{align}
Similarly to the dynamic part of the screened interaction $W$
\cite{hedin1969,aryasetiawan1998},
the dynamic part of Vxc has a bosonic spectral representation,
\begin{align}
    -iV_\mathrm{xc}^\mathrm{D}(\omega)=\int_{-\infty}^0 d\omega'
    \frac{B(\omega')}{\omega-\omega'-i\delta}
    +\int_0^\infty d\omega'
    \frac{B(\omega')}{\omega-\omega'+i\delta},
\label{eq:VxcSpec}
\end{align}
where the chemical potential is absent.

\section{Formal expression for the xc hole}

The xc hole is a fundamental quantity,
since its Coulomb potential generates Vxc.
The xc hole is expressible as a linear response of the Green function with
respect to an external probing field $\varphi(rt)$.
This can be understood by considering the two-particle Green function $G^{(2)}$
which can be expressed as
a functional derivative of the Green function with respect to
the probing field $\varphi(rt)$. It is convenient to use the notation
$1=(r_1t_1)$ etc. With this notation, the equation of motion is given by
\begin{align}\label{eq:EOMG12}
    &[i\partial_1 -h_0(1)]G(1,2)
    \nonumber\\
    &+ i\int d3\, v(1-3) 
    \langle T [\hat{\rho}(3)\hat{\psi}(1)\hat{\psi}^\dagger(2)]\rangle
=\delta(1-2).
\end{align}

From the Schwinger functional derivative method
\cite{hedin1969,aryasetiawan1998}
\begin{align}
    G^{(2)}(1,2,3)= \rho(3) G(1,2) +i
    \frac{\delta G(1,2)}{\delta\varphi(3)}.
\end{align}
It follows immediately from the definition of the xc hole in Eq. (\ref{eq:G2})
that \cite{aryasetiawan2022a}
\begin{align}
    \rho_\mathrm{xc}(1,2,3)G(1,2)=i \frac{\delta G(1,2)}{\delta\varphi(3)} 
\end{align}
or more compactly
\begin{align}
    \rho_\mathrm{xc}(1,2,3)=i \frac{\delta}{\delta\varphi(3)} \ln{G(1,2)}.
\end{align}
It is understood that the probing field is set to zero once
the derivative is taken.

The identity
\begin{align}
    GG^{-1}=1 \rightarrow \delta G = -G\delta G^{-1} G
\end{align}
yields
\begin{align}
    \frac{\delta G(1,2)}{\delta\varphi(3)}
    =-\int d4d5\, G(1,4) \frac{\delta G^{-1}(4,5)}{\delta\varphi(3)}G(5,2).
\end{align}
Together with
\begin{align}
    G^{-1}(1,2)& = (i\partial_1 - h(1)-\varphi(1))\delta(1-2)
    -\Sigma(1,2),
\end{align}
which follows from the equation of motion of $G$ in the presence
of the probing field, one arrives at \cite{aryasetiawan2022a,karlsson2023}
\begin{align}
    &\rho_\mathrm{xc}(1,2,3)G(1,2)
    \nonumber\\
    &=i\int d4\,G(1,4)\left\{ \delta(3-4) +\frac{\delta V_\mathrm{H}(4)}{\delta\varphi(3)}
    \right\} G(4,2)
    \nonumber\\
    &\qquad +i\int d4d5\,G(1,4)\frac{\delta \Sigma(4,5)}{\delta\varphi(3)} G(5,2).
\label{eq:xchole}
\end{align}
The first term on the right-hand side, $iG(1,3)G(3,2)$, 
corresponds to the exchange contribution.
The second term involving $\frac{\delta V_\mathrm{H}}{\delta\varphi}$ 
will be referred to as the density-response contribution,
and the last term with $\frac{\delta \Sigma}{\delta\varphi}$ 
as the vertex correction.
The second and third terms together constitute the correlation contribution
and neglecting the third term (vertex correction) amounts to RPA.

The quantity in the curly brackets can be identified as the inverse dielectric
function:
\begin{equation}
    \epsilon^{-1}(4,3)=\delta(4-3) +\frac{\delta V_\mathrm{H}(4)}{\delta\varphi(3)}.
\label{eq:inveps}
\end{equation}
It is convenient to define
\begin{align}
    K(4,3) &= \frac{\delta V_\mathrm{H}(4)}{\delta\varphi(3)} 
=\int d5\, v(4-5) \chi(5,3),
\label{def:K}
\end{align}
where $\chi$ is the linear density-response function
\begin{equation}
    \chi(5,3) = \frac{\delta\rho(5)}{\delta\varphi(3)}.
\label{def:chi}
\end{equation}
and
\begin{align}
    v(1-2)=v(r_1-r_2)\delta(t_1-t_2),
\end{align}
reflecting the instantaneous nature of the Coulomb interaction in the
nonrelativistic case.

\section{Exchange hole}

Considering only the exchange contribution in Eq. (\ref{eq:xchole}) 
yields the exchange hole $\rho_\mathrm{x}$:
\begin{align}
    \rho_\mathrm{x}(1,2,3) G(1,2) =i G(1,3) G(3,2). 
\end{align}
Keeping in mind that $t_2=0$ and $t_1=t_3=t$, due to the
presence of the instantaneous Coulomb interaction $v(1-3)$ in
the equation of motion in Eq. (\ref{eq:EOMG12}), the substitutions
\begin{align}
1\rightarrow (rt),\quad 2\rightarrow (r'0),\quad 3\rightarrow (r''t) ,   
\end{align}
lead to a more explicit expression for the exchange hole:
\begin{align}
    &\rho_\mathrm{x}(r,r',r'';t)G(r,r';t)
    =iG(r,r'';0^-)G(r'',r';t) .
\label{eq:x-hole}
\end{align}
On the right-hand side the rule governing an equal-time Green function,
\begin{align}
    G(rt,r''t)=G(rt,r''t^+)=G(r,r'';t-t^+=0^-),
\end{align}
has been used.
Since
\begin{align}
   iG(r,r;0^-)=-\rho(r), 
\end{align}
the exact constraint in Eq. (\ref{eq:ExactReln}) is
already fulfilled by the exchange hole.
This implies that the correlation hole fulfils
the following condition,
\begin{equation}\label{eq:rhoc=0}
    \rho_\mathrm{c}(r,r',r;t)=0.
\end{equation}

For a noninteracting $G$,
\begin{align}
    iG_0(r,r';t) &=
    -\theta(-t)\sum_k^\mathrm{occ} \phi_k(r)\phi^*_k(r')
    e^{-i\varepsilon_k t}
    \nonumber\\
    &\quad +\theta(t)\sum_k^\mathrm{unocc} \phi_k(r)\phi^*_k(r')
    e^{-i\varepsilon_k t},
\label{eq:G0}
\end{align}
so that for $t<0$
\begin{align}
    i\int dr'' G_0(r,r'';0^-)G_0(r'',r';t<0)
    &=-G_0(r,r';t<0).
\label{eq:G0G0G0}
\end{align}
This result can be shown as follows:
\begin{align}
    &i\int dr'' G_0(r,r'';0^-)G_0(r'',r';t<0)
    \nonumber\\
    &=-i\int dr'' \sum_{k}^\mathrm{occ} \phi_k(r)\phi_k^*(r'')
    \sum_{k'}^\mathrm{occ} \phi_{k'}(r'')\phi_{k'}^*(r')
    e^{-i\varepsilon_{k'}t}
    \nonumber\\
    &=-i\sum_{k}^\mathrm{occ} \phi_k(r)\phi_k^*(r')
    e^{-i\varepsilon_{k}t}
    \nonumber\\
    &=-G_0(r,r';t<0).
\end{align}
For $t>0$
\begin{equation}
    i\int dr'' G_0(r,r'';0^-)G_0(r'',r';t>0)=0
\end{equation}
since in this case $k'>k_\mathrm{F}$.
With $G=G_0$,
it follows from Eq. (\ref{eq:x-hole}) that
the exchange hole fulfils the sum rule:
\begin{align}
    \int dr'' \rho_\mathrm{x}(r,r',r'';t)&= -\theta(-t).
\end{align}

In general, if $G\neq G_0$
\begin{equation}
      -i\int dr'' G(r,r'';0^-)G(r'',r';t<0) \neq G(r,r';t<0)
\end{equation}
and
\begin{equation}
    i\int dr'' G(r,r'';0^-)G(r'',r';t>0)\neq 0.
\end{equation}
This implies that if a renormalised $G$ is used and only the exchange part
is considered,
then in general the sum rule is not fulfilled.

\subsection{Relationship with the Slater exchange hole}
\label{sec:RelnSlaterxhole}

The relationship with the Slater exchange hole can be seen by taking the limit
$t\rightarrow 0^-$, $r'\rightarrow r$, and using the noninteracting
Green function in Eq. (\ref{eq:x-hole}).
Since 
\begin{align}
    iG_0(r,r'';0^-) = -\sum_k^\mathrm{occ}\phi_k(r)\phi_k^*(r'')
\end{align}
one finds
\begin{align}
    \rho_\mathrm{x}(r,r,r'';0^-)=
    -\frac{1}{\rho(r)}
    \sum_{ik}^\mathrm{occ} 
    \phi_k^*(r)\phi_i(r) \phi_i^*(r'') \phi_k(r''),
\end{align}
which is identical with
the Slater exchange hole in Eq. (\ref{eq:Slaterx-hole}).

\section{Correlation hole}

It can be shown that there is no density-response contribution to the sum rule.
Consider
the change in charge density under a perturbation $\delta\varphi$:
\begin{equation}
   \delta\rho(1)=\int d2\, \chi(1,2) \delta\varphi(2),
   \label{eq:deltarho}
\end{equation}
where $\chi$ is the linear density-response function as defined in
Eq. (\ref{def:chi}).
Since a constant perturbation, $\delta\varphi=c$, does not affect
the density, it follows that
\begin{equation}
   \int d2\, \chi(1,2) = 0. 
   \label{eq:chi0}
\end{equation}
The RPA response function fulfils this condition when calculated
using $G_0$, as shown below.

The response function can be expanded in powers of polarisation $P$,
\begin{equation}
    \chi = P+ PvP + ...
\end{equation}
One observes that if the polarisation fulfils the condition
\begin{equation}
    \int d2\, P(1,2) = 0,
    \label{eq:intP0}
\end{equation}
then the response function $\chi$ also fulfils the condition in
Eq. (\ref{eq:chi0}).

The polarisation in RPA is given by
\begin{equation}
    P(r,r';t) = -iG(r,r';t) G(r',r;-t).
    \label{eq:PRPA}
\end{equation}
If a noninteracting $G_0$ in Eq. (\ref{eq:G0}) is used,
Eq. (\ref{eq:intP0}) and consequently
Eq. (\ref{eq:chi0}) are satisfied. This can be shown by considering
%
%
%
\begin{align}
    &G_0(r,r';t)G_0(r',r;-t)
    \nonumber\\
    &=\theta(-t)\sum_{k\leq k_\mathrm{F}} \varphi_k(r)\varphi_k^*(r')
    e^{-i\varepsilon_{k}t} 
\sum_{k'> k_\mathrm{F}} \varphi_{k'}(r')\varphi_{k'}^*(r)
    e^{i\varepsilon_{k'}t}
    \nonumber\\
    &+\theta(t) \sum_{k> k_\mathrm{F}} \varphi_k(r)\varphi_k^*(r')
    e^{-i\varepsilon_{k}t}
    \sum_{k'\leq k_\mathrm{F}} \varphi_{k'}(r')\varphi_{k'}^*(r)
    e^{i\varepsilon_{k'}t}.
\end{align}
When integrating over $r'$ one finds
\begin{equation}
    \int dr' \varphi^*_k(r')\varphi_{k'}(r') = 0,
    \label{eq:phiphi}
\end{equation}
since $\varphi_k$ and $\varphi_{k'}$ are occupied and unoccupied, 
respectively, or vice versa.
Hence, the condition in Eq. (\ref{eq:intP0}) is fulfilled.

However, it is not immediately evident that the conditions
in Eq. (\ref{eq:intP0}) and
Eq. (\ref{eq:chi0}) are satisfied when
a renormalised $G$ is used.
To see this,
consider
\begin{align}
    G(r,r';t)=\int \frac{d\omega}{2\pi} e^{-i\omega t} G(r,r';\omega).
\end{align}
Using the spectral representation,
\begin{align}
    G(r,r';\omega)=\int_{-\infty}^\mu d\omega' \frac{A(r,r';\omega')}{\omega-\omega'-i\eta}
    +\int_\mu^\infty \frac{A(r,r';\omega')}{\omega-\omega'+i\eta},
\end{align}
where $\mu$ is the chemical potential and
\begin{align}
    A(r,r';\omega) = -\frac{1}{\pi} \mathrm{sign}(\omega-\mu) \mathrm{Im}G(r,r';\omega),
\end{align}
one arrives at
\begin{align}
    & \int dr'dt' P(r,r';t-t') 
    \nonumber\\
    &= -i\int dr'dt'G(r,r';t-t') G(r',r;t'-t)
     \nonumber\\
     &=\int_{-\infty}^\mu d\omega_1 \int_\mu^\infty d\omega_2 \int dr' 
     \nonumber\\
     &\quad\left\{\frac{A(r,r';\omega_1)A(r',r;\omega_2)+A(r,r';\omega_2)A(r',r;\omega_1)}
     {\omega_1-\omega_2} \right\}.
     \label{eq:Pconst}
\end{align}
The spectral function for a noninteracting $G_0$ is given by
\begin{align}
    A_0(r,r';\omega)=\sum_k \varphi_k(r)\varphi^*_k(r') \delta(\omega-\varepsilon_k).
\end{align}
%
It is quite evident that the integral over $r'$ is zero since
it leads to integrals
between occupied and unoccupied orbitals, as
in Eq. (\ref{eq:phiphi}). 
However, it is not clear if this holds for an interacting $G$.
If the integral in Eq. (\ref{eq:Pconst}) is nonzero then
the sum rule is not fulfilled by
the exchange and density-response terms individually.
It is not impossible that the sum of these two terms would fulfill the sum rule,
but this seems unlikely.
This conjecture might explain the known fact that
a self-consistent $GW$ yields poor spectral functions \cite{holm1998}.
When a renormalised $G$ is used, inclusion of the vertex
seems necessary to preserve the sum rule.

\section{Spherical average of the xc hole}
\label{sec:LDA}

The fact that the Coulomb interaction depends only on the
spatial separation between two point charges has a very important
consequence for the xc hole:
only the spherical average
of the xc hole is needed to determine Vxc.
This follows from the well-known result
in DFT, which partially explains the success of
LDA \cite{gunnarsson1976}.

The change of variable
$\mathbf{R}=\mathbf{r}''-\mathbf{r}$ in Eq. (\ref{eq:phixc}) leads to
\begin{align}
    V_\mathrm{xc}(r,r';t) &=\sum_{\sigma_R}\int d^3R\,v(R) 
    \rho_\mathrm{xc}(r,r',\mathbf{r}+\mathbf{R};t)
    \nonumber\\
    &=\int dR\,R\, \overline{\rho}_\mathrm{xc}(r,r',R;t),
\label{eq:phixcR}
\end{align}
where 
\begin{align}
    \overline{\rho}_\mathrm{xc}(r,r',R;t)=\sum_{\sigma_R}
    \int d\Omega_R \,\rho_\mathrm{xc}(r,r',\mathbf{r}+\mathbf{R};t)
\label{eq:SphAvRhoxc}
\end{align}
is the spherical average of the exchange-correlation hole
for given $r$, $r'$, and $t$.
This result is very appealing, since it implies that
the many-electron problem of determining the Green
function is reduced to finding a spherical charge distribution,
and only its first radial moment is required.
The second radial moment is known since it corresponds to the sum rule.
This means that there are an infinite number of xc
holes that yield the exact Vxc since all radial moments
higher than two need not be exact.
This property is a consequence of the Coulomb interaction being dependent
only on the spatial separation.
This property is not utilised in the
self-energy approach in which the Coulomb interaction is 
lumped into the definition of the self-energy as can be seen in
Eq. (\ref{eq:SigmaTrad}).

\subsection{Local-density approximation (LDA)}

The reasoning by Slater described in Sec. \ref{sec:Slaterxhole}
that leads to the LDA
for the exchange potential can be readily carried over to Vxc.
As in the case of the static exchange hole,
the sum rule and the exact constraint $\rho_\mathrm{xc}(r,r',r;t)=-\rho(r)$
have a very important implication, namely
Vxc behaves approximately as $\rho^{1/3}(r)$,
independent of $r'$ and $t$. 
Following Slater \cite{slater1968}, if the xc
hole is approximated by a sphere of radius $R_0$ and
assumed to be constant, one finds
that the Coulomb potential due to the exchange hole at $r$, the centre of the
sphere, is proportional to $1/R_0$ and hence $V_\text{xc}$ is proportional to
$\rho(r)^{1/3}$ in the lowest approximation, as in Eq. (\ref{eq:SlaterVx}).
It must be emphasised again that
the $\rho(r)^{1/3}$-dependence
does not rely on the homogeneous electron gas since it is
solely based on the exact
constraint which is valid in general. 
The dependence of Vxc
on local density may be regarded as a manifestation of nearsightedness
advocated by Walter Kohn \cite{kohn1996,prodan2005,prodan2006}.

As will be described in a later section, the xc hole and potential
of the homogeneous electron gas can be calculated within RPA.
The results can be parametrised as a function of density and applied
within LDA to real systems. For the homogeneous electron gas (HEG)
the spherical average of the xc hole of a given spin $\sigma$ and density
$\rho$ has the form
\begin{align}
    \overline{\rho}_\mathrm{xc}(r,r',R;t)= 
    \overline{\rho}_\mathrm{xc}(\rho,|\mathbf{r}'-\mathbf{r}|,R;t).
\end{align}
A possible LDA for the xc hole is
\begin{align}
    \overline{\rho}^\mathrm{LDA}_\mathrm{xc}(r,r',R;t)=
    \overline{\rho}^\mathrm{HEG}_\mathrm{xc}
    (\rho(r),|\mathbf{r}'-\mathbf{r}|,R;t),
\end{align}
where $\rho(r)$ is the local density at $r$.

A more direct route is to apply LDA on Vxc directly:
\begin{align}
    V^\mathrm{LDA}_\mathrm{xc}(r,r';t)=V^\mathrm{HEG}_\mathrm{xc}
    (\rho(r),|\mathbf{r}'-\mathbf{r}|;t).
\end{align}
This approximation could be improved by taking into account the density
dependence at $r'$ as in the weighted LDA \cite{gunnarsson1979,alonso1978}.

\section{Connection with Kohn-Sham $V_\mathrm{xc}$}
\label{sec:ConnectionKS}

To establish the connection between the Kohn-Sham (KS) xc potential
$V^\mathrm{KS}_\mathrm{xc}$
and the dynamical xc field consider the equations of motion satisfied by
the two corresponding Green functions, $G^\mathrm{KS}$ and $G$.
From these two equations,
\begin{align}
    &\left(  i\frac{\partial}{\partial t}-h(r\mathbf{)}
    -V_\mathrm{xc}(r,r^{\prime};t)\right)  
    [G(r,r^{\prime};t)-G^\mathrm{KS}(r,r^{\prime};t)]
    \nonumber\\
    &=[V_\mathrm{xc}(r,r^{\prime};t)
       -V^\mathrm{KS}_\mathrm{xc}(r)] G^\mathrm{KS}(r,r^{\prime};t).
\end{align}
Using the property that the diagonal components of both
$G^\mathrm{KS}$ and $G$ give the ground-state density \cite{sham1983}, and
evaluating the equation with $r'\rightarrow r$ and $t\rightarrow 0^-$
yields the relationship between
$V_\mathrm{xc}(r,r;0^-)$ and $V^\mathrm{KS}_\mathrm{xc}(r)$:
\begin{align} \label{id0}
    &\lim_{r'\rightarrow r,t\rightarrow 0^-}\left(  \frac{\partial}{\partial t}-\frac{i}{2}\nabla^2\right)  [G(r,r^{\prime};t)-G^\mathrm{KS}(r,r^{\prime};t)]
    \nonumber\\
    &=[V_\mathrm{xc}(r,r;0^-)-V^\mathrm{KS}_\mathrm{xc}(r)] \rho(r).
   \end{align}
There does not seem to be any immediate reason that the left-hand side
vanishes so that under this assumption in general 
$V_\mathrm{xc}(r,r;0^-) \neq V^\mathrm{KS}_\mathrm{xc}(r)$. 
The first term on the left-hand side with time derivative
can be understood as the difference
in the first moment of the occupied densities of states
after integration over $r$.
The second term is the difference in the kinetic energies of the true
interacting and noninteracting systems,
which is contained in the Kohn-Sham $E_\mathrm{xc}$.

Note that
\begin{equation}
g(r,r,r'';t=0^-) \neq g^\mathrm{KS}(r,r''),
\end{equation}
since $g^\mathrm{KS}$ is defined as the average of $g_\lambda$, which is the
static correlation function corresponding to a scaled Coulomb interaction,
$\lambda v$ \cite{gunnarsson1976,langreth1975}.

The xc energy in the Kohn-Sham scheme is given by
\begin{equation}
    E^\mathrm{KS}_\mathrm{xc}= \frac{1}{2}\int dr dr' \rho(r) v(r-r')\rho_\mathrm{xc}^\mathrm{KS}(r,r'),
\end{equation}
where
\begin{equation}
    \rho_\mathrm{xc}^\mathrm{KS}(r,r') = [g^\mathrm{KS}(r,r')-1]\rho(r')
\end{equation}
is the static Kohn-Sham xc hole.
It can then be seen that the Kohn-Sham xc potential is given by
\begin{align}
    V^\mathrm{KS}_\mathrm{xc}(r)
    &=\frac{\delta E^\mathrm{KS}_\mathrm{xc}}{\delta \rho(r)}
    \nonumber\\
    &=\int dr' v(r-r')\rho_\mathrm{xc}^\mathrm{KS}(r,r') 
    \nonumber\\
    &+\frac{1}{2}\int dr' dr'' \rho(r') v(r'-r'')
    \frac{\delta g^\mathrm{KS}(r',r'')}{\delta \rho(r)}\rho(r'').
\end{align}
Thus, in addition to the Coulomb potential of the xc hole,
there is an additional
contribution arising from the dependence of the distribution function
$g^\mathrm{KS}$ on the density. This is in contrast to Vxc, which is purely
the Coulomb potential of the xc hole.
Since the xc hole
in finite systems such as atoms and molecules integrates to $-1$,
Vxc behaves asymptotically as $-\frac{1}{r}$ automatically. On the other hand,
approximate Kohn-Sham xc potentials can violate this asymptotic behaviour
\cite{vanLeeuwen1994}
even though the xc hole fulfils the sum rule, due to
the presence of $\delta g^\mathrm{KS}/\delta\rho$, which can give a spurious
asymptotic contribution.

It is well-known that
band gaps in semiconductors and insulators are systematically underestimated
within Kohn-Sham DFT.
The presence of a derivative discontinuity of the xc potential is usually
invoked to explain this underestimation \cite{perdew1982}.
The Vxc formalism offers a natural explanation for this underestimation.
The xc field arises as the Coulomb potential of the xc hole, which integrates
to $-1$ for the hole and $0$ for the electron. This implies that the
xc field for the hole (occupied state)
should be more negative than that for the electron (unoccupied state),
whereas in Kohn-Sham DFT, both hole and electron experience the same potential.
The Kohn-Sham xc potential is designed for occupied states only to reproduce
the ground-state density.

\section{Connection with the self-energy}

It is quite evident from comparison of the equations of motion in
Eqs. (\ref{eq:EOMGSigma})
and (\ref{eq:EOMGVxc}) 
that Vxc
and the self-energy are related according to
\begin{align}
    V_\mathrm{xc}(1,2)G(1,2) =\int d3\, \Sigma(1,3) G(3,2).
\label{eq:phixcSigma}
\end{align}
However, this relation by itself is too general and
rather superficial since it appears as a mere redefinition of the self-energy. 
It is more insightful to consider
the relationship between
the xc hole and the density response of the system
and how this relationship gives rise to Eq. (\ref{eq:phixcSigma}).

From Eq. (\ref{eq:xchole}) with the vertex $\delta\Sigma/\delta\varphi$ neglected
and using the definition of the inverse dielectric function
in Eq. (\ref{eq:inveps}) one finds
\begin{align}
    &\int d3\,v(1-3) \rho_\mathrm{xc}(1,2,3)G(1,2)
    \nonumber\\
    &=i\int d3\,v(1-3) 
    \int d4\,G(1,4) \epsilon^{-1}(4,3) G(4,2). 
\end{align}
The quantity $W=\epsilon^{-1}v$ is the screened interaction, and one obtains
\begin{align}
    V_\mathrm{xc}(1,2)G(1,2)&= i\int d4\,G(1,4)W(4,1) G(4,2)
    \nonumber\\
    &=\int d4\,\Sigma_{GW}(1,4)G(4,2),
\end{align}
which yields Vxc corresponding to
the $GW$ approximation for the self-energy.

\subsection{Dyson-like equation}

The equation of motion of the Hartree Green function $G^{\mathrm{H}}$
is given by
\begin{align}
    \left[ i\partial_t -h(r)
    \right] G^\mathrm{H}(r,r';t) 
    =\delta(r-r')\delta(t).
\end{align}
It then follows from the equation of motion in Eq. (\ref{eq:EOMGVxc}) that
\begin{align}
    &G(r,r';t)= G^\mathrm{H}(r,r';t) 
    \nonumber\\
    &+\int dr''dt' G^\mathrm{H}(r,r'';t-t')
    V_\mathrm{xc}(r'',r';t')G(r'',r';t'),
\end{align}
which can be verified by operating $i\partial_t-h(r)$ on both sides
of the equation. The above equation may be
regarded as the analogue of the Dyson equation in the self-energy
framework. If the Kohn-Sham Green function, $G^{\mathrm{KS}}$, which satisfies
\begin{align}
    \left[ i\partial_t -h(r) - V^\mathrm{KS}_\mathrm{xc}(r)
    \right] G^\mathrm{KS}(r,r';t) 
    =\delta(r-r')\delta(t),
\end{align}
is used as a reference instead of the Hartree Green function, 
the same Dyson-like
equation is fulfilled with $G^\mathrm{H}$ replaced by $G^{\mathrm{KS}}$
and $V_\mathrm{xc}$ replaced by $V_\mathrm{xc}- V_\mathrm{xc}^{\mathrm{KS}}$.

The Dyson equation in the
Vxc formalism is difficult to solve, since it is
not in the form of a matrix equation. This is in contrast
to the self-energy approach, in which the Dyson equation is in matrix form.
Due to its local nature in both space and time,
the Vxc formalism is well suited for solving the
Green function by point-wise propagation
in time of the equation of motion. 
On the other hand, the Green function in the
self-energy approach is difficult to propagate since the
self-energy acts as a time convolution on the Green function.
Knowledge of the Green function for all time is a prerequisite
before it can be propagated.

\section{Total spectral function}

The Fourier transform of the Green function into the frequency domain
provides information on the momentum-resolved spectra.
For many purposes, it is often sufficient to know
the total spectral function corresponding to the trace of the
Green function. 
In this case, one can take advantage of the local nature of Vxc
and as shown below only its diagonal component is needed.
A related approach is the work by Gatti \emph{et al.} \cite{gatt12007}
who introduced
an effective potential, local in space but energy dependent, defined
to reproduce the diagonal component of the Green function
from which the total spectral function can be calculated.
Another related approach is the work of
Savrasov and Kotliar \cite{savrasov2004}, who
introduced the concept of spectral-density functional theory, in which
the key variable is given by the local Green function rather than the
electron density.

\subsection{Temporal density}

The temporal density is defined as
the diagonal component of the Green function:
\begin{align}
    \rho(r,t)&=-iG(r,r;t).
\end{align}
For $t=0^-$ it is equal to the electron density.
%
%
The Fourier transform of the integral over space gives
the spectral function or density of states:
\begin{equation}
    \rho(\omega)=\frac{1}{\Omega}\int dr\int dt\, e^{i\omega t}\rho(r,t).
\end{equation}
The nomenclature "temporal density"
arises from its interpretation as the probability
amplitude that a hole or an electron created at $r$ 
is annihilated at the same position at a later time $t$. 

Note that the temporal density is \emph{not} the same as
the density fluctuation arising from the addition of a hole
or an electron \cite{aryasetiawan2023}.
Consider an electron addition ($t>0$) at $r$.
The density corresponding to the time-evolved state,
\begin{align}
    |\Psi(r,t)\rangle&= e^{-i\hat{H}t}\hat{\psi}^\dagger(r) |\Psi_0\rangle,
\end{align}
is given by 
\begin{align}
    \rho_\mathrm{fluc}(r,t)&= \langle \Psi(r,t)|\hat{\rho}(r) |\Psi(r,t)\rangle 
    \nonumber\\
    &= \langle \Psi_0|\hat{\psi}(r) e^{i\hat{H}t}
    \hat{\rho}(r) e^{-i\hat{H}t}\hat{\psi}^\dagger(r)|\Psi_0\rangle.
\end{align}
This density fluctuation is real whereas the temporal density 
\begin{align}
    \rho(r,t)&=-\langle \Psi_0 | \hat{\psi}(rt)\hat{\psi}^\dagger(r) |\Psi_0\rangle
     \nonumber\\
     &= -e^{iE_0t}\langle \Psi_0 | \hat{\psi}(r)e^{-i\hat{H}t} 
     \hat{\psi}^\dagger(r)|\Psi_0\rangle
\end{align}
is generally complex.
The temporal density is also different from the time-dependent density in
TDDFT \cite{runge1984,maitra2016}, which corresponds to the real density
that varies in time when an external time-dependent field is applied.
In the case of the temporal density,
the addition and subsequent
annihilation of an electron or a hole is performed on a system initially in its
ground state in the absence of a time-dependent external field.

Consider now the equation of motion for the Green function
in Eq. (\ref{eq:EOMGVxc}) calculated at $r'=r$ and $t\neq 0$ \cite{aryasetiawan2023}:
\begin{align}
   & \left[ i\partial_t -V_\mathrm{MF}(r)
    -V_\mathrm{xc}(r,t) \right]\rho(r,t)
  \nonumber\\
    &\qquad 
    -\left.\frac{i}{2} \nabla^2 G(r,r';t)\right|_{r'=r}=0,
\end{align}
where
\begin{align}
    V_\mathrm{xc}(r,t):=V_\mathrm{xc}(r,r;t)
\end{align}
and
\begin{align}
    V_\mathrm{MF}(r)=V_\mathrm{ext}(r)+V_\mathrm{H}(r).
\end{align}
For each $r'$ the temporal current density is defined according to
\begin{align}
    \mathbf{j}(r,r';t):= -\frac{1}{2} \nabla G(r,r';t).
\end{align}
Note that the temporal current density should be regarded
as a construct and does not necessarily correspond
to a physical current density.
The equation of motion becomes
\begin{align}
    \partial_t\rho(r,t) + \nabla \cdot \mathbf{j}(r,t)=S(r,t),
    \label{eq:continuity}
\end{align}
where
\begin{align}\label{eq:divj}
   \nabla \cdot \mathbf{j}(r,t)
   =\left. \nabla \cdot \mathbf{j}(r,r';t)\right|_{r'=r}
   =-\left.\frac{1}{2} \nabla^2 G(r,r';t)\right|_{r'=r}
\end{align}
and
\begin{align}
    S(r,t)= -i\left[V_\mathrm{MF}(r) +V_\mathrm{xc}(r,t)\right]\rho(r,t).
\end{align}

The divergence
of the temporal current density can be illustrated more concretely
by expanding $G(r,r';t)$
in a complete set of orbitals $\{\varphi_i\}$,
\begin{align}
    G(r,r';t)=\sum_{ij} \varphi_i(r) G_{ij}(t) \varphi^*_j(r').
\end{align}
The divergence of the temporal current density for a given $r$ is
then given by
\begin{align}
    \nabla \cdot\mathbf{j}(r,t)=-\frac{1}{2} \sum_{i} 
    \left[\nabla^2\varphi_i(r) \right]\psi^*_i(r,t),
\end{align}
where
\begin{align}
    \psi^*_i(r,t)= \sum_{j} G_{ij}(t) \varphi^*_j(r).
\end{align}

The equation of motion, Eq. (\ref{eq:continuity}), 
has the form of the continuity equation in electrodynamics with a source/sink
term $S$ on the right-hand side. 
Eq. (\ref{eq:divj})
implies that only the diagonal component, $G(r,r;t)$,
and the neighbouring components, $G(r\pm \delta r,r;t)$, are needed to calculate
the divergence of the temporal current density.
This is substantially less information than that
needed to calculate the momentum-resolved spectral function.
The temporal current density is not explicitly required; only its divergence
is relevant.
Note that no auxiliary system has been introduced,
and all quantities are defined in terms of exact quantities
whose existence is guaranteed.

\subsection{Practical scheme \emph{à la} Kohn-Sham}

It is useful to introduce a kinetic potential $V_\mathrm{K}$ defined as
\cite{aryasetiawan2023}
\begin{align}
    V_\mathrm{K}(r,t):= -i\frac{\nabla \cdot \mathbf{j}(r,t)}{\rho(r,t)}.
\end{align}
With this definition, the continuity equation can be rewritten as
\begin{align}
   i \partial_t \ln{\rho(r,t)} = V_\mathrm{MF}(r) +V_\mathrm{xc}(r,t)+V_\mathrm{K}(r,t)
\end{align}
with the formal solution
\begin{align}
    \rho(r,t)&=\rho(r,0^\pm) \exp{\left[-iV_\mathrm{MF}(r) t\right]}
    \nonumber\\
    &\times\exp{\left\{-i\int_0^t dt'
    \left[ {V}_\mathrm{xc}(r,t') +V_\mathrm{K}(r,t')
    \right] \right\}}.
    \label{eq:formalsoln}
\end{align}
Alternatively,
\begin{align}
    \rho(r,t)=\rho(r,0^\pm)+\int_0^t dt'\left[ S(r,t')-\nabla \cdot \mathbf{j}(r,t')\right].
\end{align}
$\rho(r,0^\pm)$ is the initial electron or hole density obtained
from, for example, density-functional calculation. 
To practically solve for $\rho(r,t)$
approximations for $V_\mathrm{xc}$ and $V_\mathrm{K}$ are required.
$V_\mathrm{xc}$ is expected to be amenable to
a local-density type of approximation, but the kinetic potential $V_\mathrm{K}$ 
is associated with the kinetic energy.
Sufficiently accurate approximations for the kinetic energy as an explicit
functional of the electron density are not yet available.

A practical scheme analogous to the
Kohn-Sham DFT
\cite{kohn1965,jones1989,becke2014,jones2015}
can be constructed by defining $\Delta V_\mathrm{K}$ as
\begin{align}
\Delta V_\mathrm{K} = -i\left\{ \frac{\nabla \cdot \mathbf{j}}{\rho} 
-\frac{\nabla \cdot \mathbf{j}^\mathrm{KS}}{\rho^\mathrm{KS}}\right\}
= V_\mathrm{K}-V_\mathrm{K}^\mathrm{KS}.
\end{align}
Here, $\rho^\mathrm{KS}$ and $\mathbf{j}^\mathrm{KS}$ are, respectively,
the temporal density and the
temporal current density associated with
the Kohn-Sham Green function.
With this definition, the continuity equation becomes
\begin{align}
    \partial_t\rho(r,t) + \frac{\nabla \cdot \mathbf{j}^\mathrm{KS}(r,t)}
                               {\rho^\mathrm{KS}(r,t)}
     \rho(r,t)=\widetilde{S}(r,t),
    \label{eq:continuity1}
\end{align}
where
\begin{align}
    \widetilde{S}(r,t)= -i\left[V_\mathrm{MF}(r) +\widetilde{V}_\mathrm{xc}(r,t)\right]\rho(r,t),
\end{align}
\begin{align}
    \widetilde{V}_\mathrm{xc} = {V}_\mathrm{xc} + \Delta V_\mathrm{K}.
\end{align}
The difference in kinetic potentials is incorporated into the xc field.
%
%
The formal solution given by
\begin{align}
    \rho(r,t)&=\rho(r,0^\pm) \exp{\left[-iV_\mathrm{MF}(r) t\right]}
    \nonumber\\
    &\times\exp{\left\{-i\int_0^t dt'
    \left[ \widetilde{V}_\mathrm{xc}(r,t') + V_\mathrm{K}^\mathrm{KS}(r,t')
    \right] \right\}}.
\end{align}
%
Note that
Eq. (\ref{eq:continuity1}) 
is the equation of motion of the \emph{true} interacting
temporal density.
Unlike the Kohn-Sham scheme, there is no auxiliary system. 

\section{Total energy: Relationship to Galitskii-Migdal formula}

The Galitskii-Migdal formula provides a way to calculate from the Green function
the total ground-state energy, and
is given by \cite{fetter-walecka}
\begin{align}
    E_0=-\frac{i}{2}\int dr \lim_{r'\rightarrow r}\lim_{t\rightarrow 0^-}
    \left( i\partial_t + h_0(r) \right) G(r,r';t),
\end{align}
where $h_0(r)=-\frac{1}{2}\nabla^2 +V_\mathrm{ext}(r)$.
Within the Vxc formalism the time derivative
of the Green function can be calculated from the equation of motion:
\begin{align}
&\left.i\partial_t G(r,r';t)\right|_{t=0^-} 
\nonumber\\
&=\left( h_0(r) + V_\mathrm{H}(r) + V_\mathrm{xc}(r,r';0^-)\right)
G(r,r';0^-).
\end{align}
Substituting the above expression into the Galitskii-Migdal formula results in
\begin{align}
    E_0&=-\frac{i}{2}\int dr \lim_{r'\rightarrow r}
    \left( 2h_0(r) + V_\mathrm{H}(r) + V_\mathrm{xc}(r,r';0^-)\right)
    \nonumber\\
&\qquad\qquad\qquad\qquad\qquad\qquad\times G(r,r':0^-)
\nonumber\\
&=-\frac{1}{2} \int dr \,\lim_{r'\rightarrow r}
\nabla^2 \Gamma^{(1)}(r,r')
+\int dr\,V_\mathrm{ext}(r) \rho(r)
\nonumber\\
&\qquad +\frac{1}{2}\int dr 
\left( V_\mathrm{H}(r) + V_\mathrm{xc}(r,r;0^-)\right)\rho(r),
\label{eq:E0}
\end{align}
where
\begin{align}
    \Gamma^{(1)}(r,r')=-iG(r,r';0^-)
\end{align}
is the one-particle reduced density matrix whose diagonal element is
the ground-state density.
One notes that
\begin{align}
    E_\mathrm{H}=\frac{1}{2}\int dr \, V_\mathrm{H}(r) \rho(r)
    =\frac{1}{2}\int drdr' \rho(r)v(r-r')\rho(r'), 
\end{align}
is the Hartree energy. The last term in $E_0$ containing $V_\mathrm{xc}$
is the xc energy,
\begin{align}
   E_\mathrm{xc}&=\frac{1}{2}\int dr \, V_\mathrm{xc}(r,r;0^-) \rho(r)
   \nonumber\\
   &=\frac{1}{2}\int drdr'' \rho(r) v(r-r'')\rho_\mathrm{xc}(r,r,r'';0^-).
\end{align}
$\rho_\mathrm{xc}(r,r,r'';0^-)$ is the static
xc hole in many-body theory. It is different from the xc hole in Kohn-Sham DFT
since the kinetic energy in Eq. (\ref{eq:E0}) is that of the interacting
system rather than the noninteracting Kohn-Sham system.
The above xc energy does not incorporate
the contribution from the kinetic energy.

The formula in Eq. (\ref{eq:E0})
offers an interesting possibility of calculating the ground-state energy
using the variational principle. The reduced density matrix
$\Gamma^{(1)}$ can be diagonalised and
expressed in terms of its eigenfunctions (natural orbitals) according to
\begin{align}
    \Gamma^{(1)}(r,r')=\sum_i n_i \phi_i(r)\phi_i^*(r'),
\end{align}
where $\phi_i$ is an eigenfunction with eigenvalue $n_i$: 
\begin{align}
    \int dr' \,\Gamma^{(1)}(r,r')\phi_i(r') =n_i\phi_i(r).
\end{align}
The eigenfunctions are orthonormal, and the eigenvalues fulfil the
conditions
\begin{align}
    \sum_i n_i = N,\quad 0\leq n_i \leq 1,
\label{eq:Nconditions}
\end{align}
where $N$ is the number of electrons in the system. The ground-state
density is given by
\begin{align}
    \rho(r)=\Gamma^{(1)}(r,r)=\sum_i n_i |\phi_i(r)|^2.
\end{align}
Note that the sum is over all orbitals, not restricted to the lowest $N$
as in the case of the Kohn-Sham system.

The ground-state energy can be regarded as a functional of the occupation
numbers $\{n_i=cos^2\theta_i\}$ and natural orbitals $\{\phi^*_i,\phi_i\}$.
Varying with respect to these parameters under the conditions in
Eq. (\ref{eq:Nconditions}) and the requirement of orthonormality of the orbitals
would lead to a set of coupled equations, which must be solved self-consistently.
This approach is known as reduced density matrix functional theory (RDMFT)
\cite{lathiotakis2005,lathiotakis2007,sharma2008,gibney2022},
regarded as
a promising way to improve the standard Kohn-Sham DFT.
In this approach, the basic variable is the one-particle reduced density matrix,
and the corresponding Hohenberg-Kohn theorem is proven by Gilbert
\cite{gilbert1975}.
There is no auxiliary system, and the kinetic energy
corresponds to the true interacting one, as indicated earlier.

\section{Quasiparticle wave function in the Vxc formalism}
\label{sec:QPWF}

In his famous phenomenological Fermi liquid theory from the mid 1950s,
Landau introduced the concept of a quasiparticle
\cite{landau1957a,landau1957b,fetter-walecka}. Landau quasiparticles are
restricted to those long-lived excitations at the Fermi level.
This concept was generalised in a later
development of Green function theory to include quasiparticles away from
the Fermi level.
In metals a quasiparticle at the Fermi level does indeed have
an infinite lifetime as predicted by Landau, but away from the Fermi level
it acquires a finite lifetime and decays with time.
This phenomenon is observed in
angle-resolved photoemission spectra of a large number of materials,
which typically exhibit a main quasiparticle peak
close to the chemical potential. In addition, there are usually
incoherent features (satellites) at higher binding energies arising from
the coupling of electrons to collective excitations such as plasmons.
In magnetic systems the electrons can be coupled to spin excitations
or magnons giving rise
to features at low binding energies such as kinks in the band dispersion.

Traditionally, the quasiparticle is defined as an eigenfunction of
the quasiparticle equation \cite{hedin1969} involving the self-energy $\Sigma$:
\begin{align}\label{eq:QPtrad}
   & \left[-\frac{1}{2}\nabla^2 + V_\mathrm{H}(r) +V_\mathrm{ext}(r)
    \right] \Psi_k(r,E_k)
    \nonumber\\
    &+ \int dr' \Sigma(r,r';E_k) \Psi_k(r',E_k)
    = E_k \Psi_k(r,E_k).
\end{align}
Since the Hamiltonian is not Hermitian, quasiparticles with different
energies are not in general orthonormal.
In the language of self-energy, the lifetime is inversely
proportional to the imaginary part of the self-energy. 
The many-electron system in Landau's phenomenological theory is qualitatively
viewed as a set of quasiparticles interacting with residual interactions.

\subsection{Quasiparticle wave function}

A different concept of a quasiparticle wave function can be constructed
from the definition of the Green function \cite{aryasetiawan2025}.
In contrast to the traditional one defined in Eq. (\ref{eq:QPtrad}),
the proposed quasiparticle wave function decays in time and contains not only
the quasiparticle mode, but also other modes arising from the coupling
to collective excitations such as plasmons. 

Consider expanding the Green function in a complete set of
orbitals $\{\varphi_k\}$:
\begin{align}
    G(r,r';t) &= \sum_{kk'} \varphi_{k}(r)G_{kk'}(t)\varphi^*_{k'}(r')
    \nonumber\\
    &= \sum_k \varphi_k(r) \psi^*_k(r',t),
\label{eq:GQP}
\end{align}
where
\begin{align}
    \psi^*_k(r',t)=\sum_{k'} G_{kk'}(t)\varphi^*_{k'}(r').
    \label{eq:QPwf}
\end{align}
It will be shown that $\sum_{k'}|G_{kk'}(t)|^2\leq 1$,
so that $G_{kk'}(t)$ can be regarded as a coefficient of
expansion of $\psi^*_k$ in orbital $\varphi^*_{k'}$.
This leads to the interpretation of $\psi^*_k$ as a quasiparticle wave function.

For a noninteracting system, $G_{kk'}(t)=G^0_{k}(t)\delta_{kk'}$ so that
\begin{align}
    G^0(r,r';t) &= \sum_{k} \varphi_{k}(r)G^0_k(t)\varphi^*_{k}(r'),
\end{align}
where
\begin{align}
    G^0_k(t) = i \theta(-t) \theta(\mu-\varepsilon_k)
                   e^{-i\varepsilon_k t} 
            - i\theta(t) \theta(\varepsilon_k-\mu)
                   e^{-i\varepsilon_k t}.
\end{align}
Here, $\varepsilon_k$ is the eigenvalue corresponding to orbital $\varphi_k$.
The noninteracting quasiparticle wave function is then
\begin{align}
    \psi^{0*}_k(r,t)=G^0_k(t)\varphi^*_{k}(r).
\end{align}
It is quite evident that
\begin{align}
    |G^0_k(t)|^2=1,
\end{align}
which implies that the noninteracting quasiparticle does not
decay, as expected.

We wish to prove that
\begin{align}
    \sum_{k'} |G_{kk'}(t)|^2 \leq 1.
\end{align}
Consider the case $t<0$
\begin{align}
    |G_{kk}(t)|^2 &= \left|
    \langle \Psi_0 | \hat{c}_k^\dagger \hat{c}_k(t) | \Psi_0\rangle \right|^2
    \nonumber\\
    &= \left|
    \langle \Psi_0 | \hat{c}_k^\dagger e^{i\hat{H}t}\hat{c}_k | \Psi_0\rangle
    \right|^2
\end{align}
Let $|\phi_k \rangle = \hat{c}_k | \Psi_0\rangle$. 
For fermions, the occupation number of a given orbital cannot exceed 1
so that
\begin{align}
    \langle \phi_k |\phi_k\rangle = \langle \Psi_0 |\hat{n}_k |\Psi_0\rangle \leq 1.
\end{align}
Since the operator $e^{i\hat{H}t}$ is unitary,
\begin{align}
    |G_{kk}(t)|^2 = \left|
    \langle \phi_k |e^{i\hat{H}t} | \phi_k\rangle
    \right|^2 \leq 1.
\end{align}
This follows from the fact that the operator $e^{i\hat{H}t}$ is unitary.
The overlap of the rotated state $|e^{i\hat{H}t} \phi_k\rangle$ with
$| \phi_k\rangle$ must be less than or equal to
$|\langle \phi_k |\phi_k\rangle|$.

For a given $t$ let $S$ be the unitary matrix that diagonalises $G$:
\begin{align}
    G_{kk'}(t) = \sum_{k_1} S_{kk_1} \widetilde{G}_{k_1}(t) S^\dagger_{k_1k'},
\end{align}
\begin{align}
    \chi_{k}(r) =\sum_{k'} \varphi_{k'}(r) S_{k'k},
\end{align}
\begin{align}
    G(r,r';t) &= \sum_{k} \chi_{k}(r)\widetilde{G}_{k}(t)\chi^*_{k}(r').
\end{align}
One finds
\begin{align}
   \sum_{k'} |G_{kk'}(t)|^2 &= \sum_{k'} \sum_{k_1k_2}
   S_{kk_1} \widetilde{G}_{k_1}(t) S^\dagger_{k_1k'} S_{k'k_2}
   \widetilde{G}^*_{k_2}(t) S^\dagger_{k_2k}
   \nonumber\\
   &=\sum_{k_1}S_{kk_1} |\widetilde{G}_{k_1}(t)|^2 S^\dagger_{k_1k}.
\end{align}
Since $|\widetilde{G}_{k_1}(t)|^2 \leq 1$ it follows that
\begin{align}
    \sum_{k'} |G_{kk'}(t)|^2 \leq 1.
\end{align}
This result shows that
\begin{align}\label{eq:QPweight}
    \int dr\, |\psi_k(r,t)|^2 = \sum_{k'} |G_{kk'}(t)|^2 \leq 1
\end{align}
and provides a justification for interpreting
$\psi^*_k(r,t)$ in Eq. (\ref{eq:QPwf})
as a quasiparticle wave function. $G_{kk'}(t)$ can be understood as an
expansion coefficient of the quasiparticle wave function
$\psi^*_k(r,t)$ in the base orbital
$\varphi_{k'}^*$. A similar analysis can be performed for $t>0$.
The result in Eq. (\ref{eq:QPweight}) could be seen from a different point
of view \cite{dvorak2021}. The extension to
nonequilibrium Green function can be found in a recent
publication \cite{aryasetiawan2025}. 

This definition of a quasiparticle
wave function is general and is unrelated to the Vxc formalism.
Unlike the standard definition, which corresponds only
to the main quasiparticle peak,
the proposed definition also contains possible
excitations arising from the coupling to collective modes.
The quasiparticle wave functions and
the Green function are equivalent since they can be constructed from
the knowledge of the other.

\subsection{Quasiparticle equation in the Vxc formalism}
\label{sec:QPeqn}

The quasiparticle equation of motion can be derived from the equation
of motion of the Green function in the Vxc formalism. It is useful
to choose a mean-field xc potential $V^0_\mathrm{xc}(r)$ and define
a deviation potential,
\begin{equation}\label{eq:DeltaV}
    \Delta V(r,r';t) = V_\mathrm{xc}(r,r';t)- V^0_\mathrm{xc}(r).
\end{equation}
$V^0_\mathrm{xc}$ may be chosen as the Kohn-Sham xc potential.
Choosing the orbitals to be those of the mean-field
Hamiltonian, 
\begin{align}
    \left[-\frac{1}{2}\nabla^2 + V_\mathrm{ext}+V_\mathrm{H}
    + V^0_\mathrm{xc} \right]\varphi_k=\varepsilon_k\varphi_k,
\end{align}
it follows from the
equation of motion for the Green function in Eq. (\ref{eq:EOMGVxc})
with $t\neq 0$
\begin{align}\label{eq:EOMQP}
    \sum_{k} \left[ i\partial_t -\varepsilon_k- \Delta V(r,r';t)\right]
     \varphi_k(r) \psi^*_k(r',t) =0.
\end{align}
Multiplying on the left by $\varphi^*_q(r)$ and integrating over $r$ yields
(renaming $r'$ as $r$ after integration),
\begin{align}
    (i\partial_t-\varepsilon_q)\psi^*_q(r,t) 
    -\sum_{k} \Delta V_{qk}(r,t) \psi^*_k(r,t)=0.
\end{align}
where
\begin{align}
    \Delta V_{qk}(r,t) = \int dr'\, \varphi^*_q(r')\varphi_k(r')\Delta V(r',r;t).
\label{eq:DeltaV}
\end{align}
%
%
This equation provides a quantitative description
of a many-electron system as a set of quasiparticles, which interact with
the (residual) interactions $\Delta V_{qk}$, as in Landau phenomenological
quasiparticle theory.

The quasiparticle equation of motion can be recast as
\begin{align}
    \left[i\partial_t-\varepsilon_q-\Xi_q(r,t)\right]\psi^*_q(r,t) =0,
\end{align}
where
\begin{align}\label{eq:Xi}
    \Xi_q(r,t) =\frac{1}{\psi^*_q(r,t)}
    \sum_{k} \Delta V_{qk}(r,t) \psi^*_k(r,t)
\end{align}
acts as a $q$-dependent effective field.
Each quasiparticle wave function experiences its own potential. 
The formal solution is given by
\begin{align}\label{eq:QPSoln}
    \psi^*_q(r,t)=\psi^*_q(r,0^\pm)
    e^{-i\varepsilon_q t 
    -i\int_0^t dt'  \Xi_{q}(r,t')}.
\end{align}
%
The formal solution displays a clear physical meaning: when the
residual effects of exchange and correlations encapsulated in $\Xi_q(r,t)$
are removed, the quasiparticle wave function returns to its mean-field
reference orbital.
Since $\Xi_q(r,t)$ is generally complex, which is a consequence of having
a nonhermitian $V_\mathrm{xc}$,
its imaginary part causes the quasiparticle
wave function to decay with time in an interacting many-electron system.

Once $\psi^*_q(r,t)$ is solved for all $q$, 
the Green function can be constructed from Eq. (\ref{eq:GQP}).
Although the quasiparticle wave functions and the Green function are equivalent, 
the former can provide a useful description of the effects of
the electron-electron interactions and a tool for analysing these effects.
$|G_{kk'}(t)|^2$
gives the weight of the quasiparticle $\psi^*_k$ in orbital $\varphi_{k'}^*$.
It provides information on how the quasiparticle is distributed among the
orbitals
and how much weight has been lost from the reference orbital
$\varphi_{k}^*$ due to residual exchange-correlation effects as a function of time.

From the formal solution of the quasiparticle wave function in
Eq. (\ref{eq:QPSoln}) it is physically suggestive to decompose
the effective field into a static (S) and a dynamic (D) term as
\begin{align}\label{eq:XiApprox}
    \Xi_{q}(r,t) \approx \Xi^\mathrm{S}_{q}(r) + \Xi^\mathrm{D}_{q}(r) e^{i\Omega t},
\end{align}
where $\Omega$ is the energy of the
main collective excitation of the system that is coupled
to the electrons. The static term corrects the reference 
orbital energy $\varepsilon_q$ while
the dynamic term induces an incoherent or satellite feature
in the spectral function.

Examples of the effective potential $\Xi_q$ are given for the Hubbard dimer
and the homogeneous electron gas in Secs. \ref{sec:HubbardDimer} and
\ref{sec:XiElgas}, respectively.

\subsection{Quasiparticle-averaged effective potential}

The effective potential $\Xi_q(r,t)$ depends on the quasiparticle states. 
A possible simplification is to take a quasiparticle average (avg)
following Slater's idea of orbital averaging,
\begin{align}
    \Xi^\mathrm{avg}(r,t)&= \sum_q 
    \frac{|\psi_q(r,t|^2}{\rho_\mathrm{QP}(r,t)}
     \Xi_{q}(r,t)
     \nonumber\\
     &= \frac{1}{\rho_\mathrm{QP}(r,t)}
     \sum_{qk} \psi_q(r,t) \Delta V_{qk}(r,t)
     \psi^*_k(r,t),
\end{align}
where
\begin{align}
    \rho_\mathrm{QP}(r,t)=\sum_q |\psi_q(r,t)|^2.
\end{align}
The equation of motion with the $q$-independent effective potential
reduces to the following,
\begin{align}
    \left[i\partial_t-\varepsilon_q-
    \Xi^\mathrm{avg}(r,t)\right]\psi^*_q(r,t) =0.
\end{align}
However, for a homogeneous system, such as an electron gas,
the $q$-dependence is important because without it
there will be no band narrowing of the free-electron dispersion.

\subsection{Self-consistency}

The dependence of the quasiparticle states on the choice of orbitals
raises an interesting question as to whether
it is possible to remove this orbital
dependency by means of a self-consistency procedure.
By definition, the quasiparticle wave functions are intrinsically
dependent on the reference orbitals.
However, the reference orbitals
can be defined self-consistently with respect to some given
conditions. One possible condition is to require
the reference orbitals to reproduce the
ground-state density. The new ground-state density obtained from the
quasiparticle wave functions is given by
\begin{align}
    \rho(r)=-iG(r,r;0^-)=-i\sum_k \varphi_k(r) \psi_k^*(r,0^-),
\end{align}
which can be used to redefine $V_\mathrm{H}$ and $V_\mathrm{xc}$.
The new reference mean-field Hamiltonian delivers an updated set of
reference orbitals, yielding a new set of quasiparticle wave functions.

\section{Extension to thermal and nonequilibrium Green function}

The Vxc formalism can be readily extended to
the thermal or Matsubara Green function:
\begin{align}
    G_\mathrm{M}(r\tau,r'\tau')=-\mathrm{Tr} \left\{
    \hat{\rho}_K T [ \hat{\psi}(r\tau) \hat{\psi}^\dagger(r'\tau') ] 
    \right\},
\end{align}
where $\hat{K}=\hat{H}-\mu \hat{N}$ and $\tau$ is an imaginary time.
Its equation of motion is given by
\begin{align}
   & \left[ \frac{\partial}{\partial\tau} + h_0(r)-\mu  
    \right]
    G_\mathrm{M}(r\tau,r'\tau') 
    \nonumber\\
    &-\int dr'' v(r-r'') 
    \langle T[ \hat{\rho}(r''\tau) 
             \hat{\psi}(r\tau)\hat{\psi}^\dagger(r'\tau') ]
    \rangle 
    \nonumber\\
    =& -\delta(\tau-\tau')\delta(r-r'),
\end{align}
where the thermal averaging is defined as
\begin{align}
    \langle \hat{O} \rangle = \mathrm{Tr}(\hat{\rho}_K\hat{O}).
\end{align}
Since $G_\mathrm{M}$ depends only on $\tau-\tau'$ it is permissible to set
$\tau'$ to zero,
\begin{align}
   & \langle T[ \hat{\rho}(r''\tau) 
             \hat{\psi}(r\tau)\hat{\psi}^\dagger(r') ]
    \rangle
    \nonumber\\
    &=-G_\mathrm{M}(r,r';\tau)[\rho(r'')+\rho_\mathrm{xc}(r,r',r'';\tau)]
\end{align}
leading to
\begin{align}
    &\left[ \frac{\partial}{\partial\tau} + h(r)+V_\mathrm{xc}(r,r';\tau)-\mu  
    \right]
    G_\mathrm{M}(r,r';\tau) 
    \nonumber\\
&\qquad\qquad\qquad= -\delta(\tau)\delta(r-r'),
\end{align}
where $h=h_0+V_\mathrm{H}$ and
\begin{align}
    V_\mathrm{xc}(r,r';\tau)=\int dr'' v(r-r'')\rho_\mathrm{xc}(r,r',r'';\tau).
\end{align}
The sum rule as well as the exact constraint for the exchange-correlation hole
in Eqs. (\ref{eq:SumRule}) and (\ref{eq:ExactReln}) are still fulfilled.


An extension to
nonequilibrium systems can also be made. 
In systems in which the Hamiltonian is time dependent,
the field operator in the Heisenberg picture is given by
\begin{align}
    \hat{\psi}(rt)=\hat{U}(0,t) \hat{\psi}(r) \hat{U}(t,0).
\end{align}
The time-evolution operator fulfils the Schrödinger equation:
\begin{align}
    i\partial_t \hat{U}(t,0) = \hat{H}(t) \hat{U}(t,0),
\end{align}
with the formal solution
\begin{align}
    \hat{U}(t,0) = T e^{-i\int_0^t dt_1 \,\hat{H}(t_1)}.
\end{align}
The conjugate equation is given by
\begin{align}
   - i\partial_t \hat{U}^\dagger(t,0) = \hat{U}^\dagger(t,0) \hat{H}(t) .
\end{align}
From the group and unitary properties,
\begin{align}
    \hat{U}(t,0) \hat{U}(0,t) = 1,\qquad \hat{U}(t,0)\hat{U}^\dagger(t,0)=1,
\end{align}
and the definition of the Heisenberg operator, it follows that
\begin{align}
    i\partial_t \hat{\psi}(rt)&= \hat{U}(0,t) [\hat{\psi}(r),\hat{H}(t)] \hat{U}(t,0)
    \nonumber\\
    &=\hat{U}(0,t) [h_0(rt)+\hat{V}_\mathrm{H}(r)]\hat{\psi}(r) \hat{U}(t,0)
    \nonumber\\
    &=[h_0(rt)+\hat{V}_\mathrm{H}(rt)]\hat{\psi}(rt),
\end{align}
where
\begin{align}
    h_0(rt) &= -\frac{1}{2}\nabla^2 + V_\mathrm{ext}(rt),
    \\
    \hat{V}_\mathrm{H}(rt)&=\int dr'' \, v(r-r'')\hat{\rho}(r''t).
\end{align}
Multiplying on the right and on the left by $\hat{\psi}^\dagger(r't')$, and
after a simple manipulation one obtains
\begin{align}
    &\left[ i\partial_t -h_0(rt)
    \right] G(rt,r't') 
    \nonumber\\
    &\qquad +i\int dr'' v(r-r'') \langle T 
    [ \hat{\rho}(r''t) \hat{\psi}(rt)\hat{\psi}^\dagger(r't')] \rangle
    \nonumber\\
    &=\delta(r-r')\delta(t-t').
\end{align}
The derivation of the xc hole and field is rather similar to
the equilibrium case. A correlator $g$ is introduced,
\begin{align}
    G^{(2)}(rt,r't',r'')&= -i\langle T 
    [ \hat{\rho}(r''t) \hat{\psi}(rt)\hat{\psi}^\dagger(r't')] \rangle
    \nonumber\\
     &=G(rt,r't') g(rt,r't',r'') \rho(r''t).
\label{eq:defG2t}
\end{align}
As in the equilibrium case, $G^{(2)}$ can be rewritten as
\begin{align}
    G^{(2)}(rt,r't',r'')= [\rho(r''t)+\rho_\mathrm{xc}(rt,r't',r'')] G(rt,r't'),
\label{eq:G2t}
\end{align}
where
\begin{align}
    \rho_\mathrm{xc}(rt,r't',r'')= [g(rt,r't',r'')-1]\rho(r''t).
\label{eq:rhoxct}
\end{align}
The sum rule and the exact constraint are given by, respectively,
\begin{align}
\int dr''\,\rho_\mathrm{xc}(rt,r't',r'') = -\theta(t'-t)\delta_{\sigma\sigma''},
\label{eq:SumRule1}
\end{align}
\begin{align}
    \rho_\mathrm{xc}(rt,r't',r) = -\rho(rt),
\label{eq:ExactConstraint}
\end{align}
for \emph{any} $r$, $r'$, $t$, and $t'$.

The equation of motion of the Green function is given by
\begin{align}
    &\left[ i\partial_t -h(rt) -V_\mathrm{xc}(rt,r't')
    \right] G(rt,r't') =\delta(r-r')\delta(t-t'),
\end{align}
where $h(rt)$ contains the Hartree potential and
\begin{align}
    V_\mathrm{xc}(rt,r't')=\int dr'' v(r-r'')\rho_\mathrm{xc}(rt,r't',r'').
\end{align}

\section{Examples of $V_\mathrm{xc}$}

After elaborating the theoretical aspects of the Vxc
formalism, some examples will now be described to illustrate some
salient features of $V_\mathrm{xc}$. 

\subsection{Hydrogen atom}

\begin{figure}[h]
\begin{center} 
\includegraphics[scale=0.4, viewport=3cm 8cm 18cm 20cm, clip, width=\columnwidth]
{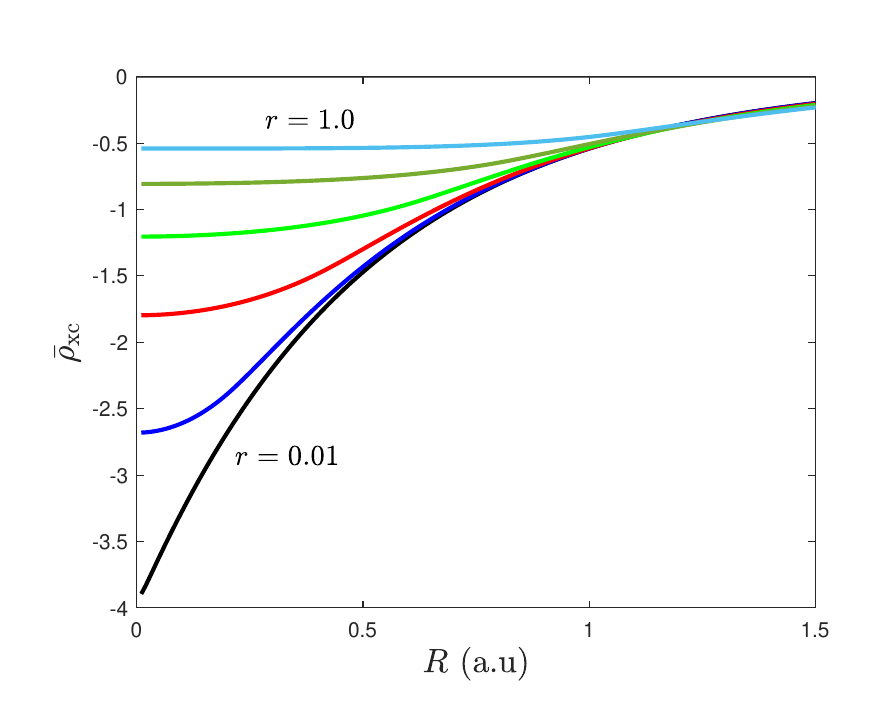}
\hfill
\includegraphics[scale=0.4, viewport=3cm 8cm 18cm 20cm, clip, width=\columnwidth]
{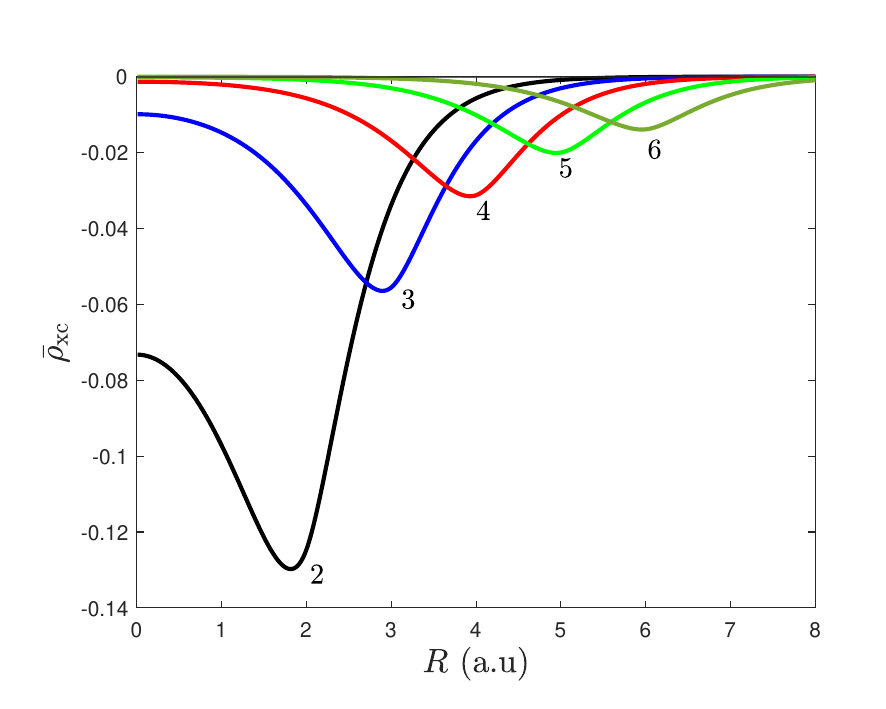}

\caption{Top: The spherical average of the 
xc hole of the hydrogen atom centred at $\mathbf{r}$ for
several values of $r=0.01$ (black), $0.2$ (blue), $0.4$ (red),
$0.6$ (green), $0.8$ (olive), and $1.0$ (light blue), as indicated in the figure
in progression from $r=0.01$ to $1.0$. 
Bottom: The same as the top figure but for $r>1$, 
beyond the Bohr radius ($r=1$) as indicated in the figure.
For large $r$ the 
xc hole is dipped around the nucleus rather than at $\mathbf{r}$,
the location where the hole is created ($R=0$).
}
\label{fig:HydrogenRhoxc}%
\end{center}
\end{figure}

The hydrogen atom provides a simple illustration for the sum rule
Eq. (\ref{eq:SumRule}) and the exact constraint Eq. (\ref{eq:ExactReln}).
It also illustrates the well-known difficulty with LDA.

For the hydrogen atom, the hole Green function is given by
\begin{equation}
G(r,r^{\prime};t<0)=i\varphi_{s}(r)\varphi_{s}(r^{\prime})\exp(-i\varepsilon
_{s}t)\theta(-t),
\end{equation}
where $\varphi_{s}$ and $\varepsilon_{s}$ are the $1s$-orbital and its energy.
The electron Green function is not considered here since it requires
eigenfunctions of the two-electron problem.
For $t<0$ $V_\mathrm{xc}$ acts to cancel the spurious self-interaction:
\begin{equation}
V_{\mathrm{xc}}(r,r^{\prime};t<0)=-V_{\mathrm{H}}(r)=-\int dr^{\prime\prime
}v(r-r^{\prime\prime})|\varphi_{s}(r^{\prime\prime})|^{2},
\end{equation}
independent of $r^{\prime}$ and $t$.
The corresponding xc hole is given by
\begin{equation}
\rho_{\mathrm{xc}}(r,r^{\prime},r^{\prime\prime};t<0)=-|\varphi_{s}%
(r^{\prime\prime})|^{2} = -\frac{1}{\pi}e^{-2r''}.
\end{equation}
The exact constraint in Eq. (\ref{eq:ExactReln}),
\begin{equation}
\rho_{\mathrm{xc}}(r,r^{\prime},r;t<0)=-\rho(r),
\end{equation}
as well as the sum rule in Eq. (\ref{eq:SumRule}),
\begin{equation}
\int dr^{\prime\prime}\rho_{\mathrm{xc}}(r,r^{\prime},r^{\prime\prime
};t<0)=-\int dr^{\prime\prime}|\varphi_{s}(r^{\prime\prime})|^{2}=-1,
\end{equation}
are clearly fulfilled.

Only the spherical average of the xc hole is needed.
Taking $\mathbf{r}$ as the centre of the hole
independent of $\mathbf{r}'$ and $t$
the spherical average as defined in Eq. (\ref{eq:SphAvRhoxc}) 
is given by
\begin{align}
    \overline{\rho}_\mathrm{xc}(r,R) &= \int d\Omega_R \rho_\mathrm{xc}(\mathbf{r}+\mathbf{R})
    \nonumber\\
    &=-2\pi \int_{-1}^1 dy \frac{1}{\pi}e^{-2\sqrt{r^2+R^2+2rRy}}.
\end{align}
The change of variable $x=\sqrt{r^2+R^2+2rRy}$ yields
\begin{align}
    \overline{\rho}_\mathrm{xc}(r,R) &=-\frac{1}{2rR}
    \left\{  (2a+1)e^{-2a} -(2b+1)e^{-2b}
    \right\},
\end{align}
where
\begin{align}
    a=|r-R|,\qquad b=r+R.
\end{align}
In the limit $R\rightarrow 0$,
\begin{align}
    \overline{\rho}_\mathrm{xc}(r,R\rightarrow 0) &= -4\pi \frac{e^{-2r}}{\pi}
    =4\pi \rho(r),
\end{align}
and in the limit $r\rightarrow 0$,
\begin{align}
    \overline{\rho}_\mathrm{xc}(r\rightarrow 0,R) &= -4\pi \frac{e^{-2R}}{\pi}
    =4\pi \rho(R).
\end{align}

Fig. \ref{fig:HydrogenRhoxc} shows the spherical average
of the xc hole centred at several values of $\mathbf{r}$.
For small $r$
it approaches $4\pi\rho(R)$ and is peaked around its centre
at $\mathbf{r}$, that is, around $R=0$,
whereas for large $r$ beyond the Bohr radius ($r=1$) the
xc hole is dipped around the nucleus rather than around its centre
as can be seen in the lower figure of Fig. \ref{fig:HydrogenRhoxc}. 
Note that $R$ is the distance from the centre at $\mathbf{r}$.
Thus, for large $r$ away from the nucleus, the xc hole is left behind
around the nucleus, whereas in LDA the hole is always centred
around $\mathbf{r}$, a well-known shortcoming of LDA.
For large $r$ the LDA is a poor approximation for the xc hole.

\subsection{The Holstein model}

The Holstein Hamiltonian describes a coupling between electrons
and a set of bosons, such as plasmons or phonons.
In its simplified version, where there is only one electron,
it is given by \cite{langreth1970}
\begin{equation}
    \hat{H} = \varepsilon \hat{c}^\dagger \hat{c}
    +\sum_q \hat{c} \hat{c}^\dagger g_q(\hat{b}_q + \hat{b}_q^\dagger)
    +\sum_q \omega_q \hat{b}_q^\dagger \hat{b}_q,
\end{equation}
where $\varepsilon$ is the core electron energy, $\omega_q$ 
is the boson energy with wave vector $q$,
and $\hat{c}$ and $\hat{b}_q$ are the core electron
and boson operators, respectively. 
Under the assumption that the boson has no dispersion
with an average energy $\omega_\mathrm{p}$, the
exact solution for the core-electron removal spectra is given by
\cite{langreth1970}
\begin{equation}
    A(\omega)=\sum_{n=0}^\infty f_n \delta(\omega-\varepsilon-\Delta\varepsilon+n\omega_\mathrm{p}),
\end{equation}
where
\begin{equation}
    f_n=\frac{e^{-a}a^n}{n!},\;\;a=\sum_q
    \left( \frac{g_q}{\omega_\mathrm{p}}\right)^2, 
    \;\; \Delta\varepsilon = a\omega_\mathrm{p}.
\end{equation}
%
The hole Green function corresponding to the above spectra is given by
\begin{equation}
    G(t<0) = i \sum_{n=0}^\infty f_n e^{-i(\varepsilon+\Delta\varepsilon-n\omega_\mathrm{p})t}\theta(-t).
\end{equation}

In the Vxc framework, the equation of motion is given by
\begin{align}
    [i\partial_t-\varepsilon -V_\mathrm{xc}(t)] G(t) = \delta(t).
\end{align}
The exact Vxc can be calculated from the Green function
yielding
\begin{equation}
    V_\mathrm{xc}(t<0)= \Delta\varepsilon \left(1-e^{i\omega_\mathrm{p} t}\right).
\end{equation}
This expression offers a very simple interpretation:
the first term corrects the noninteracting
core-electron energy, whereas the second term describes the bosonic
mode interacting with the core electron, 
exchanging multiple quanta of $\omega_\mathrm{p}$ with the field.
The result means that the simplified Holstein model can be understood
as a system driven by a periodic complex potential (Floquet system).

It is interesting to note that
the exact spectra can be obtained using the cumulant expansion
\cite{bergersen1973,hedin1980,almbladh1983} in which the cumulant is
calculated
within the $GW$ approximation of the self-energy.
Had the exact self-energy been used in the cumulant expansion,
the exact spectra would presumably be
not reproduced. This illustrates the subtle cancellation of
error in the diagrammatic approach in the traditional self-energy formalism.

\subsection{The Hubbard dimer}
\label{sec:HubbardDimer}

\begin{figure}[htp]
\centering
\includegraphics[scale=0.40, viewport=3cm 8cm 18cm 20cm, clip,
width=0.9\columnwidth]
{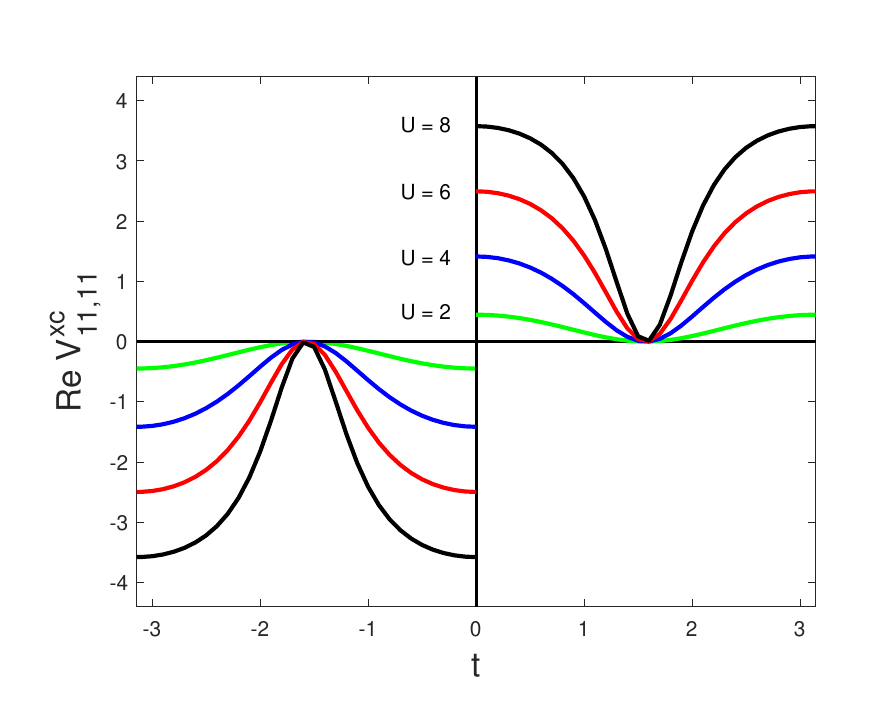}
\hfill
\includegraphics[scale=0.40, viewport=3cm 8cm 18cm 20cm, clip,
width=0.9\columnwidth]
{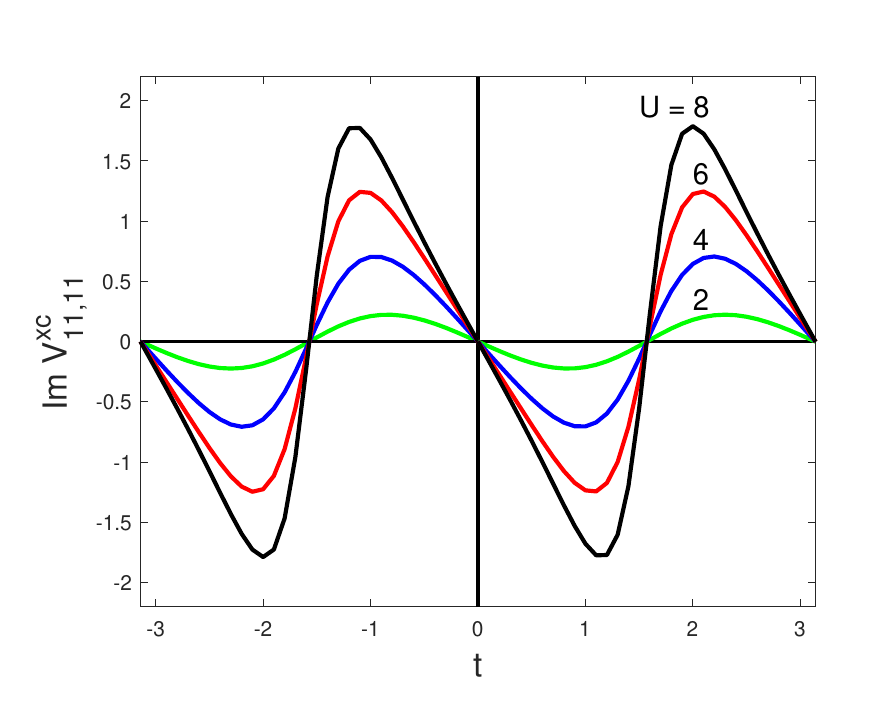}

\caption{The real and imaginary parts of 
$V^\mathrm{xc}_{11,11}$ 
of the Hubbard dimer
as functions of time for $U=2,4,6,$ and $8$ with $\Delta=1$
\cite{aryasetiawan2022a}.
Due to the
particle-hole symmetry, $V_\mathrm{xc}(-t)=-V_\mathrm{xc}(t)$.}
\label{fig:ReVxc11}%
\end{figure}

\begin{figure}[htp]
\centering
\includegraphics[scale=0.40, viewport=3cm 8cm 18cm 20cm, clip,
width=0.9\columnwidth]
{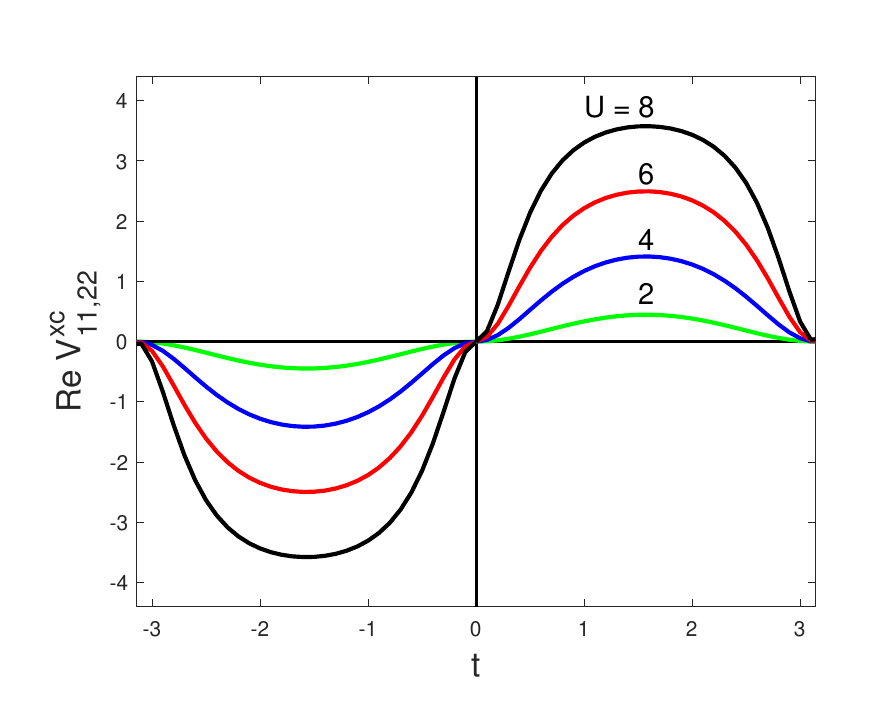}
\hfill
\includegraphics[scale=0.40, viewport=3cm 8cm 18cm 20cm, clip,
width=0.9\columnwidth]
{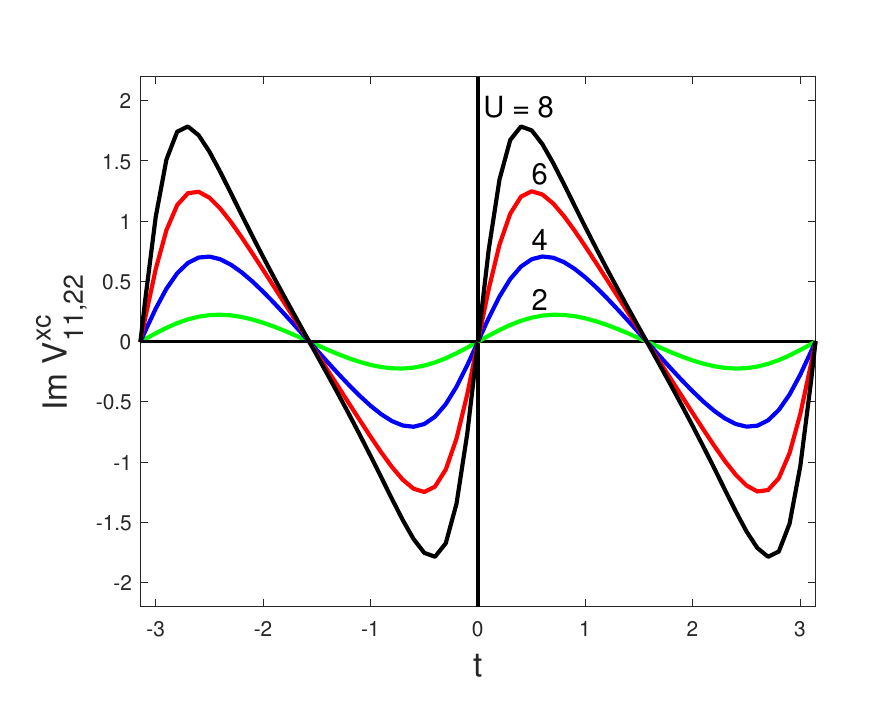}

\caption{The real and imaginary parts of 
$V^\mathrm{xc}_{11,22}$ 
of the Hubbard dimer
as functions of time for $U=2,4,6,$ and $8$ with $\Delta=1$
\cite{aryasetiawan2022a}.
Due to the
particle-hole symmetry, $V_\mathrm{xc}(-t)=-V_\mathrm{xc}(t)$.}
\label{fig:ReVxc12}%
\end{figure}

\begin{figure}[htp]
\centering
\includegraphics[scale=0.40, viewport=3cm 8cm 18cm 20cm, clip,
width=0.9\columnwidth]
{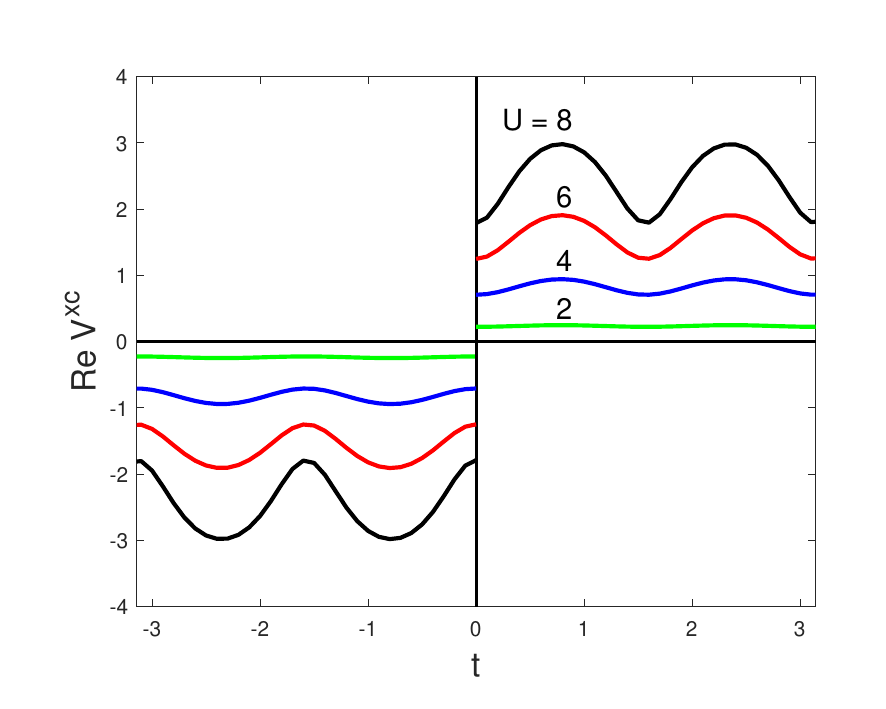}
\hfill
\includegraphics[scale=0.40, viewport=3cm 8cm 18cm 20cm, clip,
width=0.9\columnwidth]
{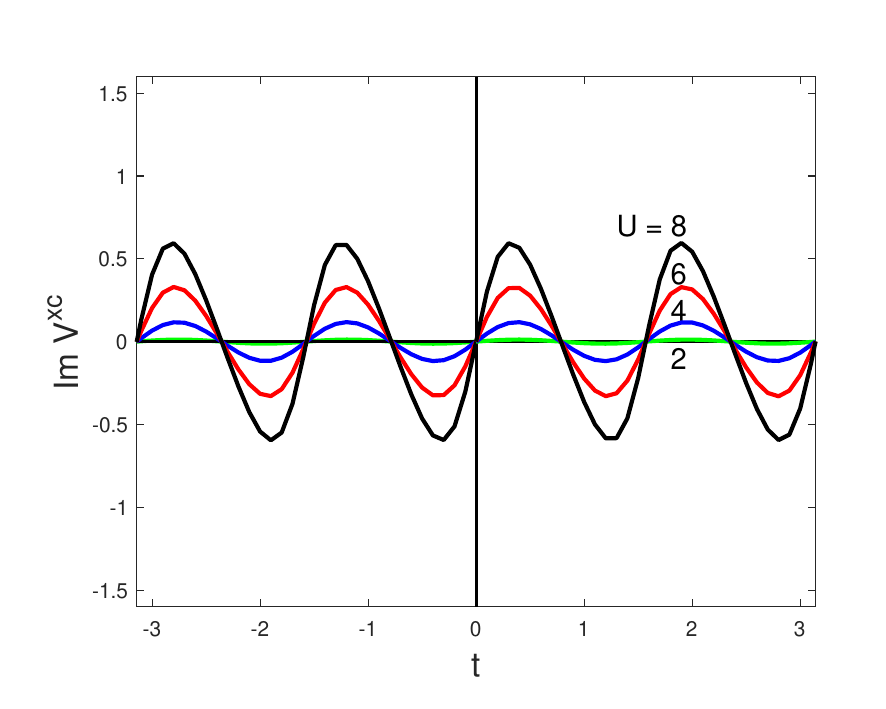}

\caption{The real and imaginary parts of
$V^\mathrm{xc}_\mathrm{BB}=V^\mathrm{xc}_\mathrm{AA}$ as functions of 
time for $U=2,4,6,$ and $8$ with $\Delta=1$ \cite{aryasetiawan2022a}. 
}
\label{fig:VxcBB}%
\end{figure}

\begin{figure}[htp]
\centering
\includegraphics[scale=0.40, viewport=3cm 8cm 18cm 20cm, clip,
width=0.9\columnwidth]
{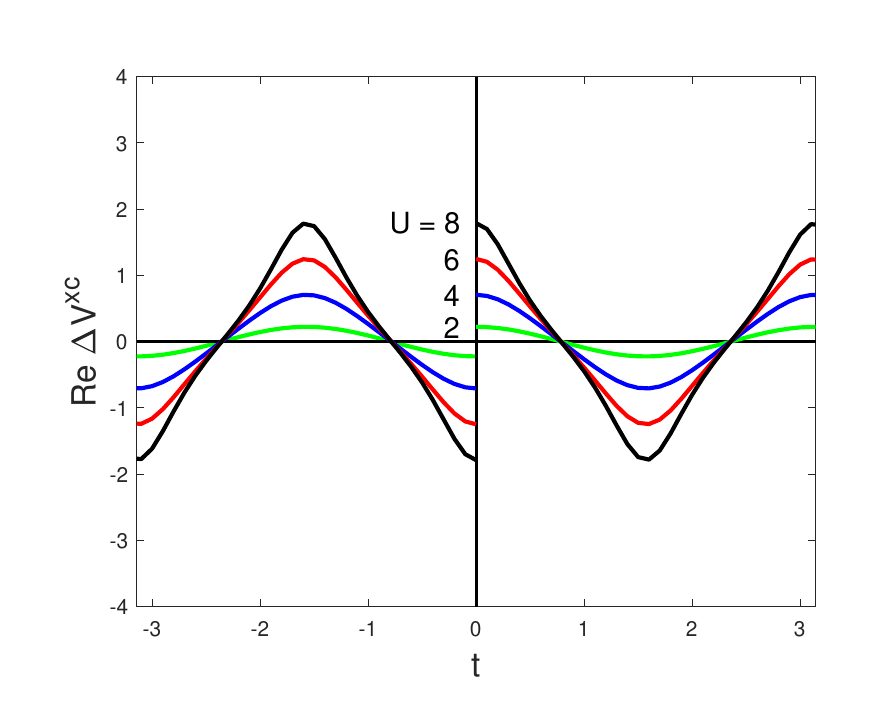}
\hfill
\includegraphics[scale=0.40, viewport=3cm 8cm 18cm 20cm, clip,
width=0.9\columnwidth]
{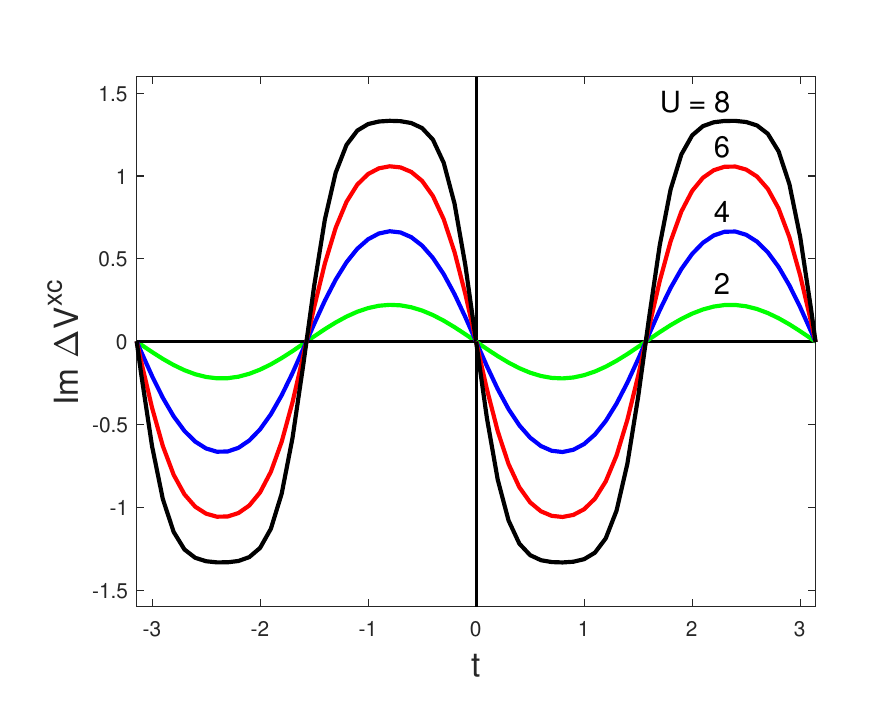}

\caption{The real and imaginary parts of
$V^\mathrm{xc}_\mathrm{AB}$ in
as functions of 
time for $U=2,4,6,$ and $8$ with $\Delta=1$ \cite{aryasetiawan2022a}. 
}
\label{fig:VxcAB}%
\end{figure}

Although it is very simple, the half-filled Hubbard dimer
with total $S_z=0$ contains some of the essential
physics of correlated electrons, and it has the merit of being
analytically solvable. It also illustrates that
the concept of xc hole is not restricted to continuous systems but
can be carried over to lattice models.

The Hamiltonian of the Hubbard dimer in standard
notation is given by
\begin{equation}
\hat{H}=-\Delta\sum_{i\neq j,\sigma}\hat{c}_{i\sigma}^{\dagger}\hat{c}_{j\sigma}%
+U\sum_{i}\hat{n}_{i\uparrow}\hat{n}_{i\downarrow}.
\end{equation}
It is convenient to define
\begin{equation}
\hat{c}_{i}=\hat{c}_{i\uparrow},\text{ \ }\hat{a}_{i}=\hat{c}_{i\downarrow
},\text{ \ }\hat{n}_{i}=\hat{c}_{i}^{\dagger}\hat{c}_{i},\text{ \ }\hat{m}%
_{i}=\hat{a}_{i}^{\dagger}\hat{a}_{i}.
\end{equation}
To compare with the continuous case, one may define the Hubbard
interaction as $U_{ij}=U\delta_{ij}$ and the Hamiltonian can be rewritten
as
\begin{align}
\hat{H}=-\Delta\sum_{i\neq j}\left(
\hat{c}_i^{\dagger}\hat{c}_j +\hat{a}_i^{\dagger}\hat{a}_j\right)
+\sum_{ij}\hat{n}_iU_{ij}\hat{m}_j.    
\end{align}

The up-spin time-ordered Green function is defined as
\begin{equation}
iG_{ij}(t)=\left\langle \Psi_{0}\right\vert T
\hat{c}_{i}(t)\hat{c}_{j}^{\dagger
}\left\vert \Psi_{0}\right\rangle ,
\end{equation}
which is identical to the down-spin one.
The Heisenberg equation of motion of $\hat{c}_{i}(t)$ is given by
\begin{equation}
i\partial_t\hat{c}_{i}(t)= \left[\hat{c}_{i}(t),\hat{H} \right]=
\sum_{k}\left( h^0_{ik}\hat{c}_{k}(t)
+ U_{ik}\hat{m}_k(t) \hat{c}_{i}(t) \right),
\end{equation}
where
\begin{equation}
h^0=\left(
\begin{array}
[c]{cc}%
0 & -\Delta\\
-\Delta & 0
\end{array}
\right)  .
\end{equation}
The equation of motion of the Green function
is then given by
\begin{align}
&  i\partial_t G_{ij}(t)
-\sum_{k}\left[h^0_{ik}G_{kj}(t) +U_{ik}G^{(2)}_{ijk}(t) \right]
=\delta_{ij}\delta(t),
\end{align}
where
\begin{align}
    G^{(2)}_{ijk}(t) &= -i\left\langle \Psi_{0}\right\vert 
T \hat{m}_{k}(t)\hat{c}_{i}(t)\hat
{c}_{j}^{\dagger} \left\vert \Psi_{0}\right\rangle
\nonumber\\
&= m_k g_{ijk}(t) G_{ij}(t),
\label{eq:defG2dimer}
\end{align}
where $g_{ijk}(t)$ is the correlation function. Separating out the mean-field
Hartree term $V^\mathrm{H}_i=\sum_k U_{ik} m_k$ 
leads to the equation of motion,
\begin{align}
&  (i\partial_t -V^\mathrm{H}_i-V^\mathrm{xc}_{ij}(t))G_{ij}(t)
-\sum_{k}h^0_{ik}G_{kj}(t) =\delta_{ij}\delta(t),
\end{align}
where $V^\mathrm{xc}_{ij}(t)$ is given by
\begin{align}\label{eq:defxchole}
    V^\mathrm{xc}_{ij}(t) = \sum_k U_{ik} \rho^\mathrm{xc}_{ijk}(t),
    \quad \rho^\mathrm{xc}_{ijk}(t)=[g_{ijk}(t)-1]m_k .
\end{align}
Note that according to Eq. (\ref{eq:Vxc_ijkl}),
$V^\mathrm{xc}_{ij}=V^\mathrm{xc}_{ii,jj}$ since the interaction
has no off-site components and there is only one orbital per site.

The interaction term of the Hamiltonian of the Hubbard dimer
explicitly takes into account exchange. For this reason, the sum rule
for the xc hole integrates to zero. Since $\hat{M}=\sum_k \hat{m}_k$ 
counts the total number of down-spin electrons, one finds from the first line
of Eq. (\ref{eq:defG2dimer})
\begin{align}
    \sum_k G^{(2)}_{ijk}(t) = MG_{ij}(t),
\end{align}
where $M$ is the total number of down-spin electrons, which is one.
This implies from the second line of Eq. (\ref{eq:defG2dimer}) that
\begin{align}
    \sum_k m_k g_{ijk}(t) = M,
\end{align}
and from Eq. (\ref{eq:defxchole}), it follows that
\begin{align}
    \sum_k \rho^\mathrm{xc}_{ijk}(t) =0.
\end{align}
This is consistent with the general sum rule in Eq. (\ref{eq:SumRule})
since $\sigma\neq \sigma''$. Similarly, the exact constraint in
Eq. (\ref{eq:ExactReln}) does not apply since
in the present case $\sigma \neq\sigma''$.

The eigenvectors and eigenvalues corresponding to one-, two-, and
three-electron fillings
can be calculated analytically.
$V_{ij}^\mathrm{xc}$
can be deduced from the equation of motion by
calculating $G^{(2)}$ or $G$. The results,
shown in Figs. \ref{fig:ReVxc11} and \ref{fig:ReVxc12}, are given by
\cite{aryasetiawan2022,aryasetiawan2022a}
\begin{equation}
V_{11,11}^\mathrm{xc}(t>0)=\frac{\alpha U}{2}\frac{1+e^{-i2\Delta t}}{1+\alpha
^{2}e^{-i2\Delta t}}.
\label{Vxc1111}
\end{equation}
\begin{equation}
V_{11,22}^\mathrm{xc}(t>0)=\frac{\alpha U}{2}\frac{1-e^{-i2\Delta t}}{1-\alpha
^{2}e^{-i2\Delta t}},
\label{Vxc1122}
\end{equation}
where
\begin{equation}
\alpha=\frac{1-x}{1+x},
\label{alpha}
\end{equation}%
\begin{equation}
x=\frac{1}{4\Delta}\left(  \sqrt{U^{2}+16\Delta^{2}}-U\right)
\end{equation}
is the relative weight of double-occupancy configurations in the ground state
with energy 
\begin{align}
E_{0}  &  =\frac{1}{2}\left(  U-\sqrt{U^{2}+16\Delta^{2}}\right) ,
\end{align}
and
\begin{equation}
2\Delta=E_{1}^{-}-E_{0}^{-}=E_{1}^{+}-E_{0}^{+}>0
\end{equation}
is the excitation energy of the $(N\pm1)$-systems. 
$E_m^\pm$ are the $m$th eigenenergies of the $(N\pm1)$-systems:
\begin{equation}
E_{0}^{-}=-\Delta,\text{ \ }E_{1}^{-}=\Delta,
\end{equation}
\begin{equation}
E_{0}^{+}=U-\Delta,\text{ \ }E_{1}^{+}=U+\Delta.
\end{equation}

From symmetry,
$V_{22}^\mathrm{xc}=V_{11}^\mathrm{xc}$ and 
$V_{21}^\mathrm{xc}=V_{12}^\mathrm{xc}$, i.e.,
$V^\mathrm{xc}$ is symmetric but not Hermitian, since it is complex. 
Due to the
particle-hole symmetry, $V^\mathrm{xc}(-t)=-V^\mathrm{xc}(t)$. 
Note that the full
$V^\mathrm{xc}$ carries four indices as in Eq. (\ref{eq:Vxc_ijkl}).
Since there is only one orbital per site, it follows that 
$V_{11,11}^\mathrm{xc}=V_{11}^\mathrm{xc}$ and 
$V_{11,22}^\mathrm{xc}=V_{12}^\mathrm{xc}$.

In some cases, it is useful to choose the bonding and antibonding orbitals
as the basis:
\begin{align}
    \phi_\mathrm{B}=\frac{1}{\sqrt{2}}\left( \varphi_1 +\varphi_2\right),
    \qquad
    \phi_\mathrm{A}=\frac{1}{\sqrt{2}}\left( \varphi_1 -\varphi_2\right).
\end{align}
In this basis the Green function is diagonal and $V^\mathrm{xc}$
is given by
\begin{align}\label{eq:defVxcBB}
    V^\mathrm{xc}_\mathrm{BB,BB}:= V^\mathrm{xc}_\mathrm{BB}&=V^\mathrm{xc}_\mathrm{AA}=
    \frac{1}{2}(V^\mathrm{xc}_{11}+V^\mathrm{xc}_{12}),\\
    \label{eq:defVxcAB}
   V^\mathrm{xc}_\mathrm{AB,BA}:= V^\mathrm{xc}_\mathrm{AB}&=V^\mathrm{xc}_\mathrm{BA}=
    \frac{1}{2}(V^\mathrm{xc}_{11}-V^\mathrm{xc}_{12}).
\end{align}
In the bonding-antibonding basis
the equation of motion of the Green function becomes
\begin{align}
    \left[ i\partial_t -\varepsilon_\mathrm{B}- V^\mathrm{xc}_\mathrm{BB}(t)
    \right] G_\mathrm{B}(t) -V^\mathrm{xc}_\mathrm{BA}(t) G_\mathrm{A}(t)
    =\delta(t),
\end{align}
\begin{align}
    \left[ i\partial_t -\varepsilon_\mathrm{A}- V^\mathrm{xc}_\mathrm{AA}(t)
    \right] G_\mathrm{A}(t) -V^\mathrm{xc}_\mathrm{AB}(t) G_\mathrm{B}(t)
    =\delta(t),
\end{align}
where
\begin{align}
    \varepsilon_\mathrm{B}=-\Delta,\quad \varepsilon_\mathrm{A}=\Delta.
\end{align}
The Green functions in the bonding and antibonding orbitals
can be solved analytically and are given by
\begin{align}\label{eq:GB+}
iG_\text{B}(t  &  >0)=a_{0}^2 (1-x)^{2}\exp(-i\varepsilon_{1}^{+}t),
\end{align}
\begin{align}\label{eq:GB-}
iG_\text{B}(t  &  <0)=-a_{0}^2 (1+x)^{2}\exp(i\varepsilon_{0}^{-}t).
\end{align}
\begin{align}\label{eq:GA+}
iG_\text{A}(t  &  >0)=a_{0}^2 (1+x)^{2}\exp(-i\varepsilon_{0}^{+}t),
\end{align}
\begin{align}\label{eq:GA-}
iG_\text{A}(t  &  <0)=-a_{0}^2 (1-x)^{2}\exp(i\varepsilon_{1}^{-}t),
\end{align}
where
\begin{equation}
    a_0^2 = \frac{1}{2(1+x^2)}.
\end{equation}
The orbital energies are defined as
\begin{equation}
    \varepsilon^\pm_m=E^\pm_m-E_0.
\end{equation}

For $t>0$,
\begin{align}\label{eq:VxcBB}
     V^\mathrm{xc}_\mathrm{BB}(t>0)=\frac{\alpha U}{2}
     \frac{1-\alpha^2 e^{-i4\Delta t}}{1-\alpha^4 e^{-i4\Delta t}},
\end{align}
\begin{align}\label{eq:VxcAB}
     V^\mathrm{xc}_\mathrm{AB}(t>0)=\frac{\alpha U}{2}
     \frac{(1-\alpha^2) e^{-i2\Delta t}}{1-\alpha^4 e^{-i4\Delta t}},
\end{align}
shown in Figs. \ref{fig:VxcBB} and \ref{fig:VxcAB}.
Note the definition of $V^\mathrm{xc}_\mathrm{BB}$ and
$V^\mathrm{xc}_\mathrm{AB}$ in Eqs. (\ref{eq:defVxcBB})
and (\ref{eq:defVxcAB}).

For a finite $\Delta$ one observes that $x=1$ when $U=0$, and $x$ approaches
zero as $U \rightarrow\infty$ so that $0<x\leq 1$, which implies that
$\alpha <1$. It is therefore legitimate to expand the denominator in powers
of $\alpha$ giving
\begin{align}
     V^\mathrm{xc}_\mathrm{BB}(t>0)&=\frac{\alpha U}{2}
     \left(1-\alpha^2 e^{-i4\Delta t}\right)
     \left(1+\alpha^4 e^{-i4\Delta t} + ...\right)
    \nonumber\\
    &=\frac{\alpha U}{2}\left\{ 1 -\alpha^2(1-\alpha^2)e^{-i4\Delta t}
    -\alpha^6 e^{-i8\Delta t} \right\},
\end{align}
\begin{align}
    V^\mathrm{xc}_\mathrm{AB}(t>0)&=\frac{\alpha U}{2}
     (1-\alpha^2) e^{-i2\Delta t}\left(1+\alpha^4 e^{-i4\Delta t}+...\right)
\nonumber\\
&=\frac{\alpha U}{2}
     (1-\alpha^2)\left( e^{-i2\Delta t}+\alpha^4 e^{-i6\Delta t}\right)
     +...
\end{align}

For a relatively strong correlation corresponding to $\frac{U}{4\Delta}=2$,
one finds $\alpha=0.62$. This yields $1-\alpha^2=0.62$ and
$\alpha^2(1-\alpha^2)=0.24$. The dominating terms in $V^\mathrm{xc}$ are
\begin{align}
     V^\mathrm{xc}_\mathrm{BB}(t>0)\approx \frac{\alpha U}{2},
\end{align}
\begin{align}
    V^\mathrm{xc}_\mathrm{AB}(t>0)&=\frac{\alpha U}{2}
    (1-\alpha^2) e^{-i2\Delta t}.
\end{align}
The first term shifts the one-particle energy, whereas the second
gives rise to an incoherent or satellite feature.
This approximate Vxc of the dimer
will be extrapolated later to the infinite chain.
The higher-energy excitations are multiples
of $2\Delta$, which is the energy difference between the bonding
and antibonding energies. This energy difference may be regarded as the bosonic
excitation (particle-hole) energy.

An interesting feature of Vxc is the presence of a discontinuity
at $t=0$, reminiscent of the derivative discontinuity of the xc energy
in DFT \cite{perdew1982}. As alluded earlier in Sec. \ref{sec:ConnectionKS},
Vxc for the negative-time
Green function is expected to be more negative than the one for positive-time 
since the former is the Coulomb potential of an xc hole that
integrates to $-1$ whereas the latter is to zero. This difference
gives rise to the discontinuity and provides a physical explanation for
the well-known band-gap underestimation in Kohn-Sham DFT.
The xc potential in Kohn-Sham DFT corresponds to occupied orbitals only,
that is, to the negative-time Green function (creation of a hole creation or
removal of an electron). The unoccupied orbitals corresponding to
the positive-time Green function should see an xc potential that is more
positive than that of the occupied orbitals.
However, the Kohn-Sham unoccupied and occupied orbitals see the same
potential, which then leads to an underestimation of the band gap in
semiconductors and insulators. Thus, even if the exact Kohn-Sham xc
potential were known, it would likely give rise to a band-gap underestimation.
In principle, there is nothing wrong with this since the Kohn-Sham scheme
is not meant to describe excited states. 

\begin{figure}[t]
\begin{center} 
\includegraphics[scale=0.6, viewport=3cm 8cm 17cm 20cm, clip,
width=0.9\columnwidth]
{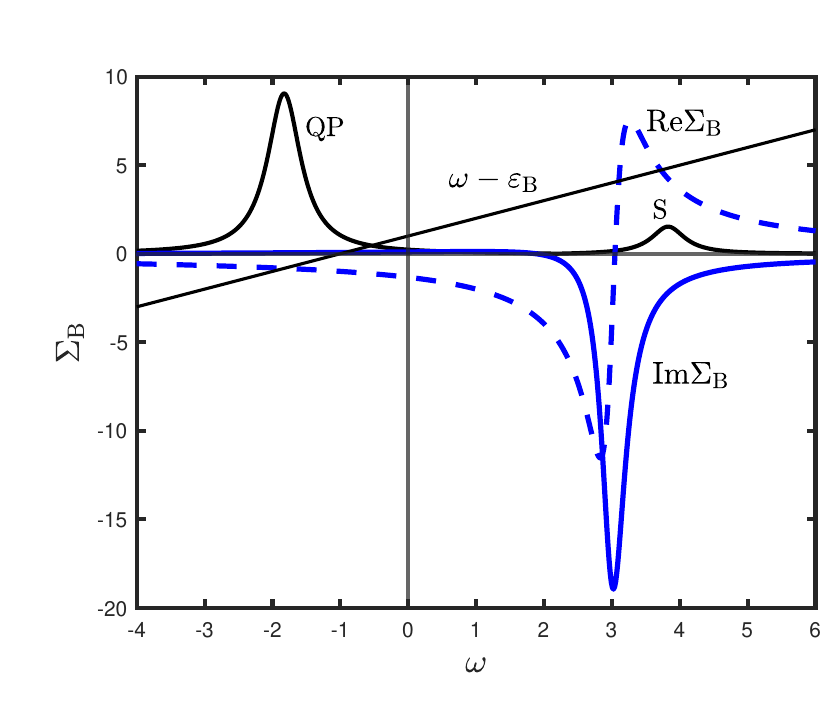}
\caption{The real and imaginary parts of the self-energy and the 
spectral function of the Hubbard dimer
for $U=4$ and $\Delta=1$ in the bonding orbital.
The energies at which the straight line $\omega-\varepsilon_\mathrm{B}$
crosses the real part of the self-energy Re$\Sigma_\mathrm{B}$
correspond to the main peak (QP) in the spectral function
and an additional structure in the
unoccupied region (S). Both of these peaks are magnified ten times for clarity
and a broadening of $0.3$ has been used for plotting purpose.
The crossing at $\omega\approx 3$ does not yield any spectral feature
since Im$\Sigma_\mathrm{B}$ at that energy is very large.
The spectral function for the antibonding orbital is the mirror image
of that of the bonding-orbital with respect to $\omega=0$.
}
\label{fig:Sigma_dimer}%
\end{center}
\end{figure}

In terms of the effective potential defined in Eq. (\ref{eq:Xi}),
the equations of motion in the bonding-antibonding basis can be recast as
\begin{align}
    \left[i\partial_t-\varepsilon_\mathrm{B}-\Xi_\mathrm{B}(t)
    \right] G_\mathrm{B}(t)
    =\delta(t),
\end{align}
\begin{align}
    \left[i\partial_t-\varepsilon_\mathrm{A}-\Xi_\mathrm{A}(t)
    \right] G_\mathrm{A}(t)
    =\delta(t),
\end{align}
where according to the definition in Eq. (\ref{eq:Xi}),
\begin{align}
    \Xi_\mathrm{B}(t)&=V_\mathrm{BB}(t)+V_\mathrm{BA}(t)
    \frac{G_\mathrm{A}(t)}{G_\mathrm{B}(t)},
\\
    \Xi_\mathrm{A}(t)&=V_\mathrm{AA}(t)+V_\mathrm{AB}(t)
    \frac{G_\mathrm{B}(t)}{G_\mathrm{A}(t)}.
\end{align}
These yield
\begin{align}
    \Xi_\mathrm{B}(t<0)    = -\frac{\alpha U}{2},
    \qquad
        \Xi_\mathrm{B}(t>0)=\frac{U}{2\alpha}.
\end{align}
\begin{align}
    \Xi_\mathrm{A}(t>0)    = \frac{\alpha U}{2},
    \qquad
        \Xi_\mathrm{A}(t<0)=-\frac{U}{2\alpha},
\end{align}
Note that the expressions for
$\Xi_\mathrm{A}(t<0)$ and $\Xi_\mathrm{B}(t>0)$ are valid for $U>0$ since
if $U=0$ then $G_\mathrm{A}(t<0)=G_\mathrm{B}(t>0)=0$. 
The effective potential $\Xi$ takes a very simple form, namely
a constant in each time segment and in stark contrast to the
traditional self-energy shown in Fig. \ref{fig:Sigma_dimer}
for the bonding orbital.
The self-energy in the frequency domain
$\Sigma_\mathrm{B}(\omega)$ is defined as
\begin{align}
    G_\mathrm{B}(\omega)=\frac{1}
    {\omega-\varepsilon_\mathrm{B}-\Sigma_\mathrm{B}(\omega)}.
\end{align}
In the Vxc formalism,
$\Xi_\mathrm{B}(t<0)$ simply shifts the
bonding energy $\varepsilon_\mathrm{B}$ by $-\frac{\alpha U}{2}$,
and as a consequence of the many-electron interactions,
$\Xi_\mathrm{B}(t>0)$ generates a new feature in the unoccupied part
of the spectrum separated in energy from $\varepsilon_\mathrm{B}$ by
$\frac{U}{2\alpha}$.
The appearance of spectral weight above the chemical potential
originating from the antibonding orbital compensates for
the loss of spectral weight of the bonding orbital for $t<0$,
as can be seen in Eqs. (\ref{eq:GA-}) and (\ref{eq:GB-}),
so that the sum of the spectral weights for $t<0$ is equal to one. 
A similar consideration applies to the case $t>0$.

\section{Exchange-correlation hole and Vxc of the homogeneous electron gas}
\label{sec:HEG}

\begin{figure}[tb]
\begin{center} 
\includegraphics[scale=0.5, viewport=7.5cm 6cm 25cm 15cm, clip,width=\columnwidth]
{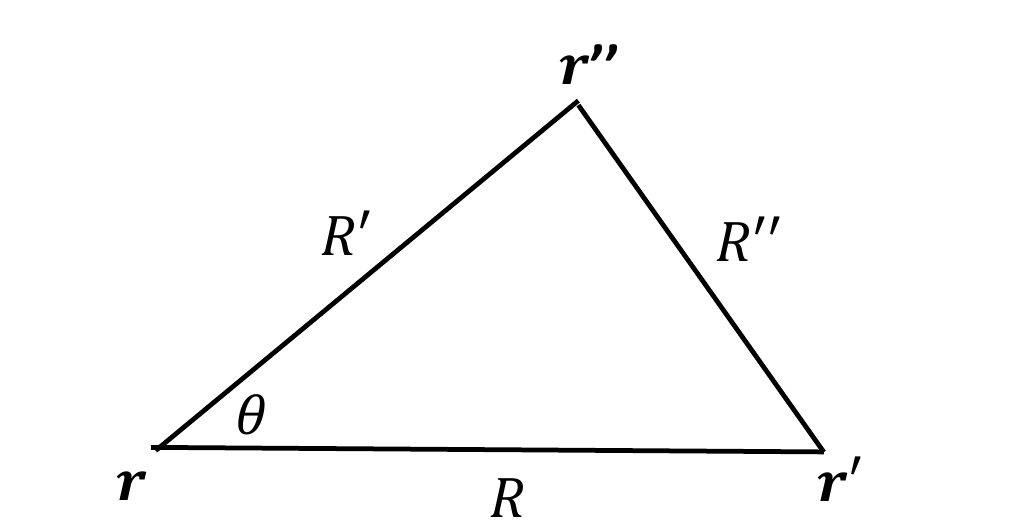}
\caption{Definition of the radial variables $R$, $R'$, and $R''$. They
are related to the angle $\theta$ by ${R''}^2=R^2-2RR'\cos{\theta} +{R'}^2$.
The corresponding vectors are defined as $\mathbf{R}=\mathbf{r}'-\mathbf{r}$,
$\mathbf{R}'=\mathbf{r}''-\mathbf{r}$, and 
$\mathbf{R}''=\mathbf{r}''-\mathbf{r}'$.
}
\label{fig:Coordinates}%
\end{center}
\end{figure}
%

%
\begin{figure}[t]
\begin{center} 
\includegraphics[scale=0.4, viewport=3cm 8cm 18cm 20cm, clip,
width=0.9\columnwidth]
{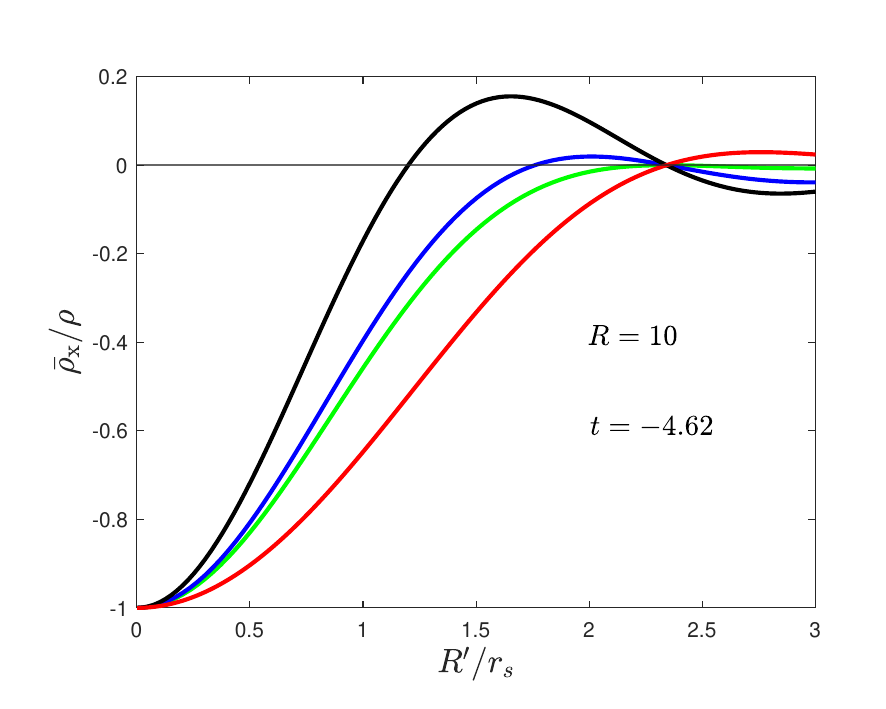}
\hfill
\includegraphics[scale=0.4, viewport=3cm 8cm 18cm 20cm, clip,
width=0.8\columnwidth]
{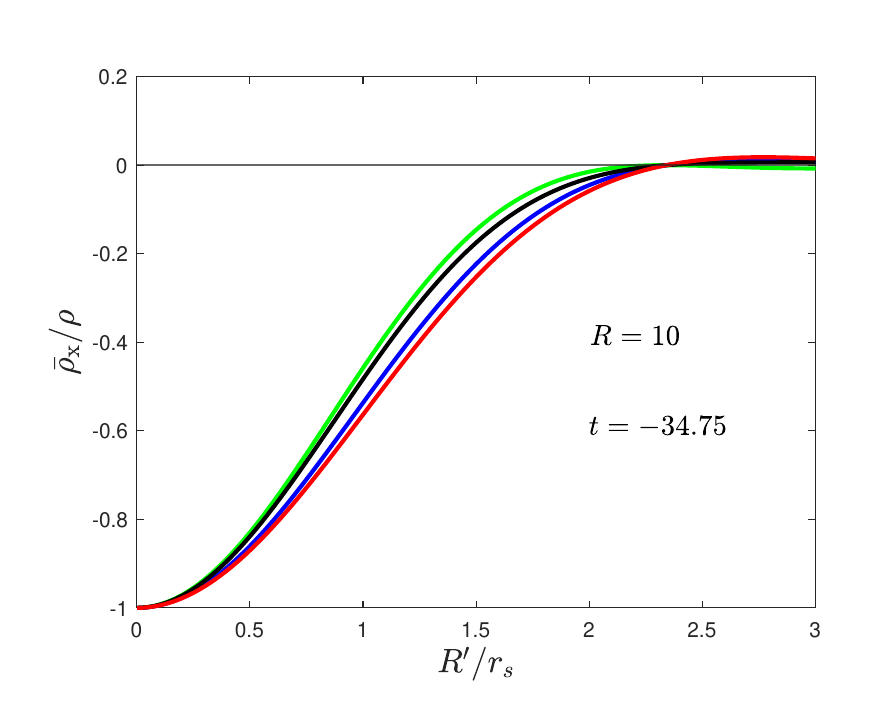}
\caption{The spherical average of the exchange hole of the electron gas scaled 
with the density with
Wigner-Seitz radius $r_s=3$ (blue), $4$ (black), and $5$ (red) plotted
against $R'/r_\mathrm{s}$ for the case
$R=10$ and $t=-4.62$ (top) and $-34.75$ (bottom),
corresponding to the inverse of the plasmon energy and the Fermi energy, 
respectively. The green curve is the static exchange hole, which is independent
of $r_\mathrm{s}$.
}
\label{fig:rhoxR10T-34}%
\end{center}
\end{figure}

\begin{figure}[h]
\begin{center} 
\includegraphics[scale=0.6, viewport=3.2cm 8cm 17.8cm 20cm, clip,width=\columnwidth]
{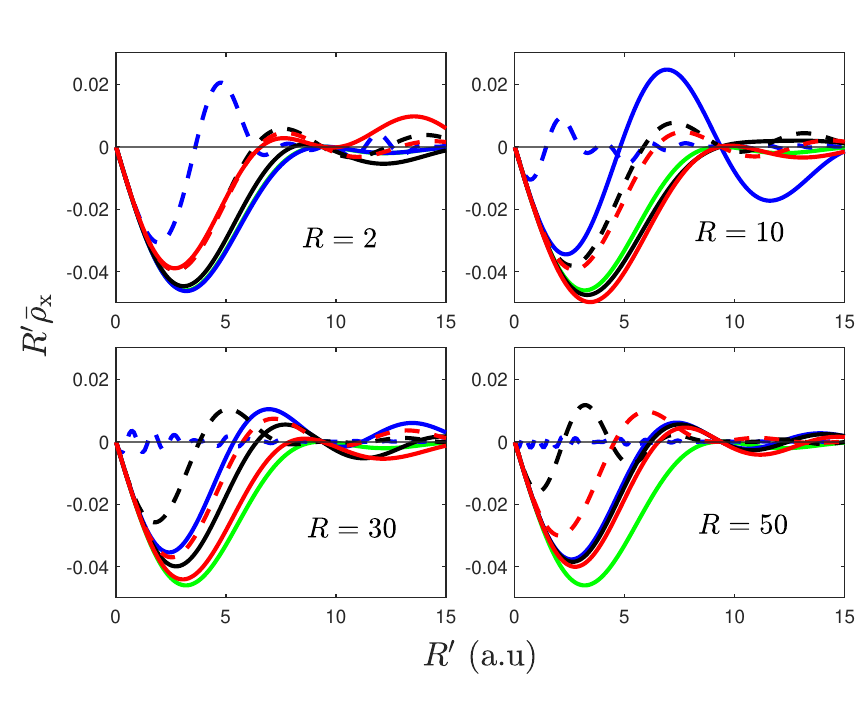}
\caption{
The real part of the spherical average of the 
exchange hole
multiplied by $R'$ for $t=\pm 4.62$ (blue), $\pm 34.75$ (black), $\pm 69.5$ (red),
and $R=2,10,30,50$.
The solid and dashed curves correspond to $t<0$ and $t>0$, respectively.
The green curve is the static exchange hole from the Hartree-Fock approximation.
For $t=-4.62$ (solid blue) and $R=2$ 
the exchange hole is virtually indistinguishable from
the static one \cite{karlsson2023}.
}
\label{fig:rhoxAv}%
\end{center}
\end{figure}
\begin{figure}[t]
\begin{center} 
\includegraphics[scale=0.6, viewport=3.2cm 8cm 17.8cm 20cm, clip,width=\columnwidth]
{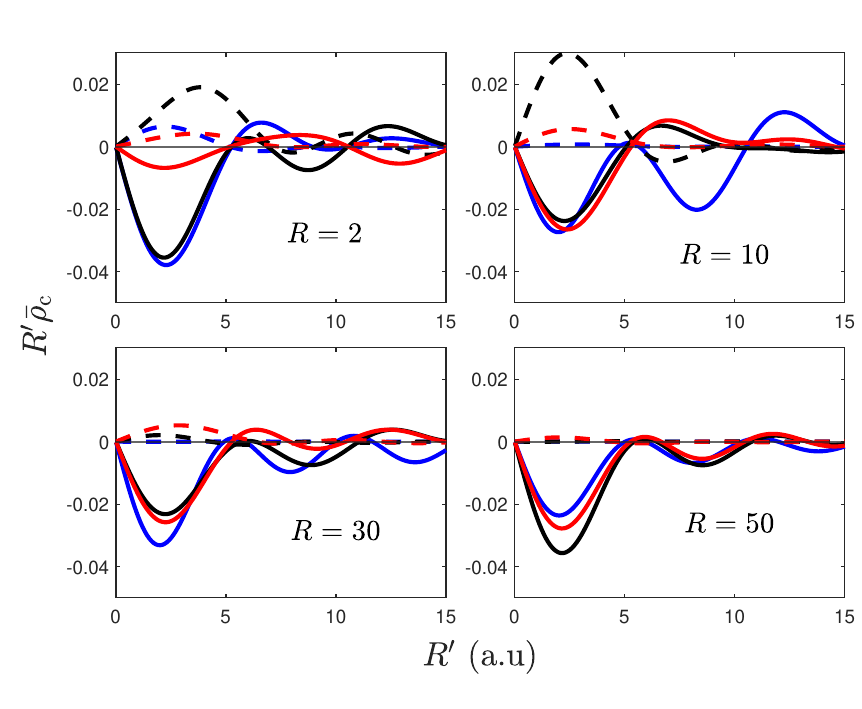}
\caption{
The real part of the spherical average of the 
correlation hole
multiplied by $R'$ for $t=\pm 4.62$ (blue), $\pm 34.75$ (black), $\pm 69.5$ (red),
and $R=2,10,30,50$.
The solid and dashed curves correspond to $t<0$ and $t>0$,
respectively  \cite{karlsson2023}.
}
\label{fig:rhocAv}%
\end{center}
\end{figure}

\begin{figure}[h]
\begin{center} 
\includegraphics[scale=0.6, viewport=3.2cm 8cm 17.8cm 20cm, clip,width=\columnwidth]
{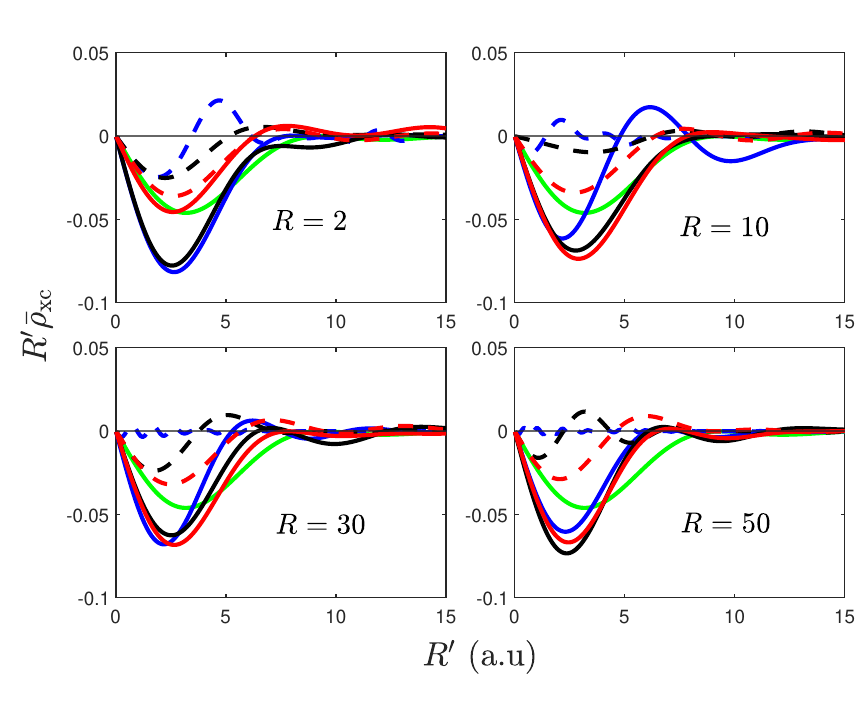}
\caption{
The real part of the spherical average of the 
xc hole
multiplied by $R'$ for $t=\pm 4.62$ (blue), $\pm 34.75$ (black), $\pm 69.5$ (red),
and $R=2,10,30,50$ \cite{karlsson2023}.
The solid and dashed curves correspond to $t<0$ and $t>0$, respectively.
The green curve is the static exchange hole from the Hartree-Fock approximation.
For $t=-4.62$ (solid blue) and $R=2$ 
the exchange hole is virtually indistinguishable from
the static one.
}
\label{fig:rhoxcAv}%
\end{center}
\end{figure}

\begin{figure}[t]
\begin{center} 
\includegraphics[scale=0.5, viewport=7.5cm 6cm 20cm 17cm, clip]
{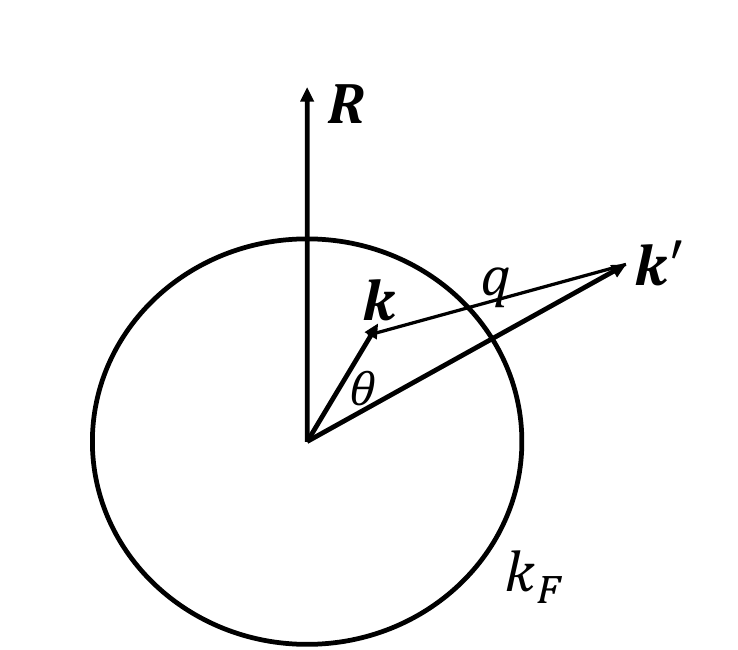}
\caption{Integration over $\mathbf{k}$ and $\mathbf{k}'$ for the correlation hole.
The angle between $\mathbf{k}$ and $\mathbf{k}'$ is labelled by $\theta$.
The angle between $\mathbf{k}'$ and $\mathbf{R}$ is $\theta'$ (not labelled in the figure).
$q=\sqrt{{k'}^2+k^2-2k'ky}$ in which $y=\cos{\theta}$.}
\label{fig:Integral}%
\end{center}
\end{figure}

\begin{figure}[h]
\centering
\includegraphics[scale=0.4, viewport=3cm 8cm 17cm 20cm, clip, width=0.8\columnwidth]
{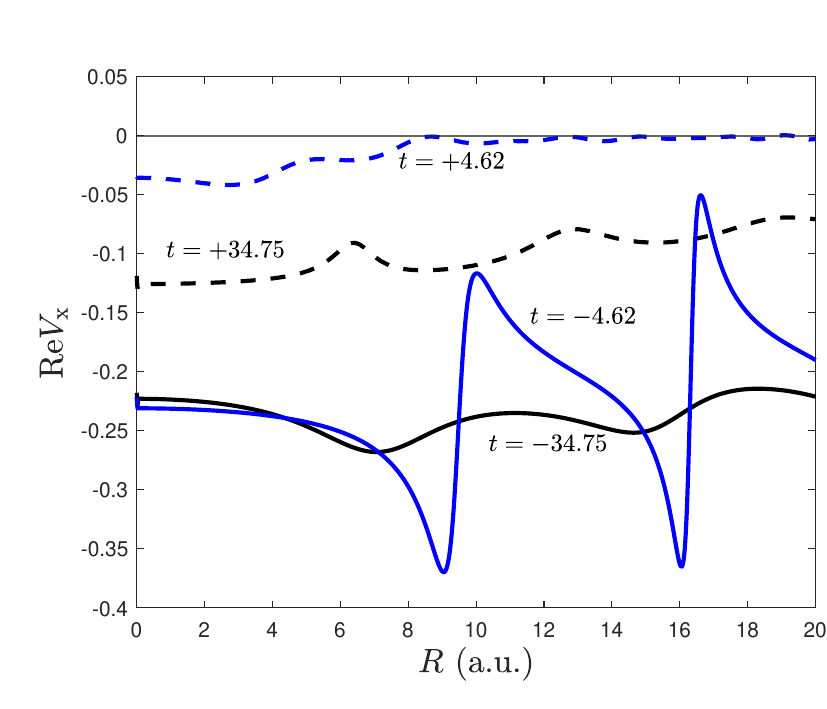}
\hfill
\includegraphics[scale=0.4, viewport=3cm 8cm 17cm 20cm, clip, width=0.8\columnwidth]
{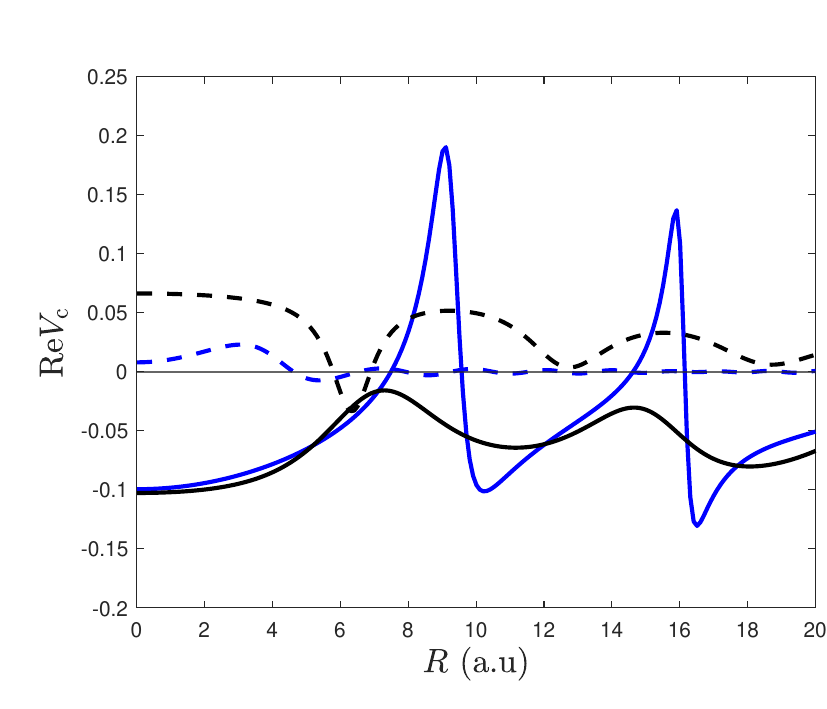}
\hfill
\includegraphics[scale=0.4, viewport=3cm 8cm 17cm 20cm, clip, width=0.8\columnwidth]
{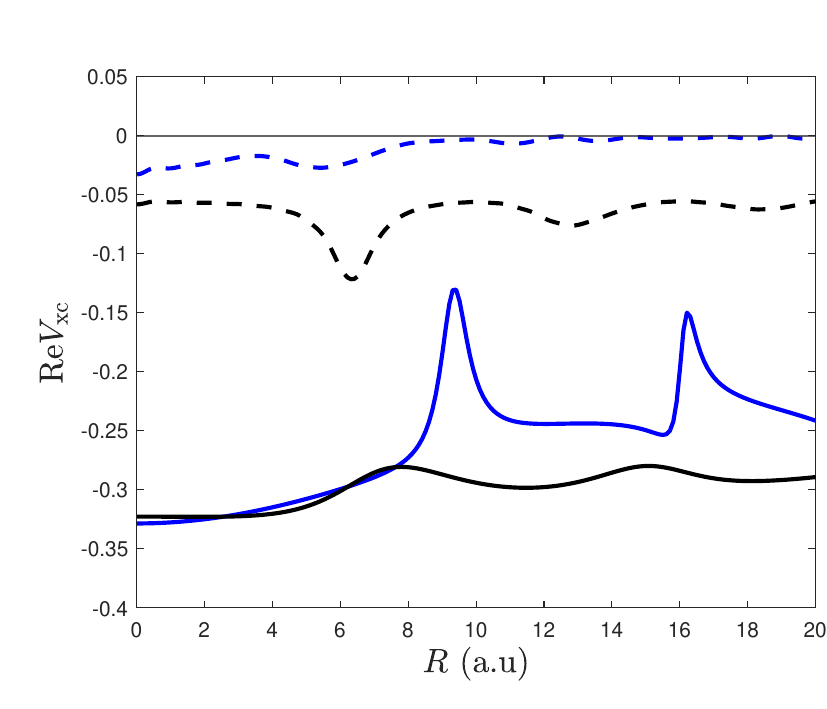}
\caption{
The real part of the exchange potential (top), the correlation potential (middle),
and Vxc for $t=\pm 4.62$ (blue) and $\pm 34.75$
(black). The solid and dashed curves correspond to $t<0$ and $t>0$,
respectively \cite{karlsson2023}.
}
\label{fig:ReVxcT4-34}%
\end{figure}
\begin{figure}[h]
\centering
\includegraphics[scale=0.4, viewport=3cm 8cm 17cm 20cm, clip, width=0.8\columnwidth]
{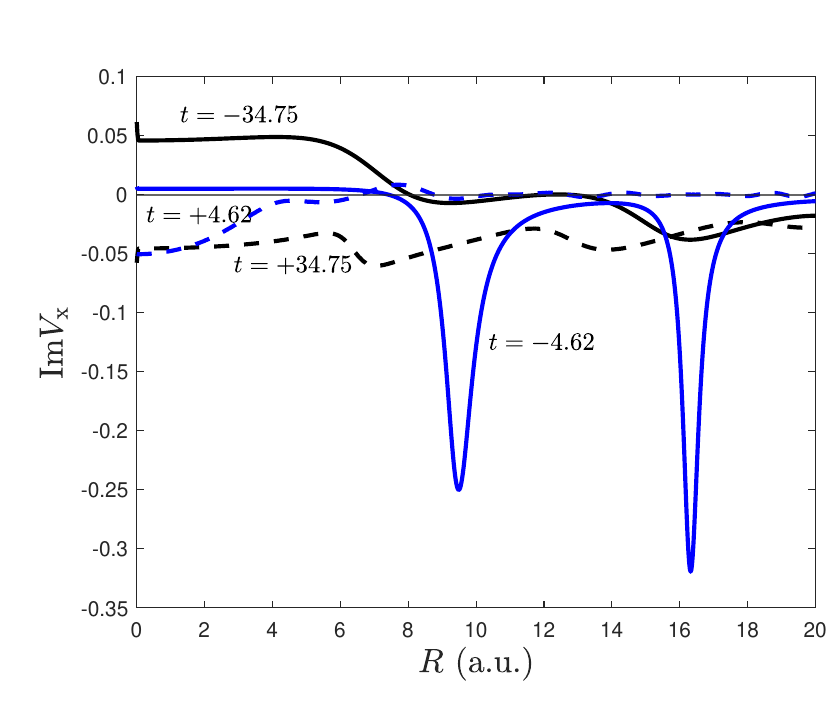}
\hfill
\includegraphics[scale=0.4, viewport=3cm 8cm 17cm 20cm, clip, width=0.8\columnwidth]
{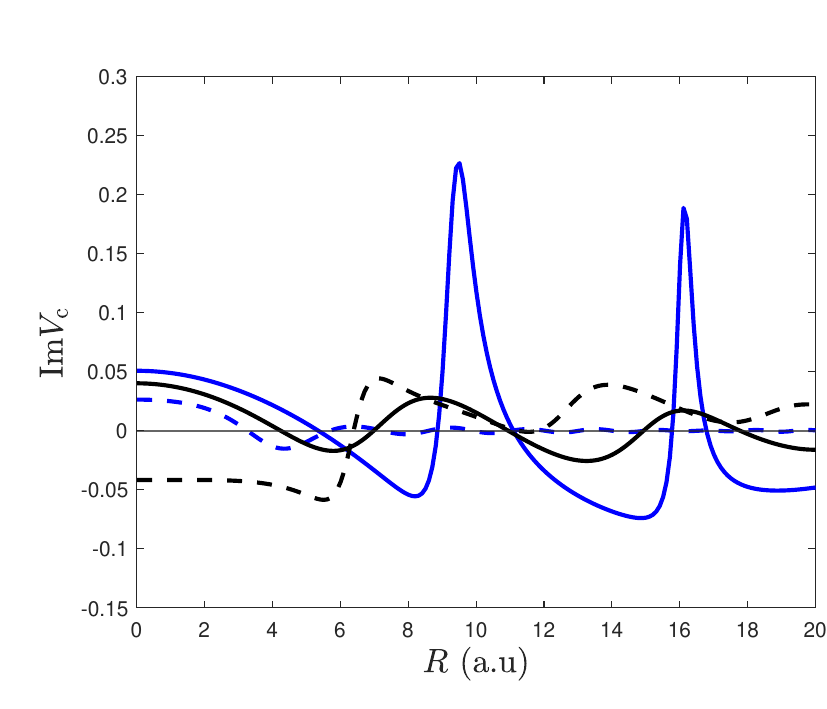}
\hfill
\includegraphics[scale=0.4, viewport=3cm 8cm 17cm 20cm, clip, width=0.8\columnwidth]
{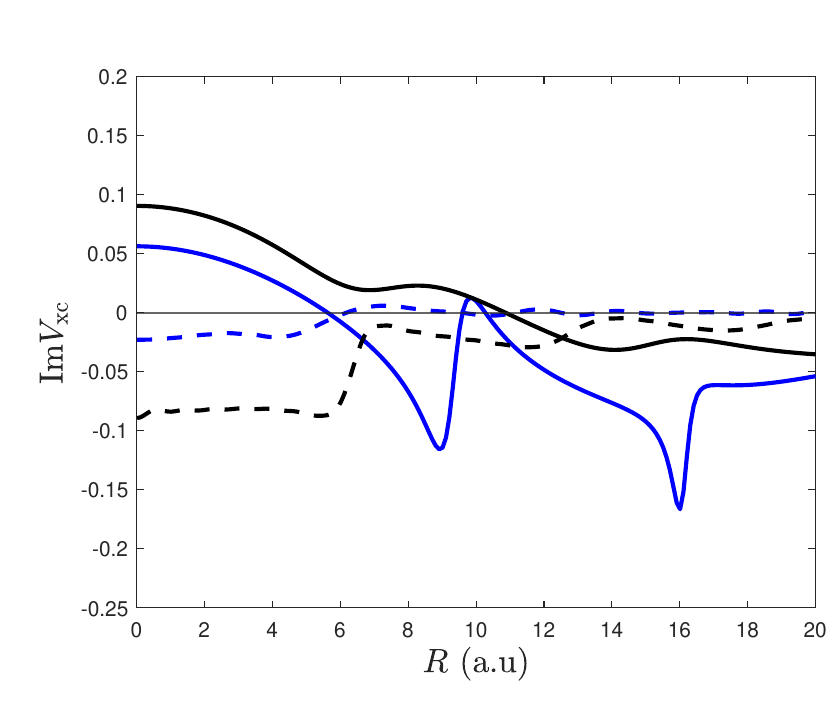}
\caption{
The imaginary part of the exchange potential (top), the correlation potential
(middle), and Vxc (bottom) for $t=\pm 4.62$ (blue) and $\pm 34.75$
(black). The solid and dashed curves correspond to $t<0$ and $t>0$,
respectively \cite{karlsson2023}.
}
\label{fig:ImVxcT4-34}
\end{figure}
\begin{figure}[t]
\begin{center} 
\includegraphics[scale=0.6, viewport=3cm 8cm 17cm 20cm, clip,
width=0.8\columnwidth]
{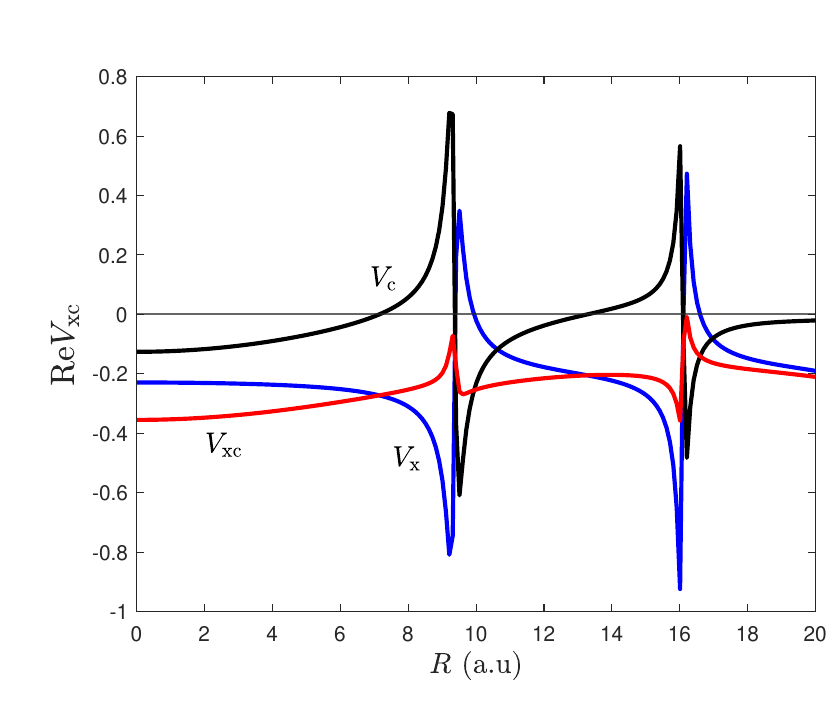}
\caption{
The real part of the exchange potential (blue), the correlation potential (black),
and Vxc (red) for $t=-1$. The large cancellation 
between exchange and correlation can be clearly seen \cite{karlsson2023}.
}
\label{fig:VxcT1}%
\end{center}
\end{figure}

The homogeneous electron gas is an invaluable model of itinerant valence
electrons in solids, which usually originate from $s$ and $p$ orbitals.
It is also an indispensable reference system when employing LDA
as an extrapolation scheme for application to real materials.
It is therefore very important to study its xc hole and potential.
A more detailed treatment of the present section can be found in Ref.
\cite{karlsson2023}.

The Green function of the
noninteracting homogeneous electron gas is given by
\begin{align}
    iG_0(r,r';t) &= \frac{1}{\Omega} \sum_{k> k_\mathrm{F}} 
    e^{i\mathbf{k}\cdot(\mathbf{r}-\mathbf{r}')} e^{-i\varepsilon_k t}\theta(t)
    \nonumber\\
    &-\frac{1}{\Omega} \sum_{k\leq k_\mathrm{F}} 
    e^{i\mathbf{k}\cdot(\mathbf{r}-\mathbf{r}')} e^{-i\varepsilon_k t}\theta(-t),
    \label{eq:G0elgas}
\end{align}
where $\varepsilon_k=\frac{1}{2}k^2$, $k_\mathrm{F}$ is the Fermi wave vector,
and $\Omega$ is
the space volume. For the paramagnetic case considered here $\sigma=\sigma'$.

For the homogeneous electron gas, it is convenient to introduce the variable
$\mathbf{R}=\mathbf{r}'-\mathbf{r}$. 
The Green function as well as Vxc depend only on the spatial separation $R$.
In spherical coordinates, the $\delta$-function is given by
\begin{align}
    \delta(\mathbf{R})=\frac{1}{R^2}\delta(R)\delta(y)\delta(\phi),
\end{align}
where $y=\cos{\theta}$ ($\theta$ is the polar angle in spherical coordinates).
For a radial function $f(R)$ without angular dependence,
\begin{align}
    \nabla^2 f= \frac{1}{R} \frac{\partial^2}{\partial R^2}(Rf).
\end{align}
After integrating over the solid angle, the equation of motion becomes
\begin{equation}
   \left(  i\frac{\partial}{\partial t}-h(R)-V_\mathrm{xc}(R,t)\right)
\widetilde{G}(R,t)= \frac{1}{4\pi R}\delta(R)\delta(t),
\label{eq:EOMelgas}
\end{equation}
where
\begin{equation}
    h(R)=-\frac{1}{2}\frac{\partial^2}{\partial R^2},\qquad \widetilde{G}(R,t)=R \,G(R,t).
\end{equation}

\subsection{Exchange hole}
\label{sec:ExchangeHole}

Using the noninteracting Green function
in Eq. (\ref{eq:G0elgas}) the exchange hole in Eq. (\ref{eq:x-hole})
can be calculated analytically,
and for the case $t<0$ one finds 
\begin{align}
   & \rho_\mathrm{x}(r,r',r'';t<0)
    \sum_{k\leq k_F} 
    e^{i\mathbf{k}\cdot(\mathbf{r}-\mathbf{r}')} e^{-i\varepsilon_k t}
    \nonumber\\
    &=-\frac{1}{\Omega} \sum_{k'\leq k_F} 
    e^{i\mathbf{k}'\cdot(\mathbf{r}-\mathbf{r}'')}
    \times\sum_{k\leq k_F} 
    e^{i\mathbf{k}\cdot(\mathbf{r}''-\mathbf{r}')} e^{-i\varepsilon_k t}.
    \label{eq:rhox}
\end{align}
For $t>0$,
\begin{align}
   & \rho_\mathrm{x}(r,r',r'';t>0)
    \sum_{k> k_F} 
    e^{i\mathbf{k}\cdot(\mathbf{r}-\mathbf{r}')} e^{-i\varepsilon_k t}
    \nonumber\\
    &=-\frac{1}{\Omega} \sum_{k'\leq k_F} 
    e^{i\mathbf{k}'\cdot(\mathbf{r}-\mathbf{r}'')}
    \times\sum_{k> k_F} 
    e^{i\mathbf{k}\cdot(\mathbf{r}''-\mathbf{r}')} e^{-i\varepsilon_k t}.
    \label{eq:rhox+}
\end{align}

It can be immediately seen that by setting $r''=r$
the exact constraint in
Eq. (\ref{eq:ExactReln}) is satisfied and
by integrating over $r''$ the sum rule in Eq. (\ref{eq:SumRule}) is fulfilled.
Expressed in terms of the radial variables $R$, $R'$, $R''$ and the angle $\theta$ 
as defined in Fig. \ref{fig:Coordinates},
\begin{align}
    \rho_\mathrm{x}(R,R',\theta;t)
    &=iG_0(R',0^-)\frac{G_0(R'',t)}{G_0(R,t)} 
\end{align}
where $R''$ depends on $\theta$.

$G_0(R,t)$ can be calculated from Eq. (\ref{eq:G0elgas})
by performing the integral over 
the solid angle in $\mathbf{k}$, yielding
\begin{align}
    iG_0(R,t<0)
    &=-\frac{1}{2\pi^2}\frac{1}{R}\int_0^{k_\mathrm{F}} dk\, k\sin{(kR)}e^{-ik^2 t/2}, 
\end{align}
\begin{align}
    iG_0(R,t>0)
    &=\frac{1}{2\pi^2}\frac{1}{R}\int_{k_\mathrm{F}}^\infty dk\, k\sin{(kR)}
    e^{-ik^2 t/2} ,   
\end{align}
\begin{align}
    iG_0(R,0^-)
    &=-\frac{1}{2\pi^2}\frac{1}{R^3}\left[ \sin{(k_\mathrm{F}R)}
    -k_\mathrm{F}R\cos{(k_\mathrm{F}R)} \right] .
\end{align}
$G_0(R,t<0)$ can be expressed in terms of the complex error function or calculated
numerically using a standard quadrature. 

To calculate $G_0(R,t>0)$ requires more care.
The integral over $k$ can be decomposed as follows:
\begin{equation}
iG_0(R,t>0)=\frac{1}{2\pi^{2}R}
\left[  I(0,\infty)-I(0,k_{\text{F}})\right]  ,
\end{equation}
where
\begin{equation}
I(a,b)=\int_{a}^{b}dk\text{ }k\sin(kR)e^{-ik^{2}t/2}.
\end{equation}
%
%
The integral $I(0,\infty)$ can be performed analytically resulting in
\begin{align}
    I(0,\infty)
    &=\sqrt{\frac{\pi}{2it}}\frac{R}{it} e^{iR^2/2t}.
\end{align}

%
%


\subsubsection{Spherical average of the exchange hole}
For the homogeneous electron gas, $\mathbf{r}$ may be chosen as the origin of
coordinate and set to zero. Defining 
$\mathbf{R}=\mathbf{r}'-\mathbf{r}=\mathbf{r}'$ and
$\mathbf{R}'=\mathbf{r}''-\mathbf{r}=\mathbf{r}''$ 
as illustrated in Fig. \ref{fig:Coordinates} 
the exchange hole in Eq. (\ref{eq:rhox}) becomes for $t<0$
\begin{align}
    &\rho_\mathrm{x}(R,R',\theta;t<0)\times iG_0(R,t<0)
    \nonumber\\
    &=-\frac{1}{\Omega^2} \sum_{k\leq k_F}
    e^{-i\mathbf{k}\cdot \mathbf{R}} e^{-i\varepsilon_k t}
    \sum_{k'\leq k_F}  
    e^{i\mathbf{q}\cdot \mathbf{R}'},
\end{align}
where $\mathbf{q}=\mathbf{k}-\mathbf{k}'$. 
The spherical averaging of $\rho_{\text{x}}$ 
amounts to performing a solid-angle
integration over $\mathbf{R}'$,
\begin{align}
\int d\Omega_{R'}e^{i\mathbf{q}\cdot
\mathbf{R}^{\prime}} 
& =4\pi\frac{\sin(qR^{\prime})}{qR^{\prime}},
\end{align}
yielding for $t<0$
\begin{align}
&\overline{\rho}_{\text{x}}(R,R^{\prime},t<0)iG_0(R,t)
\nonumber\\
&=\frac{1}{\Omega^{2}}%
\sum_{k,k^{\prime}\leq k_{\text{F}}}e^{-i\mathbf{k}\cdot\mathbf{R}%
}e^{-i\varepsilon_{k}t}\times4\pi\frac{\sin(qR^{\prime})}{qR^{\prime}}.
\label{eq:rhoxav}
\end{align}
%

Fig. \ref{fig:rhoxR10T-34} illustrates the spherical average of the
exchange hole for $r_s=3,4,5$, $R=10$, and a couple of typical times.
Fig. \ref{fig:rhoxAv} shows examples of the spherical average of the exchange
hole multiplied by the radial distance $R'$ for $r_s=4$ and for several values
of $R$.
As anticipated, the exchange holes for positive times are noticeably
more oscillatory than those for negative times. For negative times, the
exchange holes closely follow the static Hartree-Fock exchange hole.

\subsection{Correlation hole}
The density-response contribution to $\rho_\mathrm{xc}$
is given by the second term on
the right-hand side of Eq. (\ref{eq:xchole}):
\begin{align}
    &\rho_\mathrm{c}(1,2,3)G(1,2)=i\int d4\,G(1,4) K(4,3)G(4,2)
\end{align}
where $K=\delta V_\mathrm{H}/\delta \varphi=vR$ as defined earlier in 
Eq. (\ref{def:K}). 
For the homogeneous electron gas, $G$ and
$K$ depend only on the separation of the spatial and time coordinates.
Keeping in mind that $t_3=t_1=t$, $t_2=0$ and setting $r_1=r$,
$r_2=r'$, and $r_3=r''$, one finds
\begin{align}
    &\rho_\mathrm{c}(r,r',r'';t)G(r-r';t)
    \nonumber\\
    &=i\int dr_4 dt_4 G(r-r_4,t-t_4) K(r_4-r'',t_4-t)
    \nonumber\\
    &\qquad\qquad\times G(r_4-r',t_4).
\label{eq:c-hole}
\end{align}
To calculate $\rho_\mathrm{c}$, a noninteracting $G_0$ as in
Eq. (\ref{eq:G0elgas}) will be used.
%
%
%
$K$ is written in terms of its Fourier components,
\begin{align}
    &K(r_4-r'',t_4-t)
    \nonumber\\
    &=\frac{1}{\Omega}\sum_q\int \frac{d\omega}{2\pi}\,
    e^{-i\mathbf{q}\cdot(\mathbf{r}_4-\mathbf{r}'')}e^{-i\omega(t_4-t)}K(q,\omega).
\end{align}
Within RPA, $K(q,\omega)$ is known analytically.
Due to the time ordering, the product of two Green functions appearing on
the right-hand side of Eq. (\ref{eq:c-hole}) 
yields only two nonzero terms. 
Consider the case $t<0$.
The first nonzero term is (note the additional factor of $i$ from $iGKG$)
\begin{align}
    A_1
    &=\frac{1}{\Omega^2}\sum_{k\leq k_F}\sum_{k'> k_F}
    e^{-i\mathbf{k}'\cdot\mathbf{R}}
    e^{i\mathbf{q}\cdot\mathbf{R}'}
    \nonumber\\
    &\qquad\times \int \frac{d\omega}{2\pi}\, K(q,\omega)
    \frac{e^{i(\omega-\varepsilon_k)t}}
    {\omega-\varepsilon_k+\varepsilon_{k'}-i\eta},
    \label{eq:A1a}
\end{align}
where $\mathbf{R}$
and $\mathbf{R}'$ are defined in Fig. \ref{fig:Coordinates}
and $\mathbf{q}=\mathbf{k}'-\mathbf{k}$.

The spectral representation of $K$,
\begin{align}
    K(q,\omega)=\int_{-\infty}^0 d\omega'\,\frac{L(q,\omega')}{\omega-\omega'-i\delta}
    +\int_0^{\infty}d\omega'\,\frac{L(q,\omega')}{\omega-\omega'+i\delta},
\label{eq:K(q,w)}
\end{align}
where
\begin{equation}
    L(q,\omega)=-\frac{1}{\pi}\mathrm{sign}(\omega) \mathrm{Im} K(q,\omega),
    \label{eq:LkwA}
\end{equation}
is used to perform the integral over $\omega$.
The spectral function $L(q,\omega)$ is an odd function in $\omega$.

For the case
$t<0$, the contour integral for $A_1$ in the complex $\omega$ plane is closed in 
the lower quadrants, yielding
\begin{align}
&\int \frac{d\omega}{2\pi}\, \frac{1}{\omega-\omega'+i\delta}\times
    \frac{e^{i(\omega-\varepsilon_k)t}}
    {\omega-\varepsilon_k+\varepsilon_{k'}-i\eta}
    \nonumber\\
    &=\frac{-ie^{i(\omega'-\varepsilon_k-i\delta)t}}
    {\omega'-\varepsilon_k+\varepsilon_{k'}-i\eta}.
\end{align}
Using $L(q,-\omega)=-L(q,\omega)$, one finds
\begin{align}
    A_1
    &=    \frac{1}{\Omega^2}\sum_{k\leq k_F}\sum_{k'> k_F}
    e^{-i\mathbf{k}'\cdot\mathbf{R}}
    e^{i\mathbf{q}\cdot\mathbf{R}'}
    e^{-i\varepsilon_{k}t}
     \nonumber\\
&\qquad\times     \int_0^\infty d\omega'\, L(q,\omega')
    \frac{-ie^{i\omega't}}
    {\omega'+\varepsilon_{k'}-\varepsilon_k}.
\end{align}
%
%
%
%
%
The integral over $\omega'$ can be parametrised for each $t$ as follows:
\begin{align}
    M(q,\omega,t) = 
    \int_0^\infty d\omega'\, L(q,\omega')
    \frac{-ie^{i\omega't}}
    {\omega'+\omega},
    \label{eq:MqwtA}
\end{align}
$A_1$ together with the other nonzero $t<0$ term, $A_2$,
and the two nonzero terms corresponding to $t>0$, $B_1$ and $B_2$, become
\begin{align}
    A_1
    =\frac{1}{\Omega^2}\sum_{k\leq k_F}\sum_{k'> k_F}
    e^{-i\mathbf{k}'\cdot\mathbf{R}}
    e^{i\mathbf{q}\cdot\mathbf{R}'}
    e^{-i\varepsilon_{k}t}
     M(q,\varepsilon_{k'}-\varepsilon_{k},t),
    \label{eq:A1M}
\end{align}
%
%
%
%
\begin{align}
    A_2
    =\frac{1}{\Omega^2}
    \sum_{k\leq k_F}\sum_{k'>k_F}
    e^{-i\mathbf{k}\cdot\mathbf{R}}
    e^{i\mathbf{q}\cdot\mathbf{R}'}
    e^{-i\varepsilon_{k}t}
    M(q,\varepsilon_{k'}-\varepsilon_{k},0),
    \label{eq:A2M}
\end{align}
%
%
\begin{align}
    B_1
    =\frac{1}{\Omega^2}\sum_{k\leq k_F}
    \sum_{k'> k_F}
    e^{-i\mathbf{k}'\cdot\mathbf{R}}
    e^{i\mathbf{q}\cdot\mathbf{R}'}
    e^{-i\varepsilon_{k'}t}
     M(q,\varepsilon_{k'}-\varepsilon_{k},0),
     \label{eq:B1M}
\end{align}
%
%
%
\begin{align}
    B_2
    =\frac{1}{\Omega^2}
    \sum_{k\leq k_F}\sum_{k'>k_F}
    e^{-i\mathbf{k}\cdot\mathbf{R}}
    e^{i\mathbf{q}\cdot\mathbf{R}'}
    e^{-i\varepsilon_{k'}t}
    M(q,\varepsilon_{k'}-\varepsilon_{k},-t).
    \label{eq:B2M}
\end{align}
%
%
%
The correlation hole is then given by
\begin{align}
    \rho_\mathrm{c}(R,R',\theta;t<0) &=\frac{A_1+A_2}{G_0(R,t<0)}, 
    \\
    \rho_\mathrm{c}(R,R',\theta;t>0) &=\frac{B_1+B_2}{G_0(R,t>0)} .
\end{align}

The correlation hole involves coupled
integrals over momenta below and above $k_\mathrm{F}$ leading to a six-dimensional
integral, which cannot be easily performed with
standard quadratures. However, only the spherical average is needed
to determine the correlation potential and the six-dimensional
integral can be reduced to a three-dimensional one.


\subsubsection{Spherical average of the correlation hole}
%
%
%
%
%
Spherical averaging $A_1$ in Eq. (\ref{eq:A1M}) 
over $\mathbf{R}'=\mathbf{r}''$ yields
%
%
%
%
%
%
%
\begin{align}
    \overline{A}_1
    &=\frac{4\pi}{\Omega^2}\sum_{k'> k_F}  e^{-i\mathbf{k}'\cdot\mathbf{R}} \sum_{k\leq k_F}
    \frac{\sin{(qR')}}{qR'}     e^{-i\varepsilon_{k}t}
    \nonumber\\
&\qquad\qquad\qquad\qquad\times    M(q,\varepsilon_{k'}-\varepsilon_{k},t).
\end{align}
It can be seen from Fig. \ref{fig:Integral} that for a fixed $\mathbf{k}'$, 
the integration
over $\mathbf{k}$ is independent of the azimuthal angle so that
\begin{align}
    \overline{A}_1
    &=\frac{1}{\pi\Omega}\sum_{k'> k_F} 
    e^{-i\mathbf{k}'\cdot\mathbf{R}}
    \int_0^{k_\mathrm{F}} dk k^2 
    \nonumber\\
&\qquad\times    \int_{-1}^1 dy 
    \frac{\sin{(qR')}}{qR'} e^{-i\varepsilon_{k}t}
    M(q,\varepsilon_{k'}-\varepsilon_{k},t).
\end{align}
It can also be seen that the integration over $\mathbf{k}$ is independent of the
direction of $\mathbf{k}'$ so that the total integral reduces to three dimensions:
\begin{align}
    \overline{A}_1
    &=\frac{1}{2\pi^3 R} \int_{k_\mathrm{F}}^\infty  dk'k' \sin{(k'R)}
    \nonumber\\
&\qquad\times    \int_0^{k_\mathrm{F}} dk k^2 e^{-i\varepsilon_k t} Q(k,k',R',t),
\end{align}
%
%
%
%
%
\begin{align}
    \overline{A}_2
    &=\frac{1}{2\pi^3 R} \int_0^{k_\mathrm{F}}  dkk \sin{(kR)}
    e^{-i\varepsilon_{k} t}
        \nonumber\\
&\qquad\times\int_{k_\mathrm{F}}^\infty dk' {k'}^2 Q(k,k',R',0),
\end{align}
\begin{align}
    \overline{B}_1
    &=\frac{1}{2\pi^3 R} \int_{k_\mathrm{F}}^\infty  dk'k' \sin{(k'R)}
    e^{-i\varepsilon_{k'} t}
    \nonumber\\
&\qquad\times    \int_0^{k_\mathrm{F}} dk k^2 Q(k,k',R',0),
\end{align}
\begin{align}
    \overline{B}_2
    &=\frac{1}{2\pi^3 R} \int_0^{k_\mathrm{F}}  dkk \sin{(kR)}
    \nonumber\\
&\qquad\times    \int_{k_\mathrm{F}}^\infty dk' {k'}^2 e^{-i\varepsilon_{k'} t}
    Q(k,k',R',-t),
\end{align}
where
\begin{align}
    Q(k,k',R',t)= \int_{-1}^1 dy 
    \frac{\sin{(qR')}}{qR'}    M(q,\varepsilon_{k'}-\varepsilon_{k},t).
    \label{eq:Qk}
\end{align}
The dependence of the integrand
on the variable $y$ enters through $q$ as defined in Fig. \ref{fig:Integral}.

The spherical average of the correlation hole is then given by
\begin{align}
    \overline{\rho}_\mathrm{c}(R,R';t<0) 
    &=\frac{\overline{A}_1+\overline{A}_2}{G_0(R,t<0)}, 
    \label{eq:rhoch}
    \\
    \overline{\rho}_\mathrm{c}(R,R';t>0)
    &=\frac{\overline{B}_1+\overline{B}_2}{G_0(R,t>0)} .
    \label{eq:rhoce}
\end{align}

\subsection{Plasmon-pole approximation}
\label{sec:Plasmon-pole}

Great simplification results if a plasmon-pole approximation
independent of $q$ is used for $L(q,\omega)$ defined in Eq. (\ref{eq:LkwA}):
\begin{equation}
    L(q,\omega)= \frac{\omega_\mathrm{p}}{2} \left[ 
    \delta(\omega-\omega_\mathrm{p}) - \delta(\omega+\omega_\mathrm{p}) \right], 
\end{equation}
which corresponds to
\begin{equation}
    K(q\rightarrow 0,\omega)=\frac{\omega_\mathrm{p}^2}{\omega^2-\omega_\mathrm{p}^2}.
\end{equation}
The approximation is valid for $q\leq q_\mathrm{c}$ and
for $r_{s}=3,4, 5$
the critical momenta are
$q_{\text{c}}=0.86$, $0.82$, and $0.73$ $k_{\text{F}}$, respectively.
These cover most of the average valence densities in real materials,

Within the plasmon-pole approximation, the quantity $M$
defined in Eq. (\ref{eq:MqwtA}) can be calculated
analytically:
\begin{align}
    M(q,\omega,t) = \frac{\omega_\mathrm{p}}{2}
    \frac{-ie^{i\omega_\mathrm{p}t}}
    {\omega_\mathrm{p}+\omega}.
    \label{eq:Mpp}
\end{align}
The interdependence
between $\mathbf{k}$ and $\mathbf{k}'$ is partially unlock.
%
%
%
%
The plasmon-pole approximation
imparts error to the correlation contribution.
This error can be minimised by treating
the upper limit of the
integral over unoccupied states ($k'$)
as a parameter. For $r_s=4$, an upper limit of $\approx 1.5 k_\mathrm{F}$
approximately reproduces the static correlation hole \cite{wang1991}. 

Fig. \ref{fig:rhocAv} shows examples of the spherical average of the
correlation
hole multiplied by the radial distance $R'$ for $r_s=4$ and
for several values of $R$ calculated using the plasmon-pole approximation.
Fig. \ref{fig:rhoxcAv} is similar to Fig. \ref{fig:rhocAv} but for the
xc hole.
It is interesting to observe that the xc hole is more localised than
the exchange hole, as can be seen by comparing
Figs. \ref{fig:rhoxAv} and \ref{fig:rhoxcAv}. A similar behaviour is known
in DFT.

\subsection{Exchange-correlation potential}

As described in Section \ref{sec:LDA}, the change of variable
$\mathbf{R}'=\mathbf{r}''-\mathbf{r}$
reduces Vxc in Eq. (\ref{eq:phixc}) to the first radial moment of the
spherical average of $\rho_\mathrm{xc}$:
%
%
\begin{equation}
V_\mathrm{xc}(r,r^{\prime};t)=\int dR'R'\text{ }\overline{\rho}%
_\mathrm{xc}(r,r^{\prime},R';t),
\label{Vxc1mom}
\end{equation}
where $\overline{\rho}_\mathrm{xc}(r,r^{\prime},R';t)$, for given
$r$, $r'$, and $t$, depends only on the radial
distance $R'=|\mathbf{r}''-\mathbf{r}|$,
\begin{equation}
\overline{\rho}_\mathrm{xc}(r,r^{\prime},R';t)
=\int d\Omega_{R'}\rho_\mathrm{xc}(r,r^{\prime
},\mathbf{r+R'};t).
\end{equation}

\subsubsection{Exchange potential}

The exchange potential is the first moment of
$\overline{\rho}_{\text{x}}$ in $R^{\prime}$, which from Eq. (\ref{eq:rhoxav})
is given by for $t<0$
\begin{align}
V_{\text{x}}(R,t<0)&=\frac{1}{iG_0(R,t)}
\frac{4\pi }{\Omega^{2}}\sum_{k,k^{\prime}\leq k_{\text{F}%
}}e^{-i\mathbf{k}\cdot\mathbf{R}}e^{-i\varepsilon_{k}t}
\nonumber\\&\qquad\times 
\int
dR^{\prime}\frac{\sin(q R^{\prime})}{q}.%
\end{align}
Consider the integral over $R'$ with positive $\alpha\rightarrow 0$:
\begin{equation}
\lim_{\alpha\rightarrow 0}
\int_{0}^{\infty}dR^{\prime}\sin(qR^{\prime})
e^{-\alpha R'}
 =\frac{1}{q}.
\end{equation}
One finds
\begin{equation}
V_{\text{x}}(R,t<0)=\frac{1}{iG_0(R,t)}\frac{4\pi}{\Omega^{2}}\sum_{k,k^{\prime}\leq
k_{\text{F}}}e^{-i\mathbf{k}\cdot\mathbf{R}}e^{-i\varepsilon_{k}t}\frac
{1}{q^{2}}.
\end{equation}
The integral over $k^{\prime}$ is given by
\begin{align}
f(k)  & =\frac{1}{\Omega}\sum_{k^{\prime}\leq k_{\text{F}}}\frac{1}{q^{2}}
\nonumber\\
& =\frac{1}{4\pi^{2}k}\int_{0}^{k_{\text{F}}}dk^{\prime}k^{\prime}%
\ln\left\vert \frac{k+k^{\prime}}{k-k^{\prime}}\right\vert ,
\label{eq:fk}
\end{align}
which can be performed analytically yielding
\begin{align}
    f(k)&=\frac{k_\mathrm{F}}{2\pi^2} F(k/k_\mathrm{F}),
\end{align}
where
\begin{equation}
    F(x)= 
    \frac{1}{2}+\frac{1-x^{2}}{4x} \ln\left\vert \frac{1+x}{1-x}\right\vert .
\end{equation}
This function is identical to
the one appearing in the static Hartree-Fock theory for the
electron gas \cite{ashcroft1976}.
More explicitly, as a function of $k$,
\begin{align}
    f(k)
    &=\frac{k_\mathrm{F}}{2\pi^2}
    \left(
    \frac{1}{2}+\frac{k^2_\mathrm{F}-k^2}{4k_\mathrm{F}k} \ln\left\vert
    \frac{k_\mathrm{F}+k}{k_\mathrm{F}-k}\right\vert 
    \right).
\end{align}
There remains the integral over $\mathbf{k}$,
which is a one-dimensional integral over radial $k$:
\begin{align}
V_{\text{x}}(R,t<0)&=\frac{1}{iG_0(R,t)}
\nonumber\\ 
&\quad\times \frac{2}{\pi R}\int_{0}%
^{k_{\text{F}}}dk\text{ }k\sin(kR)\text{ }e^{-i\varepsilon_{k}t}f(k).
\label{eq:Vx-}
\end{align}
For $t>0$, the result is given by 
\begin{align}
V_{\text{x}}(R,t>0)&=-\frac{1}{iG_0(R,t)}
\nonumber\\ 
&\quad\times \frac{2}{\pi R}\int_{k_\text{F}}%
^{\infty}dk\text{ }k\sin(kR)\text{ }e^{-i\varepsilon_{k}t}f(k).
\label{eq:Vx+}
\end{align}

\subsubsection{Correlation potential}
%
%
%
%
%
%

Similarly, the correlation potential is given by the first moment
in $R'$ of the spherical average of the correlation hole in 
Eqs. (\ref{eq:rhoch}) and (\ref{eq:rhoce}).
The integral to be evaluated is
\begin{align}
    I(k,k',t)&=\int dR' R' Q(k,k',R',t)
    \nonumber\\
    &=\int dR'\int_{-1}^1 dy 
    \frac{\sin{(qR')}}{q}    M(q,\varepsilon_{k'}-\varepsilon_{k},t)
    \nonumber\\
    &=\int_{-1}^1 dy 
    \frac{1}{q^2}    M(q,\varepsilon_{k'}-\varepsilon_{k},t),
\end{align}
where $q=|\mathbf{k}-\mathbf{k}'|=\sqrt{k^2+{k'}^2-2kk'y}$ 
and $Q$ is defined in Eq. (\ref{eq:Qk}).
The result is given by
\begin{align}
V_{\text{c}}(R,t<0)&=\frac{C_1+C_2}{G_0(R,t<0)},
\end{align}
\begin{align}
V_{\text{c}}(R,t>0)&=\frac{D_1+D_2}{G_0(R,t>0)},
\end{align}
where
%
%
\begin{align}
    {C}_1
    &=\frac{1}{2\pi^3 R} \int_{k_\mathrm{F}}^\infty  dk'k' \sin{(k'R)}
    \int_0^{k_\mathrm{F}} dk k^2 e^{-i\varepsilon_k t} I(k,k',t),
\end{align}
\begin{align}
    {C}_2
    &=\frac{1}{2\pi^3 R} \int_0^{k_\mathrm{F}}  dkk \sin{(kR)}
    e^{-i\varepsilon_{k} t}
    \int_{k_\mathrm{F}}^\infty dk' {k'}^2 I(k,k',0),
\end{align}
\begin{align}
    {D}_1
    &=\frac{1}{2\pi^3 R} \int_{k_\mathrm{F}}^\infty  dk'k' \sin{(k'R)}
    e^{-i\varepsilon_{k'} t}
    \nonumber\\
&\qquad\qquad\times    \int_0^{k_\mathrm{F}} dk k^2 I(k,k',0),
\end{align}
\begin{align}
    {D}_2
    &=\frac{1}{2\pi^3 R} \int_0^{k_\mathrm{F}}  dkk \sin{(kR)}
        \nonumber\\
&\qquad\qquad\times  
\int_{k_\mathrm{F}}^\infty dk' {k'}^2 e^{-i\varepsilon_{k'} t}
    I(k,k',-t).
\end{align}

Within the plasmon-pole approximation, Eq. (\ref{eq:Mpp}),
\begin{equation}
  M(q,\varepsilon_{k'}-\varepsilon_{k},t'')
  =\frac{\omega_\mathrm{p}}{2}
    \frac{-ie^{i\omega_\mathrm{p}t''}}{\omega_\mathrm{p}+\varepsilon_{k'}-\varepsilon_{k}},
\end{equation}
which partially decouples the interdependence of $\mathbf{k}$ and $\mathbf{k}'$,
allowing for analytical integration over the solid angles of both variables, yielding
\begin{align}
    C_1&=P_1(R,t,0,t),\\
    C_2&=P_2(R,t,0,0),\\
    D_1&=P_1(R,0,t,0),\\
    D_2&=P_2(R,0,t,-t),
\end{align}
where
\begin{align}
P_1(R,t,t',t'')
&=
\frac{-i\omega_\mathrm{p}e^{i\omega_\mathrm{p}t''}}{4\pi^3 R}\int_0^{k_\mathrm{F}}dk\
\int_{k_\mathrm{F}}^\infty dk'\,k\sin{(k'R)}
\nonumber\\
    &\times
         \frac{e^{-i\varepsilon_{k}t}e^{-i\varepsilon_{k'}t'}}
    {\omega_\mathrm{p}+\varepsilon_{k'}-\varepsilon_{k}}
    \ln{\left\vert \frac{k+k'}{k-k'}\right\vert},
\end{align}
\begin{align}
P_2(R,t,t',t'')
&=
\frac{-i\omega_\mathrm{p}e^{i\omega_\mathrm{p}t''}}{4\pi^3 R}\int_0^{k_\mathrm{F}}dk\
\int_{k_\mathrm{F}}^\infty dk'\,k'\sin{(kR)}
\nonumber\\
    &\times
    \frac{e^{-i\varepsilon_{k}t}e^{-i\varepsilon_{k'}t'}}
    {\omega_\mathrm{p}+\varepsilon_{k'}-\varepsilon_{k}}
         \ln{\left\vert \frac{k+k'}{k-k'}\right\vert}.
\end{align}
Due to the use of the plasmon-pole approximation, the upper limit of the integral over $k'$
is restricted to $1.5 k_\mathrm{F}$ for $r_s=4$ to reproduce approximately the static
correlation hole, as described earlier in Sec. \ref{sec:Plasmon-pole}.

The real and imaginary parts of Vxc of
the homogeneous electron gas with $r_s=4$
for a couple of typical times corresponding to the plasmon energy and the Fermi
energy are shown in Figs. \ref{fig:ReVxcT4-34} and \ref{fig:ImVxcT4-34}.
The strong cancellation between exchange and correlation potentials
is illustrated in Fig. \ref{fig:VxcT1}.

\subsection{Effective potential $\Xi_q$ in the homogeneous electron gas}
\label{sec:XiElgas}

In this section, the quasiparticle equation with effective potential $\Xi_q$
derived in Sec. \ref{sec:QPeqn} is worked out for the homogeneous electron gas
\cite{aryasetiawan2025}.
Using the approximate form of the effective potential in Eq. (\ref{eq:XiApprox})
the static and dynamic terms are modelled with parameters
determined from $GW$ calculations. The parameters are functions of density,
and the model provides a realisation of LDA.

For the homogeneous electron gas, the deviation potential
defined in Eq. (\ref{eq:DeltaV}) is given by
$\Delta V=V_\mathrm{xc}$ since
$V_\mathrm{xc}^\mathrm{0}$ is a constant
and can be set to zero. 
Plane waves are the natural choice for the orbitals,
\begin{align}
    \varphi_k(r)=\frac{e^{i\mathbf{k}\cdot\mathbf{r}}}{\sqrt{\Omega}}.
\end{align}
The quasiparticle wave function
for a given momentum $k$ is given by
\begin{align}
    \psi^*_k(r,t)=G_k(t) 
    \frac{e^{-i\mathbf{k}\cdot\mathbf{r}}}{\sqrt{\Omega}}.
\end{align}
Multiplying Eq. (\ref{eq:EOMQP}) by
$e^{-i\mathbf{q}\cdot\mathbf{(\mathbf{r-r'})}}/\Omega$, 
integrating over $r$ and $r'$, and using
\begin{align}
    \int d^3R\, e^{i\mathbf{q}\cdot\mathbf{R}}=(2\pi)^3\delta(\mathbf{q}),
\end{align}
lead to the equation of motion for $t\neq 0$ and a given spin:
\begin{align}
    (i\partial_t - \varepsilon_q)G_q(t)-\frac{1}{\Omega}\sum_\mathbf{k}
    V^\mathrm{xc}_{|\mathbf{q-k}|}(t) G_k(t)=0,
\end{align}
where
\begin{align}
    V^\mathrm{xc}_{|\mathbf{q-k}|}(t)
    &=\frac{1}{\Omega^2}\int d^3r d^3r' 
    e^{-i(\mathbf{q-k})\cdot (\mathbf{r-r'})}
    V^\mathrm{xc}(|\mathbf{r-r'}|,t)
    \nonumber\\
    &=\frac{1}{\Omega} \int d^3R\, e^{-i(\mathbf{q-k})\cdot \mathbf{R}}
    V^\mathrm{xc}(R,t).
\end{align}
The equation of motion can be recast as
\begin{align}
    \left[i\partial_t - \varepsilon_q-\Xi_q(t)\right]G_q(t)=0,
\end{align}
where
\begin{align}
    \Xi_q(t)= \frac{1}{\Omega}\sum_\mathbf{k}
    V^\mathrm{xc}_{|\mathbf{q-k}|}(t) \frac{G_k(t)}{G_q(t)}.
\end{align}
For $t<0$, $\Xi_q(t)$ 
an approximate model as in Eq. (\ref{eq:XiApprox}) is
\begin{align}
    \Xi_q(t<0)\approx \Xi^\mathrm{S}_q + \Xi^\mathrm{D}_q e^{i\omega_\mathrm{p}t},
\end{align}
where
\begin{align}\label{eq:plasmon}
    \omega_\mathrm{p}=\sqrt{4\pi \rho}
\end{align}
is the plasmon energy of the electron gas with density $\rho$.
A simple model for $\Xi^\mathrm{S}_q$ is
\begin{align}\label{eq:VqS}
    \Xi^\mathrm{S}_q=(1-\gamma Z)(E_\mathrm{F}-\varepsilon_q).
\end{align}
Here, $Z$ is the quasiparticle renormalisation factor which
reduces the bandwidth and $\gamma>1$ is a factor
which takes into account band broadening due to
the momentum-dependence of the self-energy.
Both $Z$ and $\gamma$ are assumed to be momentum independent. 
The model ensures that the noninteracting occupied bandwidth
is reduced by a combined factor $\gamma Z$, and the new and the noninteracting
Fermi levels have been realigned. 

The solution is given by
\begin{align}\label{eq:Gqt}
    G_q(t)&=G_q(0)e^{-iE_q t} 
    e^{-i\int_0^t dt' \,\Xi^\mathrm{D}_q e^{i\omega_\mathrm{p}t'}}
    \nonumber\\
    &=G_q(0) e^{-iE_q t}\left( A_0 + A_1e^{i\omega_\mathrm{p}t}
    +A_2 e^{i2\omega_\mathrm{p}t}+...\right),
\end{align}
where
\begin{align}
    E_q=\varepsilon_q+\Xi_q^\mathrm{S}
\end{align}
is the quasiparticle energy. Expanding the second exponent
up to two plasmon energies and defining
\begin{align}
 \lambda=-\frac{\Xi_q^\mathrm{D}}{\omega_\mathrm{p}},   
\end{align}
one finds
\begin{align}
    A_0&= 1-\lambda + \frac{1}{2}\lambda^2,\\
    A_1&=\lambda(1-\lambda),\\
    A_2&=\frac{1}{2}\lambda^2.
\end{align}
Note that $A_0+A_1+A_2=1$. It is straightforward to include higher-order
plasmon excitations.
In this simple model, it is assumed that for $q\leq k_\mathrm{F}$ 
the spectrum does not have weight above the Fermi level.
It is possible to construct
a model that includes a weight transfer above the Fermi level \cite{karlsson2023}.
This assumption implies that $G_q(0)=1$ and
$\Xi^\mathrm{D}_q$
is then determined by the renormalisation of the quasiparticle
$Z=A_0$, resulting in
\begin{align}\label{eq:Z}
    \lambda=1-\sqrt{2Z-1}.
\end{align}
In general, $\Xi_q^\mathrm{S}$ is complex, and its imaginary part gives
a life-time broadening of the quasiparticle and the plasmon satellite.
A model for the life-time broadening,
which depends linearly on the inverse lifetime, is given by
\begin{align}\label{eq:eta}
    \eta(q)=\eta_0 + (\eta_1-\eta_0)\frac{E_\mathrm{F}-\varepsilon_q}
    {E_\mathrm{F}-\varepsilon_0}.
\end{align}
In theory, $\eta_0$ should be zero at $q=k_\mathrm{F}$ but it is taken as finite
to give a broadened $\delta$-function.

With $G_q(0)=1$ the Fourier transform of $G_q(t)$ in Eq. (\ref{eq:Gqt})
is given by
\begin{align}\label{eq:ImGq}
G_q(\omega)&=\frac{A_0}{\omega-E_q-i\eta(q)}
+\frac{A_1}{\omega-E_q+\omega_\mathrm{p}-i\eta(q)}
\nonumber\\
&\quad +\frac{A_2}{\omega-E_q+2\omega_\mathrm{p}-i\eta(q)}.
\end{align}
The spectrum, which is proportional to the imaginary part of $G_q(\omega)$,
exhibits the
expected peaks at the quasiparticle energy $E_q$ and at multiples of
the plasmon satellite energy below the quasiparticle energy, 
$E_q-\omega_\mathrm{p}$, $E_q-2\omega_\mathrm{p}$, ..., in agreement
with the result obtained from the cumulant expansion
\cite{hedin1980,almbladh1983,aryasetiawan1996,kas2014,kas2019}.
\begin{figure}[t]
\begin{center} 
\includegraphics[scale=0.6, viewport=3cm 8cm 17cm 20cm, clip]
{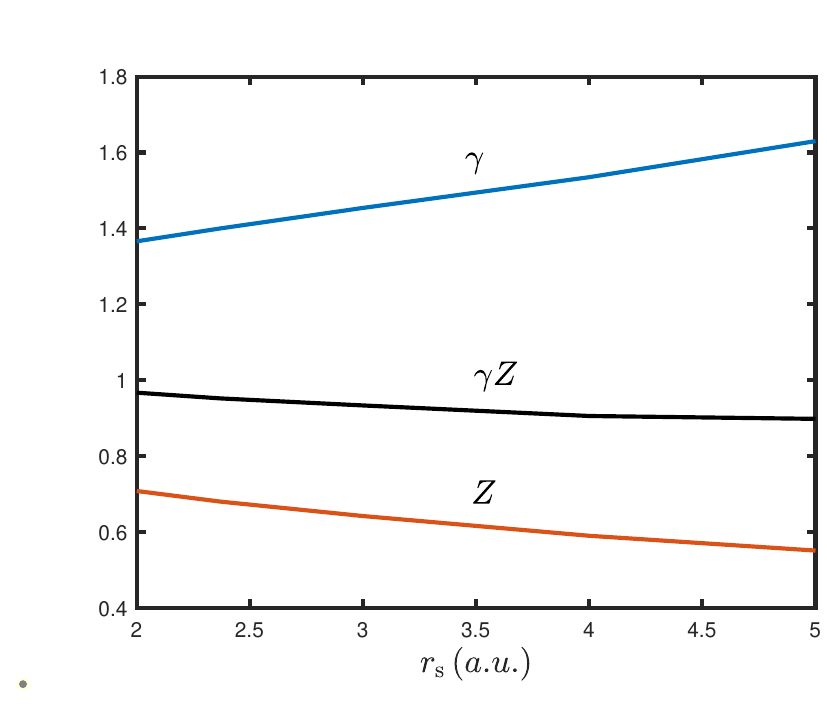}
\caption{The momentum-broadening factor $\gamma$ and
the quasiparticle renormalization factor $Z$ as described in the text
extracted from $GW$ calculations of the homogeneous electron gas.
$\gamma Z$ corresponds to the band narrowing.
The renormalization factor $Z$ is taken to be the average of the values
at $q=0$ and $q=k_\mathrm{F}$.
}
\label{fig:gammaZ}%
\end{center}
\end{figure}
\begin{figure}[t]
\begin{center} 
\includegraphics[scale=0.6, viewport=3cm 8cm 17cm 20cm, clip]
{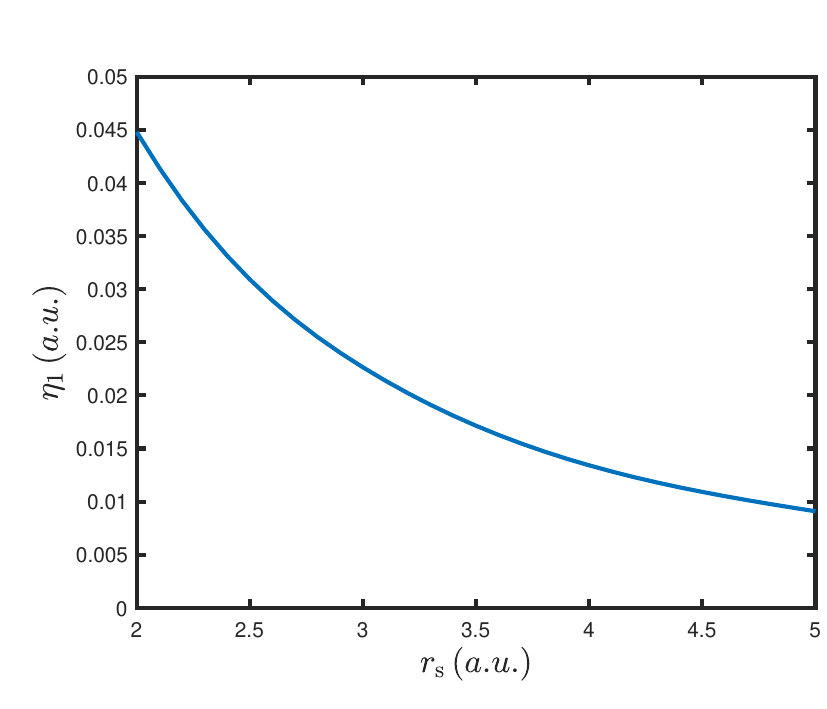}
\caption{The lifetime broadening factor ($\eta_1$ in Eq. (\ref{eq:eta}))
obtained from $GW$ calculations
of the homogeneous electron gas as a function of $r_\mathrm{s}$.
}
\label{fig:eta}%
\end{center}
\end{figure}
Fig. \ref{fig:gammaZ} and Fig. \ref{fig:eta} show
the parametrized $\gamma$ and $Z$
and the life-time broadening $\eta_1$, respectively,
as functions of $r_\mathrm{s}$.
They are extracted from one-shot $GW$ \cite{hedin1965} calculations for the 
homogeneous electron gas. The combined factor
$\gamma Z$ provides a measure of band narrowing.
As anticipated, the lower the density (larger $r_\mathrm{s})$,
the stronger the band narrowing.

\begin{figure}[t]
\begin{center} 
\includegraphics[scale=0.6, viewport=3cm 8cm 17cm 20cm, clip]
{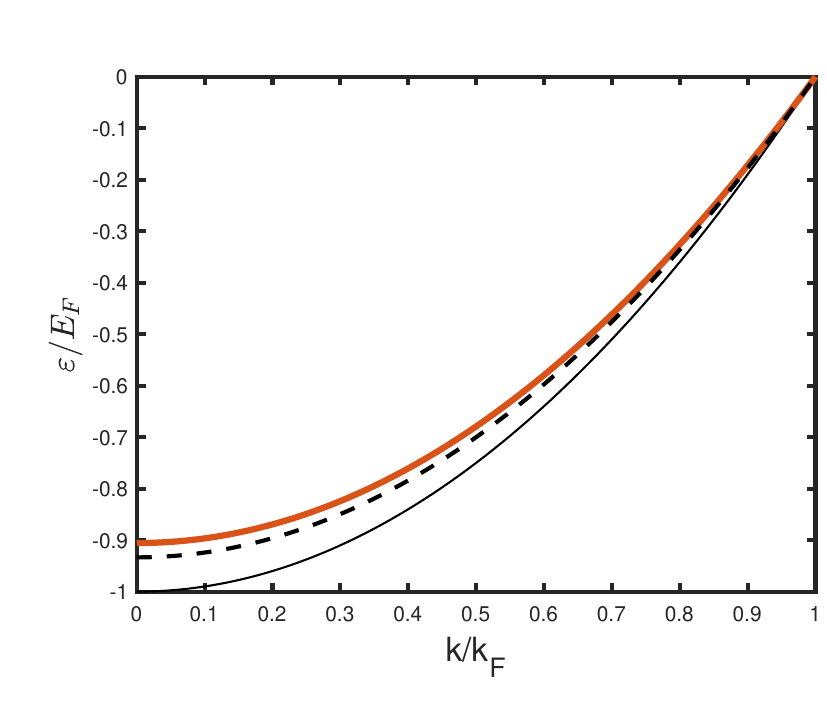}
\caption{The energy dispersion of the homogeneous electron gas
with $r_\mathrm{s}=4.0$ a.u.
corresponding to Na (thick solid line) 
and $r_\mathrm{s}=2.37$ a.u. corresponding to Al (dashed line) 
calculated using the model described in the text. For comparison,
the dispersion of the
noninteracting electron gas is also shown (thin solid line).
}
\label{fig:dispersion}%
\end{center}
\end{figure}
\begin{figure}[t]
\begin{center} 
\includegraphics[scale=0.6, viewport=5.3cm 1.5cm 20cm 16cm, clip, width=\columnwidth]
{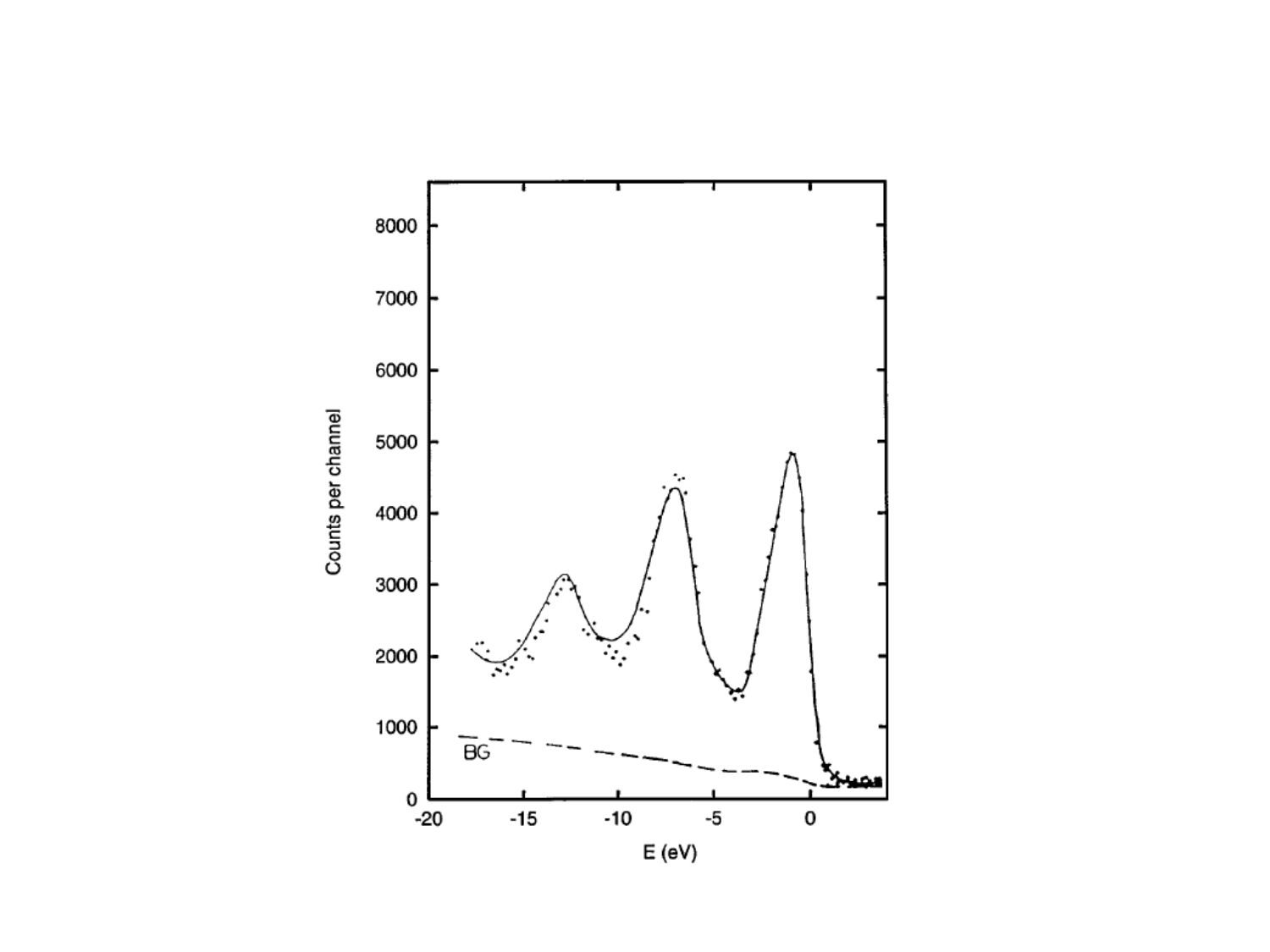}
\includegraphics[scale=0.5, viewport=3cm 8cm 17cm 19cm, clip, width=\columnwidth]
{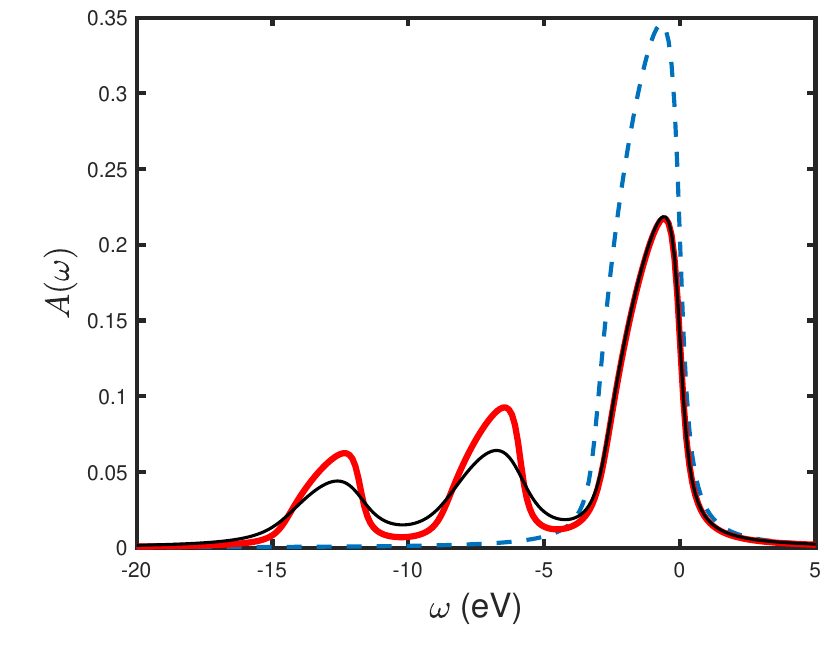}

\caption{The total spectral function of Na: experiment \cite{steiner1979} (top)
and calculated using the model described in the text
with $r_\mathrm{s}=4.0$ (bottom).
The thin solid line
includes a plasmon broadening of $0.55$ eV, which tends to shift the peak
to lower energy.
The noninteracting electron gas total 
spectrum corresponding to $r_\mathrm{s}=4.0$ is also shown (dashed line).
The figure is taken from \cite{aryasetiawan2025}.
}
\label{fig:AtotNa}%
\end{center}
\end{figure}
\begin{figure}[t]
\begin{center} 
\includegraphics[scale=0.6, viewport=3cm 8cm 17cm 20cm, clip]
{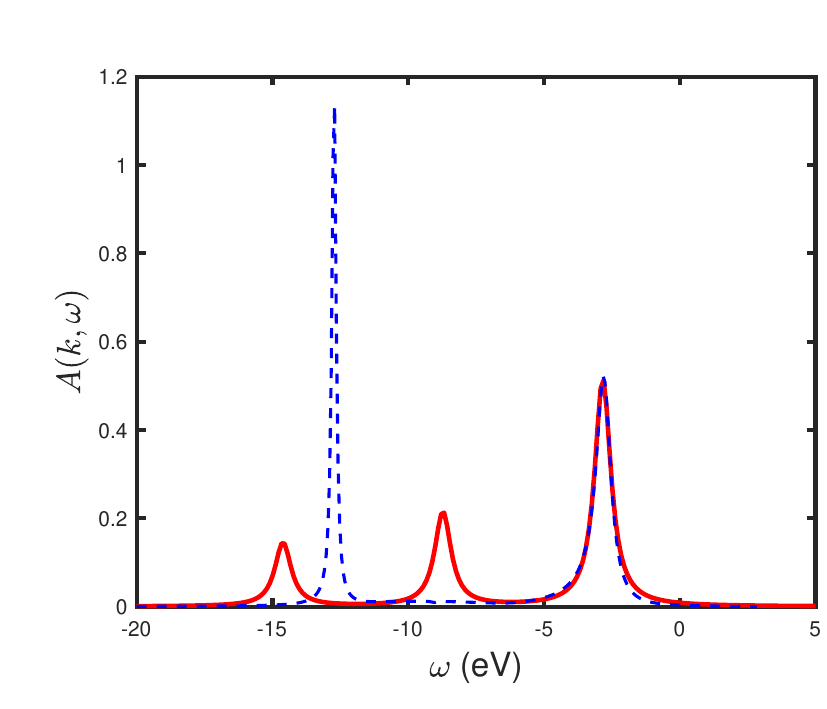}
\caption{The spectral function 
$A(k,\omega)=\frac{1}{\pi}\mathrm{Im}G(k,\omega)$ 
of Na at the $\Gamma$-point ($k=0$) calculated using the model
described in the text (solid line).
The dashed line is the $GW$ spectral function of the homogeneous electron gas
with $r_\mathrm{s}=4.0$. It shows the well-known overestimation
of the plasmon binding energy in the $GW$ approximation.
}
\label{fig:Aq0Na}%
\end{center}
\end{figure}

The model given in Eq. (\ref{eq:ImGq}) is now applied to solid Na.
Na is bcc with lattice constant $a=4.29$ \r{A} and
one $s$ valence electron
per unit cell, which corresponds to $r_\mathrm{s}=4.0$ (a.u.).
The energy dispersion corresponding
to $r_\mathrm{s}=4.0$ is
shown in Fig. \ref{fig:dispersion} and compared with the
noninteracting dispersion. The total and $q=0$ spectral functions
are shown in Figs. \ref{fig:AtotNa}
and \ref{fig:Aq0Na}.
The calculated total spectral function of Na
is in close agreement with the measured
photoemission spectrum \cite{steiner1979}
and the result obtained using the cumulant expansion
method \cite{aryasetiawan1996}. Note that
the model does not take into account the effects of band structure.
The spectrum exhibits the characteristic
main quasiparticle peak followed by plasmon satellites separated
by multiples of the plasmon energy.
An additional plasmon
lifetime tends to shift the plasmon peak to a slightly lower energy.

Fig. \ref{fig:Aq0Na} shows the calculated
spectral function at the $\Gamma$ point ($q=0$), which is
compared with the one obtained from a standard one-shot $GW$ calculation.
The $GW$ spectral function suffers from the well-known overestimation
of the plasmon binding energy.


\subsubsection{Local-density approximation}

From the ansatz in Eq. (\ref{eq:VqS}), a possible
local-density approximation \cite{kohn1965,jones1989,becke2014} 
for simple metals is
%
\begin{align}
    \Xi_q^\mathrm{S}(r)= \left[1-\gamma(\overline{\rho})
    Z(\overline{\rho})\right]  
    \left( \frac{1}{2}\left[ 3\pi^2\rho(r)\right]^{2/3}
    -\varepsilon_q\right),
\end{align}
where
\begin{align}
    \overline{\rho}=\frac{1}{\Omega}\int dr \,\rho(r)
\end{align}
is the average density.
Both $\gamma$ and $Z$ can be calculated as functions of the
electron gas density $\overline{\rho}$ within
the $GW$ approximation \cite{hedin1965} or
using more accurate approximations \cite{huotari2010,li2024}.
Alternatively, they can also
be treated as fitting parameters. The plasmon energy
$\omega_\mathrm{p}$ is given by
Eq. (\ref{eq:plasmon}) with $\rho=\overline{\rho}$
which together with $Z$ determines $\Xi_q^\mathrm{D}$ using Eq. (\ref{eq:Z}) .

\section{Applications to model Hamiltonians}
\label{sec:Applications}
\subsection{One-dimensional Hubbard model}

\begin{figure}[htp]
\centering
\includegraphics[scale=0.5, viewport=3cm 8cm 17cm 20cm, clip]
{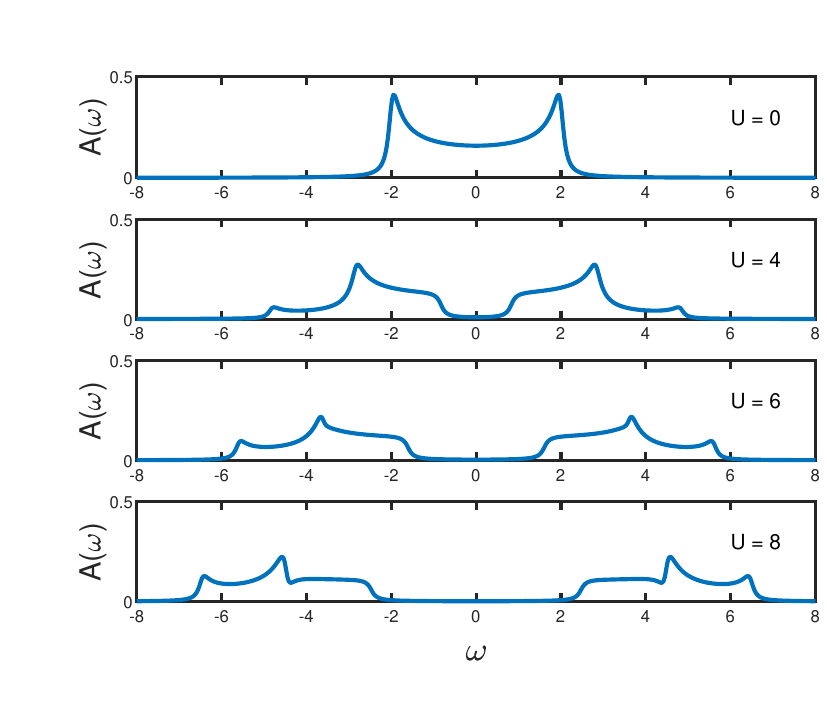}
\caption{The calculated total spectral functions of
the 1D Hubbard model for $U=0,4,6,$ and $8$
using approximations described in the text.
A broadening of $0.1$ has been used \cite{aryasetiawan2022a}.}
\label{fig:Aw}%
\end{figure}

\begin{figure}[htp]
\centering
\includegraphics[scale=0.5, viewport=3cm 8cm 17cm 20cm, clip]
{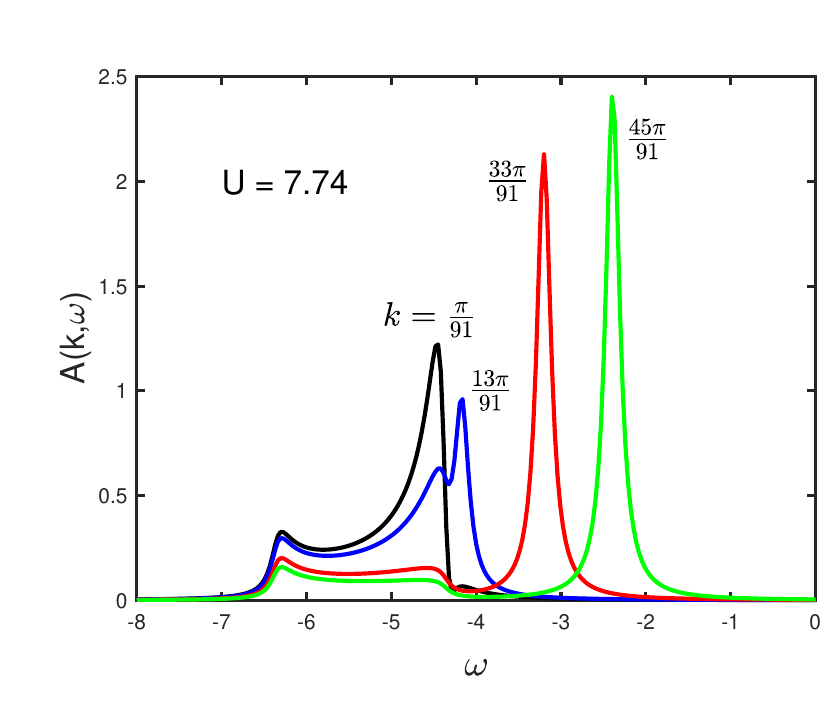}
\hfill
\includegraphics[scale=0.5, viewport=8cm 4cm 26cm 15cm, clip]
{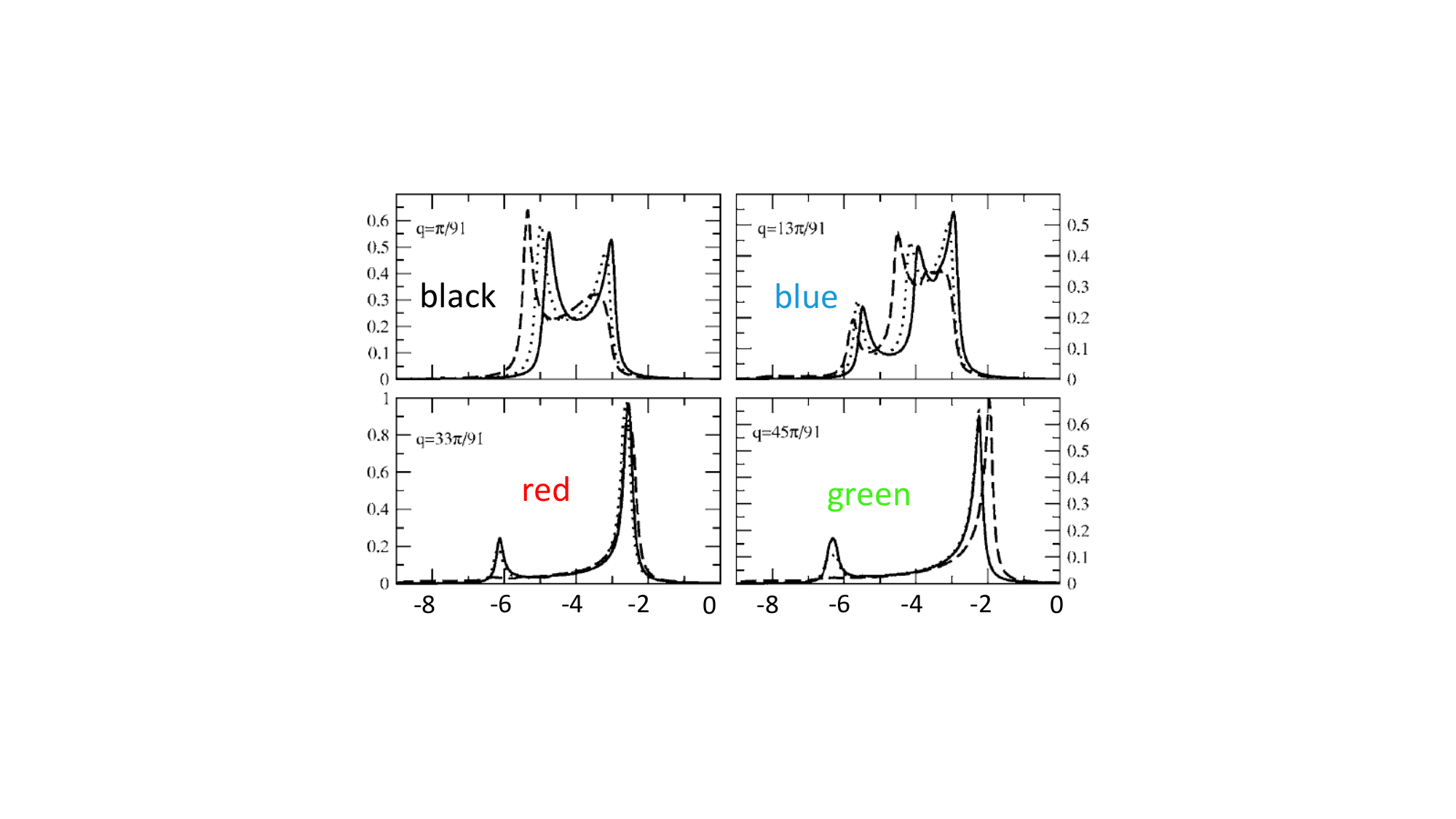}
\caption{The $k$-resolved spectral functions of the 1D Hubbard chain for $U=7.74$
(upper figure) \cite{aryasetiawan2022a}.
The $U$ and $k$ values have been chosen to facilitate comparison with Fig. 11 of 
Benthien and Jeckelmann \cite{benthien2007}, reproduced in the lower figure. 
For ease of comparison, the corresponding colour of each curve in the upper figure 
is indicated in the lower figure.
A broadening of $0.1$ has been used.}
\label{fig:AkwU7-74}%
\end{figure}

\begin{figure}[htp]
\centering
\includegraphics[scale=0.5, viewport=3cm 8cm 17cm 20cm, clip]
{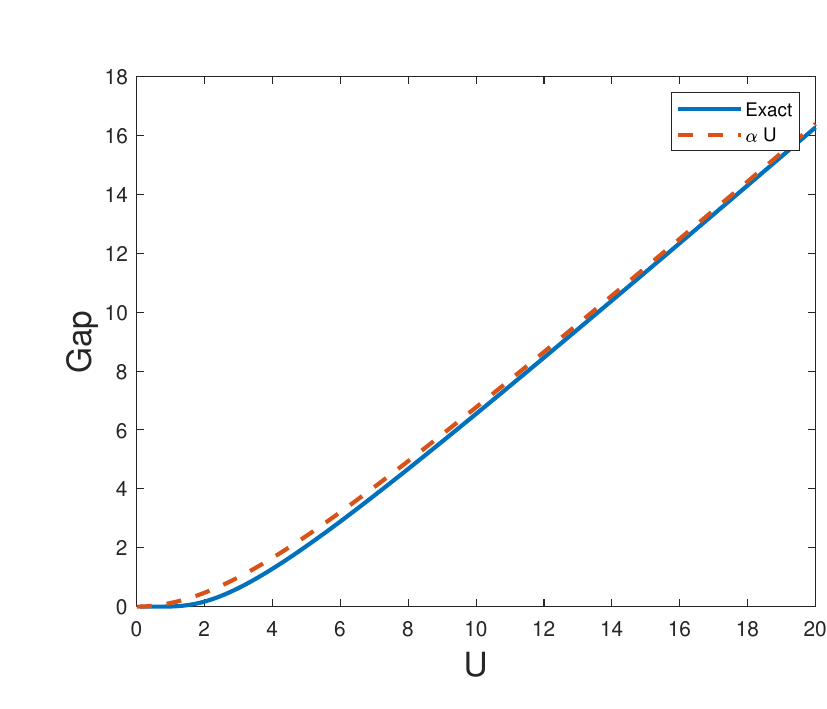}
\caption{The calculated band gap, $\alpha U$, 
compared with the exact gap obtained from the Bethe ansatz 
as a function of $U$.
The calculated gap approaches the exact result as $U$ increases
\cite{aryasetiawan2022a}.}
\label{fig:Gap}%
\end{figure}

The first application of the Vxc formalism is
to calculate the Green function for the half-filled
one-dimensional (1D) Hubbard chain \cite{aryasetiawan2022a}
with the Hamiltonian given by
%
\begin{equation}
\hat{H}=-\Delta\sum_{i\neq j}\hat{c}_{i\sigma}^{\dagger}\hat{c}_{j\sigma}%
+U\sum_{i}\hat{n}_{i\uparrow}\hat{n}_{i\downarrow}.
\label{eq:HubbardChain}%
\end{equation}
Vxc derived analytically for
the Hubbard dimer will be used and extrapolated to the lattice case. 
Since Vxc of the Hubbard dimer only contains
information about the onsite and nearest-neighbour sites, this
approximation clearly 
neglects components of $V_{\mathrm{xc}}$ beyond nearest neighbours.

For the Hubbard chain which possesses lattice translation symmetry, 
it is convenient to introduce Bloch base
functions,
\begin{equation}
\varphi_{k}(r)=\frac{1}{\sqrt{N}}\sum_{T}e^{ikT}\varphi(r-T),
\end{equation}
where $T$ denotes a lattice site.
The Green function expressed in these Bloch functions takes the form
\begin{equation}
G(r,r^{\prime};t)=\sum_{k}\varphi_{k}(r)G(k,t)\varphi_{k}^{\ast}(r^{\prime
}),\label{eq:Gkt}%
\end{equation}
where%
\begin{equation}
G(k,t)=\int drdr^{\prime}\varphi_{k}^{\ast}(r)G(r,r^{\prime};t)\varphi
_{k}(r^{\prime}).
\end{equation}

The equation of motion in the Bloch base functions is obtained by
inserting the expansion in Eq. (\ref{eq:Gkt}) in Eq. (\ref{eq:EOMGVxc}),
yielding
\begin{equation}
    (i\partial_t -\varepsilon_q)G(q,t) -F(q,t) = \delta(t),
    \label{eq:EOMqt}
\end{equation}
where
\begin{align}
    \varepsilon_q &= \int dr\, \varphi^*_q(r) h(r) \varphi_q(r)
    \nonumber\\
    &=\frac{1}{N}\sum_{TT'} e^{-ik(T-T')} 
    \int dr\, \varphi^*(r-T) h(r) \varphi(r-T')
    \nonumber\\
    &=-2\Delta \cos{q}
\end{align}
and
\begin{align}
    &F(q,t) = \sum_k \int drdr' 
    \nonumber\\
    &\times \varphi_q^*(r)\varphi_k(r)V_{\mathrm{xc}}(r,r';t)
    \varphi_k^*(r')\varphi_q(r')
    \times G(k,t).
\end{align}
Because the site occupation number is uniform, 
the Hartree potential is constant and can be absorbed into the chemical potential.
The terms $k\neq q$ in the summation over $k$ in
$F(q,t)$ describe the coupling between the propagator of momentum $q$
and the rest of the propagators. 

To utilise $V^\mathrm{xc}$ deduced for the Hubbard
dimer, consider the equations of motion of the dimer Green function
in the bonding and antibonding orbitals, which are given by
\begin{equation}
    \left[i\partial_t - \varepsilon_\mathrm{A} - V^\mathrm{xc}(t)\right]
    G_\mathrm{A}(t) -\Delta V^\mathrm{xc}(t)G_\mathrm{B}(t)=\delta(t),
\label{eq:GA}
\end{equation}
\begin{equation}
    \left[i\partial_t - \varepsilon_\mathrm{B} - V^\mathrm{xc}(t)\right]
    G_\mathrm{B}(t) -\Delta V^\mathrm{xc}(t)G_\mathrm{A}(t)=\delta(t),
\label{eq:GB}
\end{equation}
where
\begin{equation}
    \varepsilon_\mathrm{A}=\Delta,\;\;\;\varepsilon_\mathrm{B}=-\Delta,
\end{equation}
are the one-particle antibonding and bonding energies.
$V^\mathrm{xc}=V^\mathrm{xc}_\mathrm{BB}$ 
and $\Delta V^\mathrm{xc}=V^\mathrm{xc}_\mathrm{AB}$ are given in Eqs.
(\ref{eq:VxcBB}) and (\ref{eq:VxcAB}), respectively.
Eqs. (\ref{eq:GA}) and (\ref{eq:GB}) can be thought of as a special case
of Eq. (\ref{eq:EOMqt}) with only two $k$ points corresponding to the bonding
and antibonding states. These two states may be regarded as
the centra of the occupied and unoccupied bands of the one-dimensional
Hubbard chain.

Comparison between Eq. (\ref{eq:EOMqt}) and
Eq. (\ref{eq:GA}) or (\ref{eq:GB})
offers a physically motivated approximation:
$G_\mathrm{A}$ or $G_\mathrm{B}$ is replaced
by $G(q,t)$ and $G_\mathrm{B}$ or $G_\mathrm{A}$ by the average
$\frac{1}{N}\sum_k G(k,t)$:
\begin{equation}
    \left[i\partial_t - \varepsilon_q - V^\mathrm{xc}(t)\right]
    G(q,t) -\frac{1}{N}\sum_k \Delta V^\mathrm{xc}(t)G(k,t)=\delta(t).
\end{equation}
The above equation can be rewritten as
\begin{equation}
    \left[i\partial_t - \varepsilon_q - V^\mathrm{xc}(t)
    -\Delta {V}^\mathrm{xc} (q,t)\right]
    G(q,t) =\delta(t),
\end{equation}
where 
\begin{equation}
    \Delta {V}^\mathrm{xc} (q,t) = \frac{1}{N}\sum_k \Delta V^\mathrm{xc}(t)
    \frac{G(k,t)}{G(q,t)}. 
    \label{DVxcApprox}
\end{equation}
In terms of the effective potential $\Xi$ defined in Eq. (\ref{eq:Xi}), it
can be recognised that
\begin{align}
    \Xi(q,t)=V^\mathrm{xc}(t)+\Delta {V}^\mathrm{xc} (q,t). 
\end{align}
The solution for the electron case ($t>0$) is assumed to be given by 
\begin{equation}
    G^e(q,t)=-i\theta(t)e^{-i \varepsilon_q t - i\int_0^t dt'\left[V^\mathrm{xc}(t')
    +\Delta V^\mathrm{xc} (q,t') \right]}.
\end{equation}
This solution assumes that the weight is unity whereas in reality it should be
renormalised to the quasiparticle weight.
A similar result can be derived for the hole Green function, keeping in mind that 
$V^\mathrm{xc}(-t)=-V^\mathrm{xc}(t)$.

To proceed further, the following approximation is proposed:
\begin{equation}
    \Delta V^\mathrm{xc} (q,t) \approx \frac{1}{N}\sum_k \Delta V^\mathrm{xc}(t)
    e^{-i(\varepsilon_k-\varepsilon_q)t},
\end{equation}
which corresponds to replacing $G$ by a noninteracting $G^0$.
Furthermore, 
to facilitate analytical calculations 
$V^\mathrm{xc}=V^\mathrm{xc}_\mathrm{BB}$ 
and $\Delta V^\mathrm{xc}=V^\mathrm{xc}_\mathrm{AB}$ 
in Eqs. (\ref{eq:VxcBB}) and (\ref{eq:VxcAB}) are
truncated as follows: 
\begin{equation}
    V^\mathrm{xc}(t) \approx \frac{\alpha U}{2},
\end{equation}
\begin{equation}
    \Delta V^\mathrm{xc}(t) \approx \frac{\alpha U}{2} (1-\alpha^2)
     e^{-i2\Delta t}.
\end{equation}
The first term, which is a constant, shifts the one-particle energy
whereas the second term, corresponding to the coupling between the propagator
$G(q,t)$ and the rest of the system, gives rise to the main satellite.
The higher excitation term $\exp{(-i4\Delta t)}$ is neglected.

To first order in $\Delta V^\mathrm{xc}$,
\begin{align}
   & G^e(q,t) = -i\theta(t) e^{-i(\varepsilon_q+\frac{\alpha U}{2}) t}
    \nonumber\\
    &\times\left\{1  -i\frac{\alpha U}{2}(1-\alpha^2)\frac{1}{N}\sum_k \int_0^t dt' 
    e^{-i(\varepsilon_k-\varepsilon_q+2\Delta)t'}
     \right\}.
\end{align}
Performing the time integral,
\begin{align}
    &G^e(q,t) = -i\theta(t) e^{-i(\varepsilon_q+\frac{\alpha U}{2}) t}
    \nonumber\\
    &\times\left\{1+\frac{\alpha U}{2}(1-\alpha^2)\frac{1}{N}\sum_k 
    \frac{e^{-i(\varepsilon_k-\varepsilon_q+2\Delta)t }-1}{\varepsilon_k-\varepsilon_q+2\Delta} 
     \right\}.
\end{align}
Its Fourier transform is given by
\begin{align}
    G^e(q,\omega) &= 
    \frac{A^e_0(q)}{\omega-\left(\varepsilon_q+\frac{\alpha U}{2}\right)+i\eta}
    \nonumber\\
    &+\frac{1}{N}\sum_k \frac{A^e(q,k)}{\omega-(\varepsilon_k+\frac{\alpha U}{2}+2\Delta)+i\eta},
    \label{Geqt}
\end{align}
where
\begin{equation}
    A^e_0(q)=1-\frac{\alpha U}{2}(1-\alpha^2)
    \frac{1}{N}\sum_k \frac{1}{\varepsilon_k-\varepsilon_q+2\Delta},
\end{equation}
\begin{equation}
    A^e(q,k)=\frac{\alpha U}{2} (1-\alpha^2)\frac{1}{\varepsilon_k-\varepsilon_q+2\Delta}.
\end{equation}

A similar derivation for the hole Green function yields
\begin{align}
    &G^h(q,t) = i\theta(-t) e^{-i(\varepsilon_q-\frac{\alpha U}{2}) t}
    \nonumber\\
    &\times\left\{1-\frac{\alpha U}{2}(1-\alpha^2)\frac{1}{N}\sum_k 
    \frac{e^{-i(\varepsilon_k-\varepsilon_q-2\Delta)t }-1}{\varepsilon_k-\varepsilon_q-2\Delta} 
     \right\},
     \label{Ghqt}
\end{align}
\begin{align}
    G^h(q,\omega)& = 
    \frac{A^h_0(q)}{\omega-\left(\varepsilon_q-\frac{\alpha U}{2}\right)-i\eta}
    \nonumber\\
    &-\frac{1}{N}\sum_k \frac{A^h(q,k)}{\omega-(\varepsilon_k-\frac{\alpha U}{2}-2\Delta)-i\eta},
    \label{Ghqw}
\end{align}
where
\begin{equation}
    A^h_0(q)=1+\frac{\alpha U}{2}(1-\alpha^2)
    \frac{1}{N}\sum_k \frac{1}{\varepsilon_k-\varepsilon_q-2\Delta},
    \label{Ah0}
\end{equation}
\begin{equation}
    A^h(q,k)=\frac{\alpha U}{2}(1-\alpha^2) \frac{1}{\varepsilon_k-\varepsilon_q-2\Delta}.
    \label{Ahk}
\end{equation}
%
For the electron case, both $\varepsilon_k$ and
$\varepsilon_q$ correspond to unoccupied states while for the hole case,
to occupied states.

Fig. \ref{fig:Aw} shows the total spectral functions for $U = 0,4,6,$ and  $8$.
Fig. \ref{fig:AkwU7-74} displays
$k$-resolved spectra for a number of $k$-points for $U=7.74$. 
For the purpose of comparison,
the chosen values of $k$ and $U$ coincide with those of the benchmark results
obtained using the dynamical density-matrix renormalisation group
method \cite{benthien2007}.
Since in Eq. (\ref{DVxcApprox}) the Green function is approximated
by the noninteracting one, the electron Green function does not contribute
to spectral weight below the chemical potential.
Similarly, the hole Green function makes no contribution to
spectral weight
above the chemical potential. 
Furthermore, the use of the noninteracting Green function neglects
life-time effects so that the peaks are
sharp $\delta$-functions with a renormalised weight due to weight transfer to
lower or higher energy.
The angle-resolved spectra consists of
two branches; a spinless holon (antiholon) dispersion
with a hole (electron) charge and a charge-neutral spinon dispersion of 
spin $\frac{1}{2}$ \cite{giamarchi}.
Such an interpretation would probably require analysis in terms of
state vectors, which are not accessible in Green function theory.

Despite the simplicity of the approximation used, the structure of the spectra
of the benchmark results \cite{benthien2007} is correctly reproduced
as can be seen in Fig. \ref{fig:AkwU7-74}. 
A major discrepancy is that for small $k$
the peak positions are lower than
the benchmark results, resulting in a wider dispersion for both the spinon
and holon branches. 
This discrepancy can be traced back to the
use of a noninteracting $G^0$. 
A further factor that contributes to the discrepancy is
the neglect of long-range correlations, which are expected to narrow the
band dispersion. Evidently, the Hubbard dimer $V^\mathrm{xc}$ used in the
approximation does not contain these long-range correlations
beyond the nearest neighbours.

The analytic expression for the hole Green function in Eq. (\ref{Ghqw})
provides a valuable means of understanding
the structure of the calculated $k$-resolved
hole spectra in Fig. \ref{fig:AkwU7-74}.
The first term gives rise to
the main peak centred at $\omega=\varepsilon_q-\frac{\alpha U}{2}$
weighted by $A^h_0(q)$ and this peak corresponds to the spinon excitation. The
second term is the noninteracting occupied
density of states shifted by $-(\frac{\alpha U}{2}+2\Delta)$ and weighted by
$A^h(q,k)$.
Taking the spectrum corresponding to $q=\frac{45\pi}{91}$ (green curve)
as an example, the peak at around $-2.4$ is the spinon excitation. 
The low-energy spectra from $-6.5$ to $-4.5$ is the shifted occupied
density of states weighted by $A^h(q,k)$, which indeed mimics the
curve in Fig. \ref{fig:Aw} for $U=0$.
The feature at $-6.5$ is interpreted as the holon excitation.
The structure at around $-4.5$ is largest for small $q$ close to
the bottom of the noninteracting band since
the main peak merges with the shifted density of states.  
For $q=\frac{13\pi}{91}$ the main peak separates from
the shifted density of states resulting in
a three-peak structure, which is also
found in the benchmark results \cite{benthien2007}. 

The calculated band gap, $\alpha U$, is shown Fig. \ref{fig:Gap}
and in strikingly good agreement with the exact result obtained from
the Bethe ansatz \cite{ovchinnikov1970},
\begin{equation}
    E_\mathrm{gap} = \frac{16\Delta^2}{U} \int_1^\infty dy
    \frac{\sqrt{y^2-1}}{\sinh{\left(\frac{2\pi\Delta y}{U}\right)} }.
    \label{ExactGap}
\end{equation}
In the large-$U$ limit the calculated gap reproduces the exact gap.
The calculated gap in nonzero for any finite $U$ as in the exact case.
At small values of $U$, it is larger than the exact one.
This discrepancy could be attributed to the lack of long-range correlations
in the dimer Vxc discussed earlier, resulting in
a too large effective $U$ and consequently a larger band gap.

It is interesting to make a comparison with the result obtained from
the dynamical mean-field theory (DMFT) \cite{georges1996}. 
Within the single-site approximation, the system remains metallic until
$U > 6$. This is not inconsistent with the present result since
Vxc is based on the Hubbard dimer.
Indeed, a cluster DMFT
with an even number of sites produces a gap for any finite $U$.
With an odd number of sites, there exists a coexistence region
in which the system remains metallic before
entering the Mott insulating regime after exceeding
a certain value of $U$ \cite{go2009}. 

\begin{figure}[htp]
\centering
\includegraphics[scale=0.5, viewport=9cm 0.5cm 24cm 18cm, clip]
{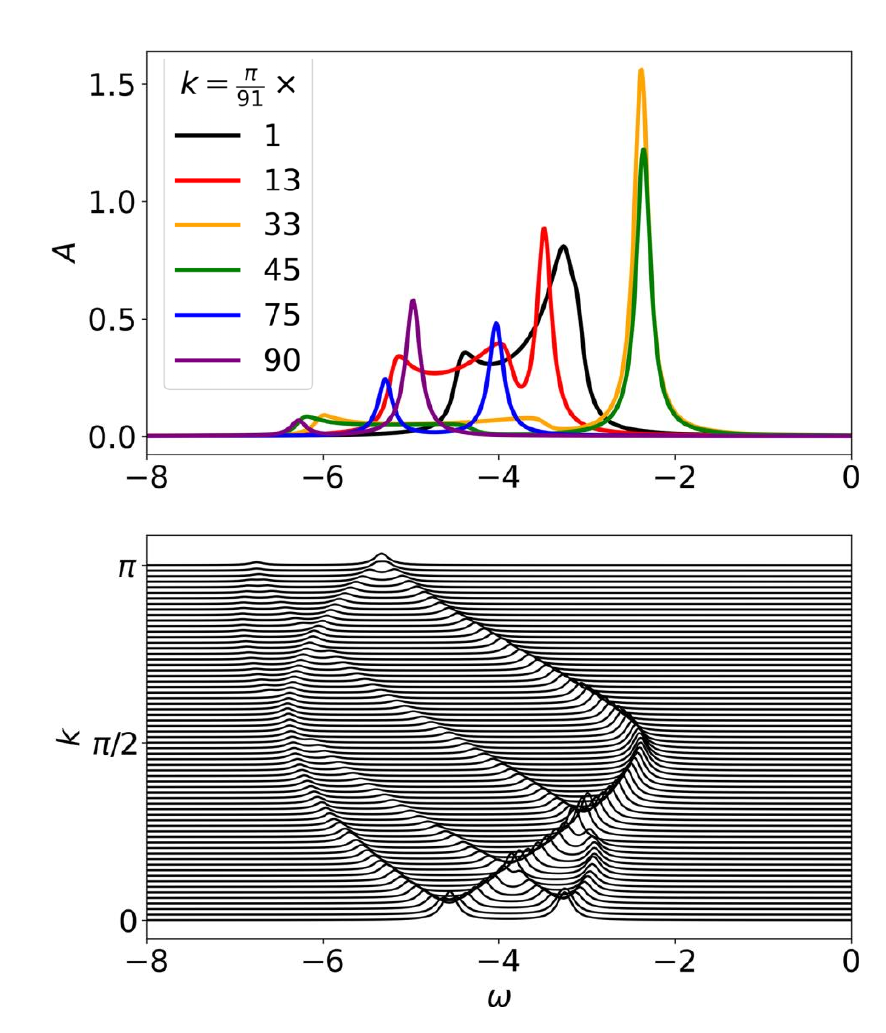}
\caption{The calculated angle-resolved spectra with improved 
$V_\mathrm{xc}$ obtained from a six-site cluster. The bottom figure
shows a two-dimensional plot of $k$ versus $\omega$. The agreement 
with the benchmark DMRG result is clearly much improved compared with
the one based on the dimer $V_\mathrm{xc}$
\cite{zhao2023}.}
\label{fig:Ak1D_Improved}%
\end{figure}

To improve the calculated angle-resolved spectra, $V_\mathrm{xc}$
derived from a larger cluster with six sites is used \cite{zhao2023}. 
The results are shown in
Fig. \ref{fig:Ak1D_Improved}. The agreement with the benchmark 
density matrix renormalisation group (DMRG) results
shown in Fig. \ref{fig:AkwU7-74} is markedly improved compared with the
one based on the dimer $V_\mathrm{xc}$.

\subsection{One-dimensional Heisenberg model}
\label{sec:HeisenbergModel}

The Vxc formalism has also been
extended to spin systems and applied to the paradigmatic 1D
antiferromagnetic (AF) Heisenberg model with encouraging results \cite{zhao2023}.

The 1D Heisenberg spin model
with a nearest-neighbour coupling is given by
\begin{align}
    \hat{H}&=-J\sum_i \hat{\mathbf{S}}_i \cdot\hat{\mathbf{S}}_{i+1}
    \nonumber\\
    &=-J\sum_i \left\{ \frac{1}{2}
    \left(  \hat{S}_i^+ \hat{S}_{i+1}^- + \mathrm{h.c.} \right) 
    +\hat{S}_i^z \hat{S}_{i+1}^z
    \right\},
\end{align}
where
\begin{align}
    \hat{S}_i^\pm = \hat{S}_i^x \pm i \hat{S}_i^y.
\end{align}
Experimentally of particular interest is the transverse-spin dynamical
structure factor,
\begin{align}
    S^{+-}(k,\omega)=\frac{1}{N}
    \sum_{mn} \int dt \langle \hat{S}_m^+(t)\hat{S}_m^-(0) \rangle
    e^{i\omega t} e^{-ik(m-n)},
\end{align}
where the expectation value is taken with respect to the ground state,
and $\hat{S}_m^\pm(t)$ are the spin field operators in the Heisenberg picture.
Here, $m$ and $n$ are site labels, and $k$ is a momentum label.
The transverse-spin Green function is defined as
\begin{align}
    iG_{ij}(t) = \langle \hat{S}_i^+(t)\hat{S}_j^-(0) \rangle \theta(t)
    +\langle \hat{S}_j^-(0) \hat{S}_i^+(t) \rangle \theta(-t).
\end{align}
Its equation of motion reads
\begin{align}
    i\partial_t G_{ij}(t)-F_{ij}(t)=2\delta_{ij} \delta(t)
    \langle \hat{S}_i^z \rangle,
\end{align}
where the interaction term is given by
\begin{align}
    F_{ij}(t) = -i\sum_m J_{im}\left[ 
    \langle m,ij\rangle -\langle i,mj \rangle
    \right].
\end{align}
Note that this definition differs slightly from the one in Ref. \cite{zhao2023}.
Here
\begin{align}
    \langle m,ij\rangle &= \langle T \hat{S}_m^z(t)
    \hat{S}_i^+(t)\hat{S}_j^-(0) \rangle,
    \\
    J_{im}&= J(\delta_{m,i+1}+\delta_{m,i-1}).
\end{align}
In analogy with the charge case, the spin Vxc can be
defined as
\begin{align}
    V^\mathrm{xc}_{ii,jj}(t) G_{ij}(t)
    =F_{ij}(t)-V^\mathrm{H}_i G_{ij}(t) -\sum_m V^\mathrm{F}_{im}G_{mj}(t),
\end{align}
where $V^\mathrm{H}$ and $V^\mathrm{F}$ are the analogue of the
Hartree and Fock potentials:
\begin{align}
    V^\mathrm{H}_i&=\sum_m J_{im} \langle \hat{S}^z_m \rangle,\\
    V^\mathrm{F}_{im}&=-J_{im} \langle \hat{S}^z_i \rangle.
\end{align}
A spin correlator $g_{mij}(t)$ can now be defined such that
\begin{align}
    -i\langle m,ij\rangle &= \langle \hat{S}^z_m \rangle 
    g_{mij}(t) G_{ij}(t)
    \nonumber\\
    &=[\rho^\mathrm{xc}_{mij}(t) 
    + \langle \hat{S}^z_m \rangle ]G_{ij}(t),
\end{align}
where the spin xc hole is given by
\begin{align}
    \rho^\mathrm{xc}_{mij}(t)= \langle \hat{S}^z_m \rangle 
    [g_{mij}(t)-1].
\end{align}
Since
\begin{align}
    \langle m,ij\rangle &= \theta(t)\langle \hat{S}_m^z(t)
    \hat{S}_i^+(t)\hat{S}_j^-(0) \rangle
    \nonumber\\
    &\quad +\theta(-t)\langle \hat{S}_j^-(0) \hat{S}_m^z(t)
    \hat{S}_i^+(t) \rangle,
\end{align}
it follows that
\begin{align}
    -i\sum_m \langle m,ij\rangle = [\theta(-t)+ S^z] G_{ij}(t),
\end{align}
where $S^z=\sum_m \langle\hat{S}^z_m\rangle$ is the total $z$ component
of the spin. The sum rule for the spin xc hole is then given by
\begin{align}
    \sum_m \rho^\mathrm{xc}_{mij}(t)= \theta(-t).
\end{align}
Note that the xc hole is defined with opposite sign to that
in Ref. \cite{zhao2023}.

\subsubsection{A four-site spin chain}

Before considering the infinite lattice, it is instructive
to consider a minimal cluster with an even number of sites for which
the $V^\mathrm{xc}$ is nonzero and can be calculated analytically.
This simplest cluster reveals features which are likely to be generic.
As an illustration, one of the diagonal elements is \cite{zhao2023}
\begin{widetext}
\begin{eqnarray}\label{wideq}
V^\mathrm{xc}_{11,11}(t>0)
=-J\frac{(\frac{(xy+x)(xy+x+2y)}{a_+^2})f_1+(x^2+x)f_2
+(\frac{(xy-3x)(xy-3x+2y-4)}{a_-^2})f_3}
{(\frac{xy+x+2y}{a_+})^2f_1+x^2f_2
+(\frac{xy-3x+2y-4}{a_-})^2f_3}
\end{eqnarray}
\end{widetext}
where
$ x=1+\sqrt{3}, y=1+\sqrt{2}, a_\pm=\sqrt{8\pm4\sqrt{2}}$, 
and $f_{i=1,2,3}$ are time oscillation factors  
determined by the difference between
the spin excitation energies and the ground state energy
The rest of the matrix elements of $V^\mathrm{xc}$
can be found in Ref. \cite{zhao2023}.

As in the dimer case, it is fruitful to work in the bonding-antibonding-like 
orbitals $\phi_\mu$ in view of an extrapolation to the infinite lattice case.
These are linear combinations of the site-orbitals
\begin{align}
 \phi_\mu = \sum_i \varphi_i U_{i\mu},\qquad  
\end{align}
where $\mu=A,B,C,D$,  $i=1,2,3,4$, and the unitary transformation matrix
$ U$ is
\begin{eqnarray}
U=\frac{1}{2}\left(
\begin{array}{cccc}
1 & \;\;\;1 & \;\;\;1 & \;\;\;1\\
1 & -1 & \;\;\;1 & -1\\
1 &  \;\;\;1 & -1 & -1\\
1 & -1 & -1 & \;\;\;1
\end{array}
\right).
\end{eqnarray}
For the Green function, the transformation reads
\begin{align}
 G_{\mu\nu}=\sum_{ij}U^\dagger_{\mu i}G_{ij}U_{j\nu },
\end{align}
The corresponding $V^\mathrm{xc}$ in the new basis is given by
\begin{eqnarray}
V^\mathrm{xc}_{\mu\alpha,\beta\nu}(t)
:=\sum_{mn}U^\dagger_{\mu m}U_{m\alpha}V^\mathrm{xc}_{mm,nn}(t)
U^\dagger_{\beta n}U_{n \nu}
\end{eqnarray}
so that the equation of motion is now
\begin{eqnarray}
i\partial_tG_{\mu\nu}(t)-
\sum_{\alpha\beta}V^\mathrm{xc}_{\mu\alpha,\beta\nu}(t)G_{\alpha\beta}(t)
=s_{\mu\nu}\delta(t),
\end{eqnarray}
where
\begin{align}
  s_{\mu\nu}=2\sum_{i}U^\dagger_{\mu i}\langle S_i^z \rangle 
  U_{i\nu}.  
\end{align}

To compute the Green function for positive time,
\begin{eqnarray}
G_{ij}(t)&=\langle{\Psi|e^{iHt}\hat{S}_i^+e^{-iHt}\hat{S}_j^-|\Psi}\rangle,
\end{eqnarray}
where $|\Psi\rangle$ is the ground state, 
one needs to use a complete set of eigenstates $|n\rangle$, which give nonzero
weight elements $\langle{n|\hat{S}_j^-|\Psi}\rangle$. 
For an even number of sites and AF coupling, the total 
$z$-spin of $|\Psi\rangle$
is zero, which means that the states $\{|n\rangle\}$ are in the $S^z=-1$ sector. 
Labeling the eigenenergy corresponding to state $|n\rangle$ with $E^{-}_n$, the
Green function can be written as
\begin{align}
    G_{ij}(t>0)=\sum_n e^{-i(E_n^- -E^0)t}
\langle{\Psi|\hat{S}_i^+|n\rangle\langle n|\hat{S}_j^-|\Psi}\rangle,
\end{align}
and the high-order term for positive time is
%
%
\begin{align}
    \langle{l,ij}\rangle_{t>0}
=\sum_n e^{-i(E_n^--E^0)t}
\langle\Psi|\hat{S}_l^z\hat{S}_i^+|n\rangle\langle n|\hat{S}_j^-|\Psi\rangle.
\end{align}
By diagonalising the Hamiltonian in the $S^z=0$ and $S^z=-1$ sectors, 
one gets $\{|\Psi\rangle; E^0\}$ and
$\{|n\rangle;E_n^-\}$,  respectively, and thus the weight elements
$\langle{n|\hat{S}_j^-|\Psi}\rangle$
and $\langle\Psi|\hat{S}_l^z\hat{S}_i^+|n\rangle$, respectively.
Among the four states of $|n\rangle$, only three give nonzero
$\langle{n|\hat{S}_q^-|\Psi}\rangle$. Explicitly, the time factors are
\begin{eqnarray}
f_1&=&e^{-i(E_0^--E^0)t}=e^{iJ(\frac{\sqrt{3}-\sqrt{2}+1}{2})t},\\ 
f_2&=&e^{-i(E_1^--E^0)t}=e^{iJ(\frac{\sqrt{3}+1}{2})t},\\
f_3&=&e^{-i(E_2^--E^0)t}=e^{iJ(\frac{\sqrt{3}+\sqrt{2}+1}{2})t}.
\end{eqnarray}
The independent elements of $V^\mathrm{xc}$ in orbital basis can be calculated
using $V^\mathrm{xc}_{ii,jj}(t)=\frac{F_{ij}(t)}{iG_{ij}(t)}$.

\begin{figure}
\centering
\includegraphics[scale=0.4, viewport=7cm 0cm 25cm 20cm, clip]
{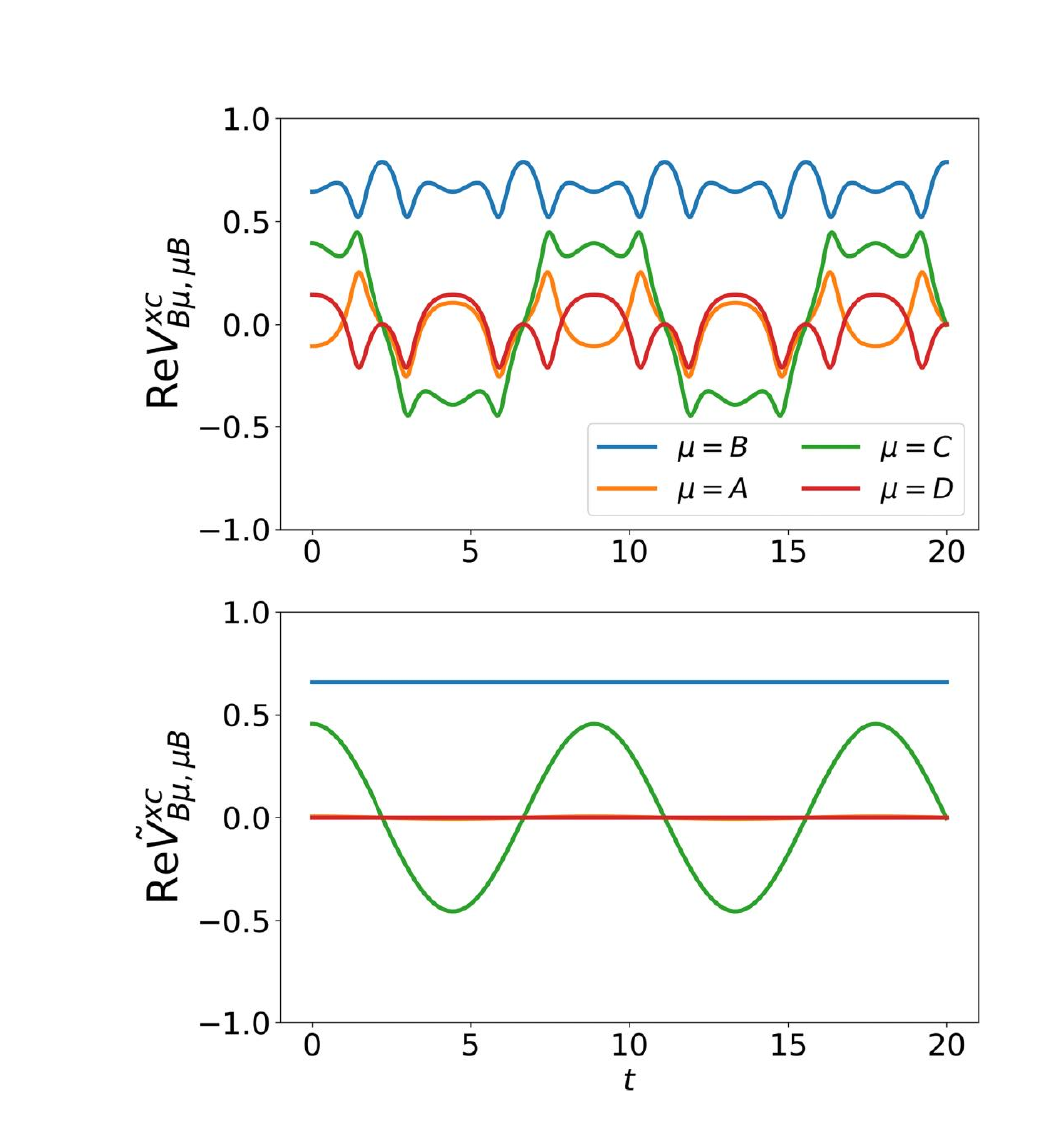}
\caption{Real part of $V^\mathrm{xc}$ of 
four-site spin-$\frac{1}{2}$ AF Heisenberg chain, in the unit of $|J|$
\cite{zhao2023}. 
Top panel: exact. Bottom panel: 
results when the high-excitation contribution is ignored 
(see Eqs. (\ref{eq:Vxc_heis_apprx}-\ref{beta_eq}) and related discussion).}
\label{fig:Vxc4sites}
\end{figure}

Consider the equation of motion of the diagonal term $G_{\mu\mu}$:
\begin{eqnarray}
\label{eq:eom_heis_mu}
[i\partial_t-V^\mathrm{xc}_{\mu\mu,\mu\mu}]G_{\mu\mu}(t)-
\sum_{\gamma\neq\mu}V^\mathrm{xc}_{\mu\gamma,\gamma\mu}(t)
G_{\gamma\gamma}(t)
\nonumber\\
-\sum_{\gamma\neq\delta}V^\mathrm{xc}_{\mu\gamma,\delta\mu(t)}
G_{\gamma\delta}(t)=s_{\mu\mu}\delta(t).
\end{eqnarray}
The contribution from fully off-diagonal terms 
$V^\mathrm{xc}_{\mu\nu,\delta\mu}$ should be negligible and 
the higher excitation term $f_3$ can be neglected since the weight
is relatively small. 
With these in mind, one arrives
at an approximate expression for the matrix elements of 
$V^\mathrm{xc}_{\mu\gamma,\gamma\mu}$:
\begin{align}
 V_{BD,DB}^{\mathrm{xc}}(t>0) &\approx 0,\\
V_{BA,AB}^{\mathrm{xc}}(t>0) &\approx 0,\\
V_{BB,BB}^\mathrm{xc}(t>0)&\approx -J\alpha, \\
V_{BC,CB}^\mathrm{xc}(t>0)&\approx -J\beta \exp\left[\frac{iJt}{\sqrt{2}}\right],
\label{eq:Vxc_heis_apprx}   
\end{align}
where
\begin{eqnarray}
&\alpha :=\frac{xy+x}{xy+x+2y}=\frac{2x+2}{xy+x+2},\\
&\beta :=\frac{1}{4}(\frac{a_+}{xy+x+2y}
+\frac{a_+}{xy+x+2})^2(x^2+x-\alpha x^2). 
\label{beta_eq}
\end{eqnarray}
As in Eq. (\ref{wideq}), $a_+=\sqrt{8+4\sqrt{2}}$.
The analytic spinon $V^\mathrm{xc}$ 
in the bonding-like basis and its
approximation are shown in Fig. \ref{fig:Vxc4sites}. 
Ignoring the high-excitation factor $f_3$ reduces the fine-structure details 
in $V^\mathrm{xc}$. 
Consequently, $V^\mathrm{xc}_{BB,BB}$ simplifies to a constant whereas
$V^\mathrm{xc}_{BC,CB}$ oscillates with a single frequency
and a constant magnitude, and all other components are negligible. 

\subsubsection{Extrapolation to the infinite lattice}

For the infinite lattice it is natural to use a Bloch basis:
\begin{align}
    G(k,t)=\frac{1}{N}\sum_{ij} G_{ij}(t) e^{-ik(i-j)},
\end{align}
\begin{align}
    V^\mathrm{xc}(k,t)=\frac{1}{N^2}\sum_{ij} 
    V^\mathrm{xc}_{ii,jj}(t) e^{-ik(i-j)}.
\end{align}
The equation of motion becomes
\begin{align}
    i\partial_t G(k,t) - \sum_q V^\mathrm{xc}(k-q,t)G(q,t)=2 s\delta(t),
\end{align}
where $s=\langle \hat{S}^z_i\rangle$, which is independent of the lattice site
due to translational symmetry.

The effect of $V^\mathrm{xc}$ on $G$ 
can be decomposed into a static (S) and a dynamic (D) part:
\begin{align}
    \sum_q V^\mathrm{xc}(k-q,t)G(q,t)=
    \left[V^\mathrm{S}(k) + V^\mathrm{D}(k,t)\right] G(k,t). 
\end{align}
The effective potential $\Xi$ defined in Eq. (\ref{eq:Xi})
can be identified as
\begin{align}
    \Xi(k,t)=V^\mathrm{S}(k) + V^\mathrm{D}(k,t).
\end{align}
The solution to the equation of motion for $t>0$ is then
\begin{align}\label{eq:G_sp}
    G(k,t)= G(k,0^+) e^{-iV^\mathrm{S}(k)t}
    e^{-i\int_0^t dt' V^\mathrm{D}(k,t')}.
\end{align}
The first term on the right-hand side 
of the above expression can be interpreted as the main quasiparticle 
excitation whereas the second to an incoherent or satellite feature.

For the antiferromagnetic case ($J<0$) that will be considered here,
it is known that the spinon dispersion has a lower (L) and an upper (U)
boundary:
\begin{align}
    \Omega^\mathrm{L}(k)&=-J\frac{\pi}{2}|\sin{k}|,\\
    \Omega^\mathrm{U}(k)&=-J\pi\sin{\frac{k}{2}}.
\end{align}
A reference Green function is constructed to reproduce the lower boundary
by choosing $V^\mathrm{S}(k)=\Omega^\mathrm{L}(k)$ yielding
\begin{align}
    G^\mathrm{ref}(k,\omega)=
    \frac{1}{\omega-\Omega^\mathrm{L}(k)}.
\end{align}
To solve for the full $G$ the dynamical component $V^\mathrm{D}(k,t)$
is needed, which is obtained by extrapolation from a finite cluster.

\begin{figure}
\centering
\includegraphics[scale=0.4, viewport=7cm 0cm 25cm 20cm, clip]
{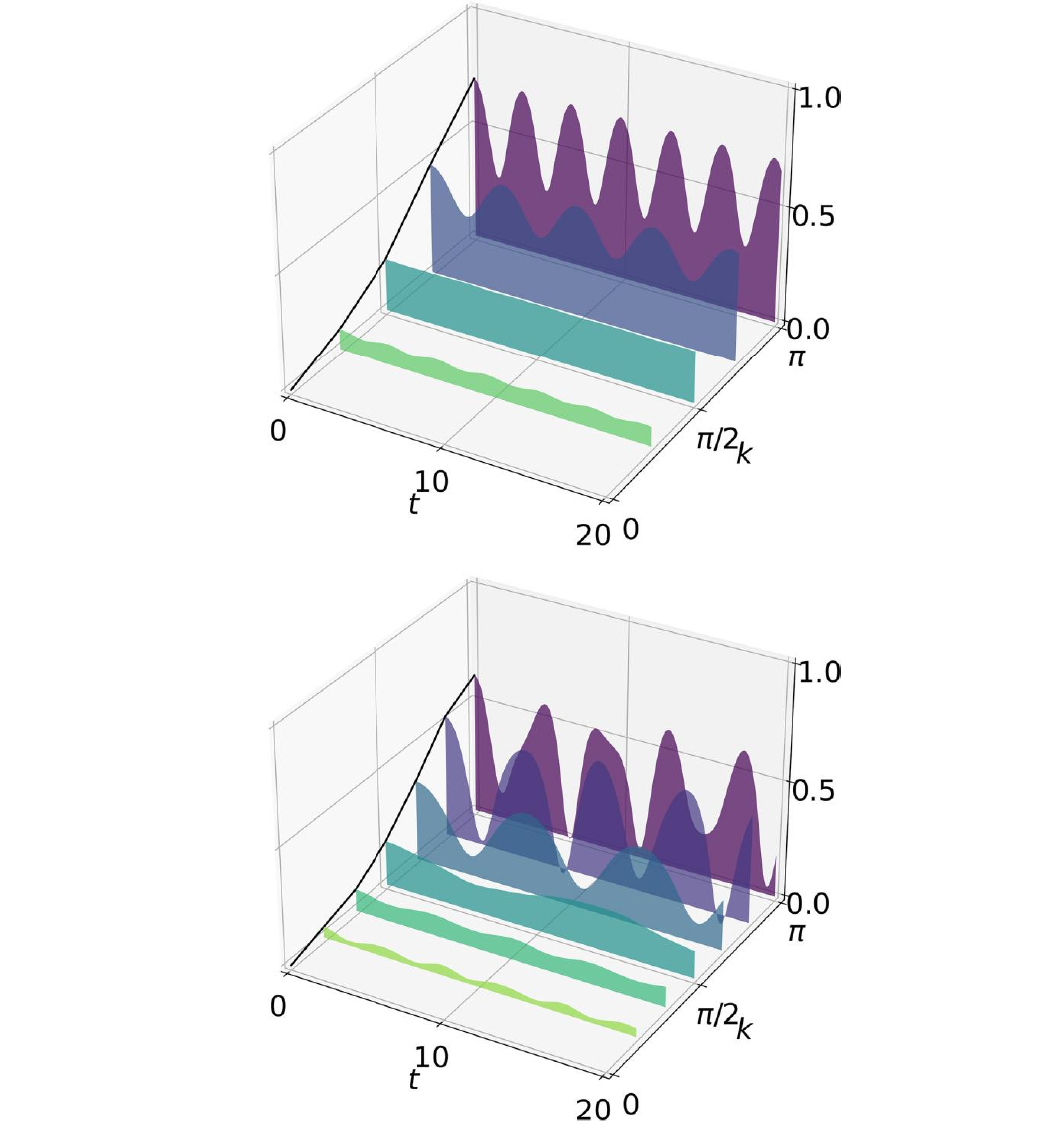}
\caption{Real part of $V^\text{D}$ from a spin-$\frac{1}{2}$ 
AF Heisenberg ring \cite{zhao2023}.
Top (bottom) panel: results for a ring with 8 (12) sites.
The number of available $k$-points is equal to half the number of sites.}
\label{fig:Z_hei}
\end{figure}

Fig. \ref{fig:Z_hei} displays the real part of $V^\mathrm{D}(k,t)$,
which reveals that
for each $k$, $\text{Re}V^\text{D}(k,t)$ 
oscillates in time around a momentum-dependent term. 
This behaviour reflects a single quasiparticle-like main excitation. 
Based on this observation, a physically motivated ansatz
for $V^\text{D}(k,t)$ in the infinite-chain case is
\begin{eqnarray}
\label{eq:sp_quasip}
V^\text{D}(k,t)=\mathcal{A}(k)e^{-i\omega^\text{sp}(k) t}+\mathcal{B}(k),
\end{eqnarray}
where the amplitude $\mathcal{A}$, 
the spinon excitation energy $\omega^\text{sp}$, 
and the shift term $\mathcal{B}$ increase monotonically as 
$k$ increases from $0$ to $\pi$. 
Inserting this ansatz into Eq.\eqref{eq:G_sp} yields
\begin{align}\label{eq:G_sp_Z}
    G(k,t>0)=G(k,0^+)&e^{-i[V^\text{S}(k)+\mathcal{B}(k)]t}
    \nonumber\\
    &\times 
e^{\frac{\mathcal{A}(k)}{\omega^\text{sp}(k)}(e^{-i\omega^\text{sp}(k)t}-1)},
\end{align}
where the static potential is
$V^\text{S}(k)=-J\pi|\sin k|/2$.

Expanding the last term on the RHS of Eq.\eqref{eq:G_sp_Z} to first order in 
$e^{-i\omega^\text{sp}(k)t}$ leads to
\begin{eqnarray}
G(k,t>0)\approx 
G(k,0^+)e^{-i[V^\text{S}(k)+\mathcal{B}(k)]t}\nonumber\\
\times\big[1+\frac{\mathcal{A}(k)}{\omega^\text{sp}(k)}
(e^{-i\omega^\text{sp}(k)t}-1)\big],
\end{eqnarray}
which in the frequency domain becomes
\begin{widetext}
\begin{eqnarray}\label{eq:FinalVxc}
G(k,\omega)=G(k,0^+)\Big[\frac{1-\frac{\mathcal{A}(k)}{\omega^\text{sp}(k)}}
{\omega-[V^\text{S}(k)+\mathcal{B}(k)]}+\frac{\frac{\mathcal{A}(k)}
{\omega^\text{sp}(k)}}{\omega-[V^\text{S}(k)+\mathcal{B}(k)
+\omega^\text{sp}(k)]}\Big].
\end{eqnarray}
\end{widetext}
The above equation shows that the dynamical structure factor
consists of a main peak centred at $\omega=V^\text{S}+\mathcal{B}$
followed by a satellite peak separated from the main peak by
the spinon energy $\omega^\text{sp}$.
Spectral weight in the amount of $\frac{\mathcal{A}(k)}
{\omega^\text{sp}(k)}$ is transferred from the main peak to the satellite.
Due to finite-size effects, the finite cluster solution gives nonzero
$\mathcal{B}$ at $k=\pi$, 
which opens a spin gap that is not present
for the spin$-\frac{1}{2}$ lattice. 
To account for these finite-size effects, $\mathcal{B}$ is adjusted to 
a smaller value in our extrapolation.

\begin{figure}
\centering
\includegraphics[scale=0.4, viewport=7cm 0cm 25cm 20cm, clip]
{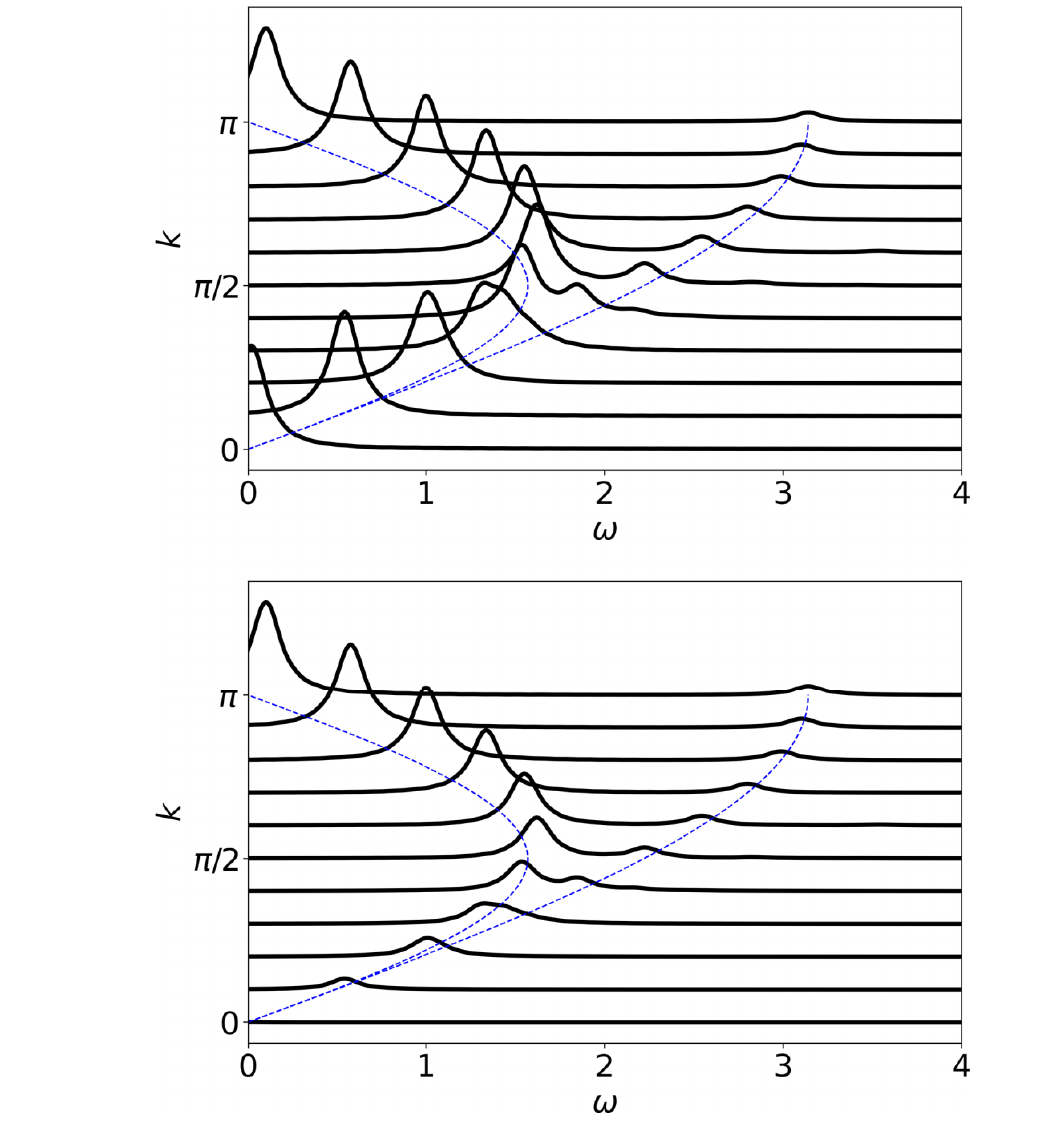}
\caption{Dynamic structure factor of 1D spin-$\frac{1}{2}$ AF Heisenberg lattice 
calculated with an extrapolated $V^\mathrm{xc}$, 
with broadening 0.1 \cite{zhao2023}.
Top panel: weight factor $G(k,t=0)$ 
considered as unit. Bottom panel: weight renormalised with cluster $G(k,0^+)$. 
Blue dashed curves are the boundaries for two-spinon processes.}
\label{fig:Sqw_hei}
\end{figure}

Based on the preceding discussion, the lattice $V^\mathrm{xc}$ is
obtained by extrapolation from the cluster result. 
With $V^\text{D}$ obtained by an exact diagonalisation
of a cluster of 12 sites,
the parameters $\mathcal{A},\mathcal{B}$ and $G(k,0^+)$ are estimated
by linear interpolation. 
The spinon excitation energy is estimated by fitting the cluster
$\omega^\text{sp}$ to the two-spinon spectrum boundary,
\begin{eqnarray}
\omega^{\text{sp}}\rightarrow(-J)\pi
\left(\sin\frac{k}{2}-\frac{1}{2}|\sin k|\right).
\end{eqnarray}
The longitudinal and transverse spin dynamical structure factors are then
calculated from the spinon Green function. 
For a spin-isotropic system $S^{zz}$ and $S^{+-}$ 
differ by a constant factor. For this reason
only the spectral function of the Green function in 
Eq. (\ref{eq:G_sp_Z}) is calculated and the result
is shown in Fig. \ref{fig:Sqw_hei}.
%
%
Comparison with the spin dynamical structure factor measured in
inelastic neutron scattering for the 1D compound KCuF$_3$
\citep{lake2013} shows close agreement in
both the peak locations and the relative weights.

\begin{figure}
\centering
\includegraphics[scale=0.5, viewport=8cm 5cm 25cm 14cm, clip]
{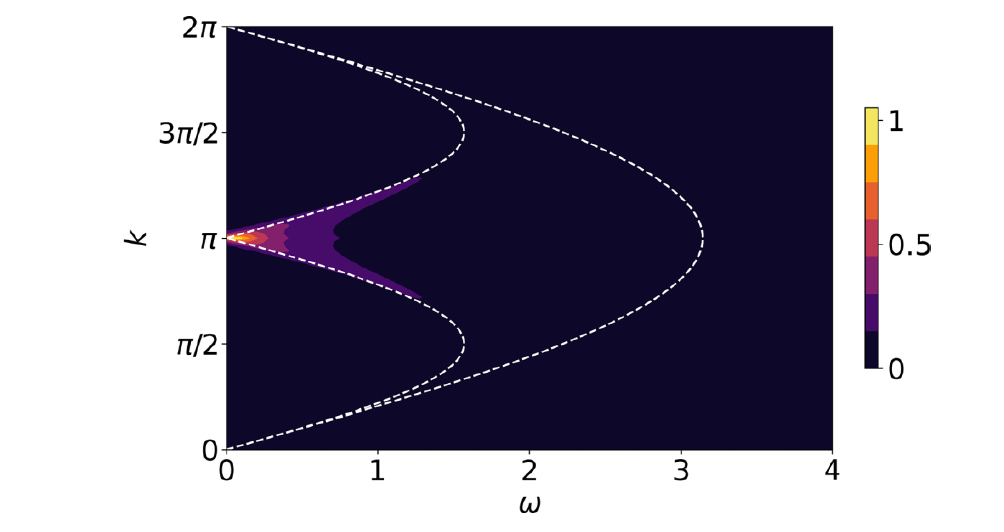}
\caption{Dynamic structure factor of a 100-site 
1D spin-$\frac{1}{2}$ AF Heisenberg chain calculated within DMRG
\cite{zhao2023}.
The weight is renormalised to be in the range $0$-$1$.}
\label{fig:Sqw_DMRG}
\end{figure}

A notable feature of the
dynamical structure factor $S(k,\omega)$ shown in
the bottom panel of Fig. \ref{fig:Sqw_hei} is that
the weight of the main peak is close to zero at small $k$ and
increases with $k$ reaching a maximum at $k\rightarrow\pi$. 
The spectrum with a broadening factor of 0.1 is gapless.
The result agrees well with that calculated using DMRG shown in
Fig. \ref{fig:Sqw_DMRG}.

Although the extrapolation procedure may leave out fine details of the
spectra, it captures most of the essential physical
characteristics of the 1D AF Heisenberg model
with a very low computational load.
This appealing feature of the
method is expected to also apply in more challenging situations, 
for example, in higher dimensions and realistic materials,
where rigorous references such as the Bethe ansatz are not available.

\subsection{The single-impurity Anderson model}

The single-impurity Anderson model describes an impurity with a few
discrete levels embedded
in a bath of noninteracting conduction electrons with a continuous degree
of freedom. Although it is seemingly a simple model,
it is relevant in a wide range
of physical phenomena such as the Kondo effect and quantum transport.
In particular, it plays an important role as an auxiliary system
in dynamical mean-field theory, a method of choice for treating
strongly-correlated materials from first principles in combination
with DFT and $GW$ method.

The treatment in this section follows closely the
work of Zhao \cite{zhao2025}.
The single-impurity Anderson model (SIAM) is given by
\begin{align}\label{eq:H_SIAM}
\hat{H}_\mathrm{SIAM}
&=\epsilon_f(\hat{n}_{f\uparrow}+\hat{n}_{f\downarrow})
+U\hat{n}_{f\uparrow}\hat{n}_{f\downarrow}
\nonumber\\ 
&\quad 
+\sum_{k\sigma}\Big[\epsilon_k\hat{c}_{k\sigma}^\dagger\hat{c}_{k\sigma}
+(v_k\hat{f}_{\sigma}^\dagger\hat{c}_{k\sigma}+\text{h.c.})\Big].
\end{align}
Here $\hat{f}_\sigma^\dagger$ ($\hat{f}_\sigma$) creates (annihilates)
an electron with spin $\sigma$ on the impurity site,
$\hat{n}_\sigma=\hat{f}_\sigma^\dagger\hat{f}_\sigma$ is
the corresponding number operator, $\hat{c}_{k\sigma}^\dagger$
($\hat{c}_{k\sigma}$) creates (annihilates) a bath electron with energy
$\epsilon_k$.
The coupling between the impurity and bath electrons is
given by the hybridisation amplitude $v_k$ assumed to be spin independent,
and $\epsilon_f$ and $U$ are the impurity onsite energy
and Coulomb interaction, respectively.

We consider a symmetric SIAM at half-filling, which means
that the ensemble average
\begin{align}
 n_{f\sigma}=\text{Tr}\{\hat{\rho}_\beta\hat{n}_{f\sigma}\}=0.5 ,  
\end{align}
where the statistical operator and the partition function
are given by, respectively,
\begin{align}
    \hat{\rho}_\beta=Z_\beta^{-1}e^{-\beta(\hat{H}-\mu\hat{N})},
    \quad Z_\beta=\text{Tr}[e^{-\beta(\hat{H}-\mu\hat{N})}].
\end{align}
$\beta$ is the inverse temperature.
The impurity level $\epsilon_f$ is chosen so that
\begin{align}\label{eq:half-filled}
 \epsilon_f+\frac{U}{2}=0.   
\end{align}
In the limit of no hybridisation, the removal
and addition impurity spectra consists of two peaks at $\epsilon_f$ and
$\epsilon_f+U$. Setting the chemical potential in the middle of
these two levels
to zero leads to the choice in Eq. (\ref{eq:half-filled}) yielding
two peaks at $\pm \frac{U}{2}$. The hybridisation shifts and broadens the
peaks and induces the Kondo peak at the chemical potential.
The number of fermionic sites (impurity + bath) $L$ is chosen to be even.

The equilibrium finite-temperature time-ordered Green function is given by
\begin{align}
    i{G}_{ff,\sigma}(t,\beta)
    &=\langle T\hat{f}_{\sigma}(t) 
    \hat{f}_{\sigma}^\dagger(0) \rangle
:=\text{Tr}\{\hat{\rho}_\beta[T
\hat{f}_{\sigma}(t) 
    \hat{f}_{\sigma}^\dagger(0)]\}, 
\end{align}
where $T$ is the time-ordering symbol.
Its spectral decomposition is given by
\begin{align}\label{eq:SpecRepGff}
    &i{G}_{ff,\sigma}(t,\beta)
    \nonumber\\
    &=\theta(t)
    \frac{1}{Z_\beta}
    \sum_{mn}e^{-\beta E_m}
    e^{-i(E_n^+-E_m)t}
    \big|\langle n^+|\hat{f}_{\sigma}^\dagger|m\rangle\big|^2 
    \nonumber\\
&-\theta(-t)\frac{1}{Z_\beta}
\sum_{mn}e^{-\beta E_m}e^{i(E_n^- -E_m)t}
\big|\langle n^-|\hat{f}_{\sigma}|m\rangle \big|^2,
\end{align}
where
$|n^\pm\rangle$ are the $(L\pm 1)$-electron eigenstates with
eigenenergies $E^\pm_n$.
Due to particle-hole symmetry of half-filling, it is only
necessary to consider the case $t>0$.
The equation of motion of the Green function in the Vxc formalism
is given by
\begin{align}\label{eq:EOMGff}
    \big[i\partial_t-\epsilon_f-V^\text{H}-V^\mathrm{xc}_\sigma(t,\beta)\big]
    G_{ff,\sigma}(t,\beta)=\delta(t),
\end{align}
where the Hartree term
\begin{align}\label{eq:VH}
 V^\text{H}=Un_{f\bar{\sigma}}=\frac{U}{2}   
\end{align}
is due to the impurity electron with opposite spin
$\bar{\sigma}=-\sigma$. The xc field $V^\mathrm{xc}_\sigma$
in the above equation of motion includes the hybridisation effects.
The formal solution is given by
\begin{align}\label{eq:FormalSolnG}
    G_{ff,\sigma}(t,\beta)=G_{ff,\sigma}(0,\beta)
    e^{-i(\epsilon_f+V^\text{H})t 
    -i\int_0^t dt'V^\mathrm{xc}_\sigma(t',\beta)}.
\end{align}

To investigate the effects of the hybridisation,
it is useful to
consider the noninteracting case ($U=0$) at zero temperature.
In this case the Green function can be solved analytically yielding
\begin{align}
    G_{ff,\sigma}(\omega)=\frac{1}{\omega-\epsilon_f-\Delta(\omega)},
\end{align}
where
\begin{equation}
\Delta(\omega)=\sum_k\frac{|v_k|^2}{\omega-\epsilon_k}
\end{equation}
is the hybridisation function.
By modeling the continuous
bath as a tight-binding ring with $N_c$ sites and hopping strength $t_h$,
and the impurity site couples to one site with strength $V$,
$\Delta(\omega)$ can be calculated analytically.
In this model, the SIAM parameters are given by
\begin{align}
    \epsilon_k=2t_h\cos(k),\quad v_k=\frac{V}{\sqrt{N_c}}.
\end{align}
When $|\epsilon_f|,V\ll 2|t_h|$,
we approach the so-called wide-band limit (WBL)
so that the hybridisation function can be treated as a constant,
\begin{equation}
\Delta(\omega)=i\Gamma=i\frac{\pi V^2}{4t_h}.
\end{equation}
This yields
\begin{equation}
V_\text{nonint,WBL}^\mathrm{xc}(t)=i\Gamma\theta(-t).
\end{equation}
The infinitely wide bath band results in a purely imaginary
hybridisation field, leading to a broadening of the impurity level
$\epsilon_f$. 
This hybridisation effect persists in non-WBL or interacting cases.

Consider now the case of a finite but low temperature.
Vxc can be obtained from Eqs. (\ref{eq:SpecRepGff})
and (\ref{eq:EOMGff}), yielding for $t>0$
\begin{align} \label{eq:Vxc_SIAM}
V^\mathrm{xc}_{\sigma}(t>0,\beta)
=\frac{\sum_m e^{-\beta E_m}\sum_{n}a^+_{n,m}
\omega^+_{n,m}e^{-i\omega^+_{n,m}t}}{\sum_m e^{-\beta E_m}
\sum_{n}a^+_{n,m}e^{-i\omega^+_{n,m}t}},
\end{align}
where
\begin{align}
    \omega^+_{n,m}=E^+_n-E_m,\quad 
    a^+_{n,m}=\big|\langle n^+|\hat{f}_{\sigma}^\dagger|m\rangle\big|^2.
\end{align}
At low temperature $e^{-\beta E_m}$ is negligible except for
the two lowest eigenstates $m=1,2$.
Vxc can be written as \cite{zhao2025}
\begin{align}
\label{eq:Vxc_lowT}
V^\mathrm{xc}_{\sigma}(t,\beta)
=V^\mathrm{xc}_{\sigma}(t,T=0)+\widetilde{V}(t)e^{-\beta(E_2-E_1)}.
\end{align}
The second term on the right-hand side is a thermal correction.

\subsubsection{Impurity model with a single-site bath (dimer)}

Before considering the full SIAM, it is instructive to study a dimer
with $U$ only on one site and the other site treated as a bath.
This model can be solved analytically, providing valuable physical insights
into some of the key features of the SIAM.
In this highly simplified model, in which the bath is represented by one site, 
the Hamiltonian is given by
\begin{align}
\hat{H}_\text{dimer}=\epsilon_f
\sum_\sigma \hat{n}_{f\sigma}
+U\hat{n}_{f\uparrow}\hat{n}_{f\downarrow}
+V\sum_{\sigma}(\hat{f}_{\sigma}^\dagger\hat{c}_{\sigma}+\text{h.c.}).
\end{align}
Consider first the case of $T=0$. 
In the Kondo regime ($U\gg V$), the particle part of Vxc
($t>0$) has an approximate form \cite{zhao2025}
\begin{eqnarray}
\label{eq:dimer_Vxc_0}
V^\mathrm{xc}_{\sigma}(t>0,T=0)
\approx\omega_{1}-\lambda\Omega e^{i\Omega t},
\end{eqnarray}
where
\begin{align}
    \omega_1&=\sqrt{\frac{U^2}{16}+4V^2}+\sqrt{\frac{U^2}{16}+V^2},
    \label{eq:omega1}
    \\
    \lambda&\approx\frac{36V^2}{U^2}, \label{eq:lambda}
    \\
    \Omega&=\sqrt{\frac{U^2}{4}+4V^2}.\label{eq:Omega}
\end{align}
Keeping in mind the condition in Eq. (\ref{eq:VH}) and that
for half-filling $V^\mathrm{H}=\frac{U}{2}$,
the impurity Green function
according to Eq. (\ref{eq:FormalSolnG}) is given by
\begin{align}
    G_{ff,\sigma}(t,T=0)&=g^+ 
    e^{-i\omega_{1}t+\lambda\Omega i\int_0^t dt' e^{i\Omega t'} }
    \nonumber\\
    &\approx g^+ e^{-i\omega_1 t}\left[ 1 
    +i\lambda\Omega\int_0^t dt' e^{i\Omega t'} \right]
    \nonumber\\
    &\approx g^+ \left[ 
    (1-\lambda) e^{-i\omega_1 t}+\lambda e^{-i\omega_0 t} 
    \right],
\end{align}
where
\begin{align}
    g^+&=G_{ff,\sigma}(0,T=0)=-0.5 i,\\
    \omega_0&=\omega_1-\Omega\approx \frac{6V^2}{U}.
\end{align}
The particle-hole symmetric spectral function is then given by
\begin{align}
    A_\text{dimer}(\omega,T=0)
    &=\frac{1-\lambda}{2}\left[
    \delta(\omega+\omega_1)+\delta(\omega-\omega_1) \right]
    \nonumber\\
    &\quad +\frac{\lambda}{2} \left[ 
    \delta(\omega+\omega_0)+\delta(\omega-\omega_0) \right].
\end{align}

Some important
features of SIAM can be deduced: for large $U$, there are two peaks
($\omega=\pm\omega_1$) corresponding to impurity levels
$\epsilon_f$ and $\epsilon_f+U$. 
The excitation associated with energy $\Omega$
induces two central peaks at
$\omega=\pm\omega_0\approx0$, which may be regarded as a harbinger of
the Kondo resonance. However, for the dimer the spectral weight
given in Eq. (\ref{eq:lambda}),
$\frac{\lambda}{2}\sim (\frac{V}{U})^2$, vanishes as $U$ increases.
This can also be seen from Vxc
in Eq. (\ref{eq:dimer_Vxc_0}): with increasing $U$,
the exponential term with amplitude $\lambda\Omega\sim\frac{V^2}{U}$
becomes negligible.
The vanishing of the Kondo resonance
is a consequence of having a single site instead of a continuous bath.

Consider now the case of a finite but low temperature.
At low temperature, the term that oscillates with time in
Eq. (\ref{eq:Vxc_lowT}) can be written as \cite{zhao2025}
\begin{eqnarray}
\frac{\widetilde{V}(t)}{V^\mathrm{xc}_{\sigma}(t,T=0)}
\approx\lambda_1e^{i\Omega_1t}-\lambda_2e^{i\Omega_2t},
\end{eqnarray}
where
\begin{align}
 \lambda_1,\lambda_2\sim\frac{V^2}{U^2}, \quad
 \Omega_1\sim U, \quad \Omega_2\sim\frac{V^2}{U}.
\end{align}
Vxc is then
\begin{align}
    \label{eq:dimer_Vxc_T}
V^\mathrm{xc}_{\sigma}(t,\beta)
\approx\omega_{1}-\lambda\Omega e^{i\Omega t}
+e^{-\beta\Delta_0}\omega_1(\lambda_1 e^{i\Omega_1 t}
-\lambda_2e^{i\Omega_2 t}),
\end{align}
where $\Delta_0\sim\frac{V^2}{U}$.
At low temperature, the condition $e^{-\beta\Delta_0}\ll 1$ is valid.
The particle ($\omega>0$) spectral function is given by
\begin{align}\label{eq:A_dimer_T}
&A_\text{dimer}(\omega>0,\beta)
\nonumber\\
&\approx\frac{1-\lambda-e^{-\beta\Delta_0}\omega_1(
\frac{\lambda_2}{\Omega_2}-\frac{\lambda_1}{\Omega_1})}{2}\delta
(\omega-\omega_1)+\frac{\lambda}{2}\delta(\omega-\omega_0)
\nonumber\\
&\quad +\frac{e^{-\beta\Delta_0}\omega_1}{2}
\left[ \frac{\lambda_2}{\Omega_2}
\delta(\omega-\widetilde{\omega}_1)
-\frac{\lambda_1}{\Omega_1}
\delta(\omega-\widetilde{\omega}_0) \right],    
\end{align}
where
\begin{align}
    \widetilde{\omega}_0=\omega_1-\Omega_1,\quad 
    \widetilde{\omega}_1=\omega_1-\Omega_2.
\end{align}
The first two terms on the right-hand side of Eq.~\eqref{eq:A_dimer_T}
correspond to the original zero-temperature peaks, while the last two terms,
with weights proportional to $e^{-\beta\Delta_0}$, are
referred to as thermal peaks, which are
close to the original zero-temperature peaks.
The mixture of a zero-temperature peak and a thermal peak
which are close in energy induces a broadening of
the zero-temperature peak with width given effectively by
the energy difference between the two peaks.

\subsubsection{Impurity coupled to a finite cluster}
\begin{figure}[h]
\begin{center} 
\includegraphics[scale=0.36, viewport=5.5cm 7cm 28cm 15cm, clip,
width=\columnwidth]
{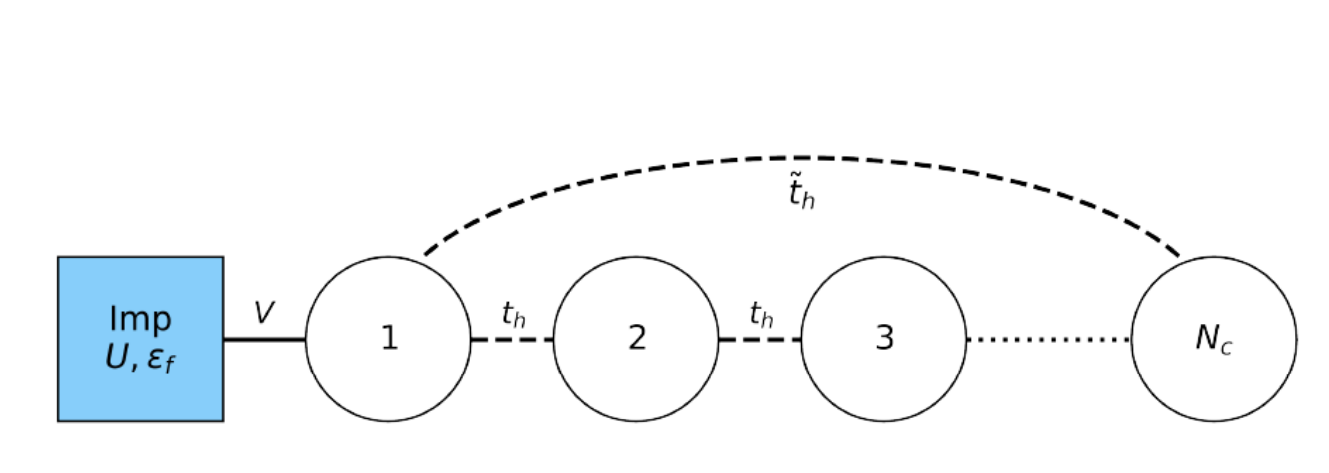}
\caption{Illustration of the single-impurity Anderson model with a bath of
$N_c$ sites. When a periodic boundary condition is applied ($\widetilde{t}_h=t_h$)
the SIAM parameters become
$\epsilon=2t_h \cos{k}$ and $v_k=\frac{V}{\sqrt{N_c}}$.
The figure is taken from \cite{zhao2025}.
}
\label{fig:SIAM}%
\end{center}
\end{figure}

\begin{figure}[h]
\begin{center} 
\includegraphics[scale=0.5, viewport=8cm 0.2cm 25cm 18cm, clip]
{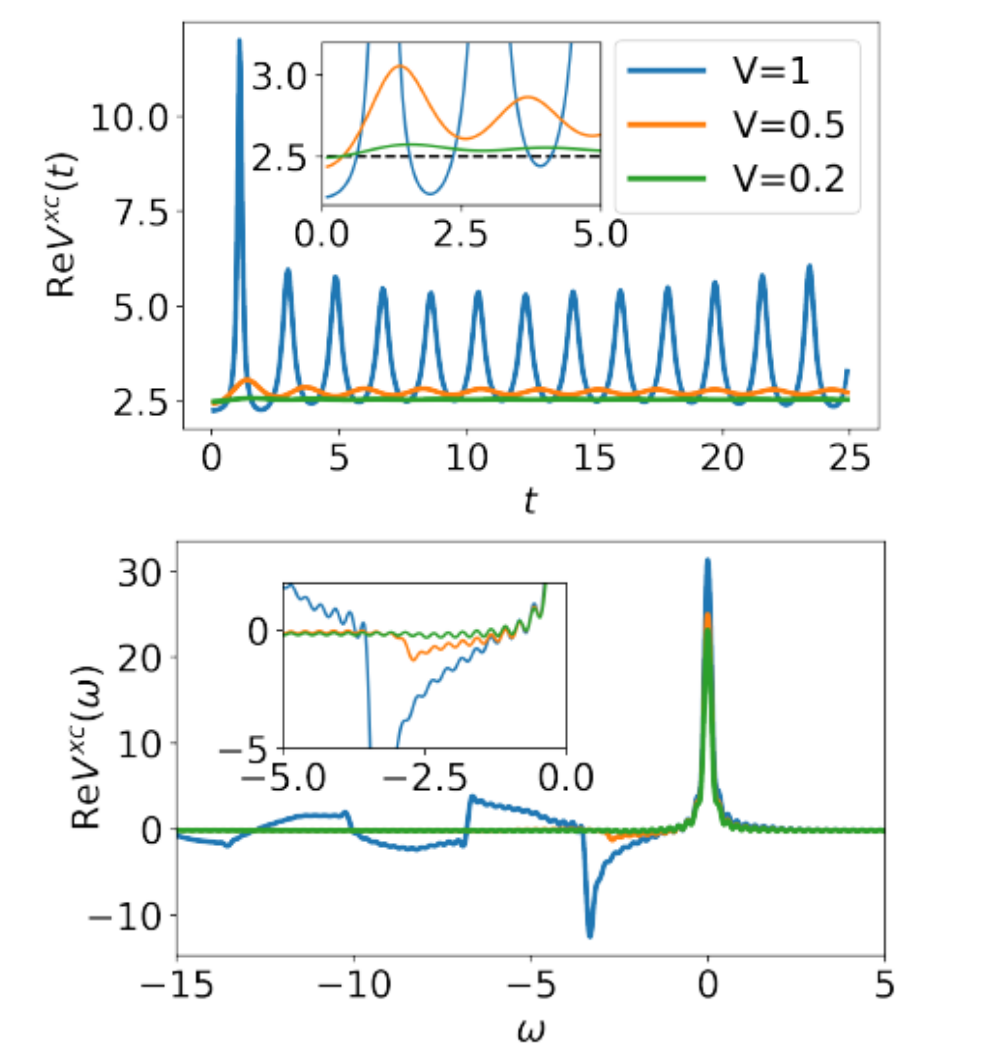}
\caption{Top: The real part of Vxc as a function of time. 
Bottom: The real part of Vxc as a function of frequency calculated
with a broadening factor $\eta=0.1$.
The calculations have been performed at zero temperature with parameters
$\epsilon_f=-2.5$, $U=5$, and $t_h=-1$.
The figure is taken from \cite{zhao2025}.
}
\label{fig:VxcSIAM50sites}%
\end{center}
\end{figure}

After studying the simplest impurity model, a dimer, we are in the position
to consider a more realistic model in which the impurity is coupled
to a bath with a finite number $N_c$ of noninteracting sites as illustrated
in Fig. \ref{fig:SIAM}.
The SIAM defined in Eq. (\ref{eq:H_SIAM}) with
\begin{align}
    \epsilon_k = 2t_h \cos{k},\quad v_k =\frac{V}{\sqrt{N_c}}
\end{align}
is reproduced in the limit $N_c\rightarrow\infty$. 

Fig. \ref{fig:VxcSIAM50sites} shows the numerically exact Vxc
of the SIAM obtained with a bath consisting of a
cluster of 50 sites \cite{zhao2025}.
Re$V^\mathrm{xc}(t)$ exhibits the characteristic oscillating behaviour,
of the form
\begin{align}\label{eq:VxcCluster}
 V^\mathrm{xc}(t)=\mathcal{A}e^{-i\omega_1 t}+\mathcal{C}   
\end{align}
The hybridisation between the impurity and the bath requires
a complex $\mathcal{C}$: Re $\mathcal{C}$ and Im $\mathcal{C}$ determine
the peak position and the width of the Hubbard band, respectively.
Moreover, unlike the dimer, $\omega_1$ can be complex.

To develop an ansatz for Vxc, it is necessary to consider
its limiting behaviour for large $t$. Since time and frequency
are reciprocal variables, the large-time behaviour is determined by the
low-frequency structure of the Green function. If $\Gamma_\mathrm{K}$ is the
half-width of the Kondo resonance at low frequency,
then at large time one expects
\begin{align}
    G_{ff,\sigma}(t>0,\beta) \propto e^{-i\Gamma_\mathrm{K} t}.
\end{align}
According to the formal solution in Eq. (\ref{eq:FormalSolnG}),
the corresponding large-$t$ Vxc
is proportional to $\Gamma_\mathrm{K}$.
This together with the finite-cluster result leads to
\begin{equation}
   V^\mathrm{xc}(t>0,\beta)\approx \left\{
   \begin{array}[c]{c l}
\lambda\omega_1 e^{-i\omega_1 t}+\mathcal{C},&\quad \mathrm{small} \,\,t\\
-i\Gamma_\mathrm{K},&\quad \mathrm{large}\,\, t
\end{array}
\right.
\end{equation}

\subsubsection{Ansatz for SIAM Vxc}
\begin{figure}[h]
\begin{center} 
\includegraphics[scale=1, viewport=2cm 4cm 32cm 14cm, clip, width=\columnwidth]
{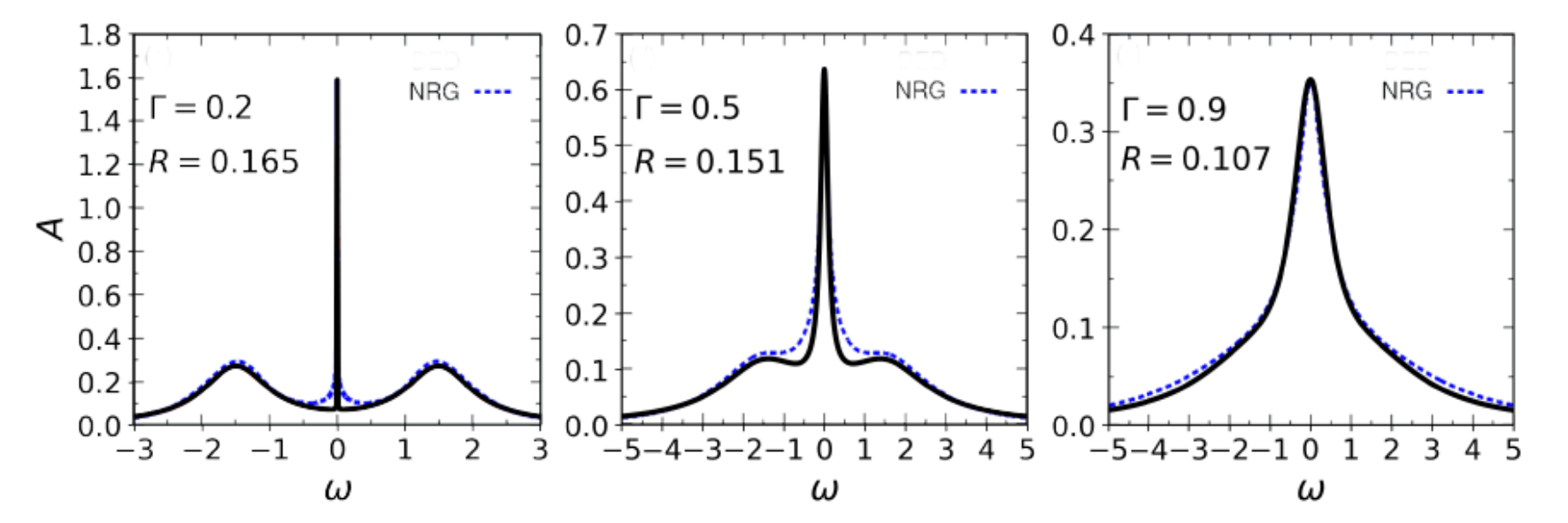}
\end{center}
\caption{The zero-temperature particle-hole symmetric
SIAM spectra calculated 
with $U=3$ and $t_h=50$ to approach the wide-band limit and compared with
benchmark results obtained using the numerical renormalisation group
(NRG) method \cite{zhao2025}.
$\Gamma$ is related to the height of the Kondo resonance as
$A(\omega=0)=\frac{1}{\pi\Gamma}$ and $R$ is the ratio between the Hubbard
and Kondo peaks. For details of the determination of the parameters
defining the ansatz see Zhao \cite{zhao2025}.
}
\label{fig:SIAM_Spectra}%
\end{figure}

An ansatz that fulfils the large- and small-time behaviour of
Vxc is \cite{zhao2025}
\begin{align}
    V^\mathrm{xc}(t>0,\beta)=
    \frac{\lambda(\omega_1+\mathcal{C})+(1-\lambda)\mathcal{C}e^{i\omega_1t}}
    {\lambda+(1-\lambda)e^{i\omega_1t}}.
\end{align}
Here, $\lambda$ is real, $\omega_1$ and $\mathcal{C}$ are complex,
and $\omega_1+\mathcal{C}=-i\Gamma_\mathrm{K}$ is temperature dependent.
The fractional form follows the general solution in Eq. (\ref{eq:Vxc_SIAM}).
The particle-hole symmetry implies that
\begin{align}
    V^\mathrm{xc}(-t,\beta)=-V^\mathrm{xc}(t,\beta).
\end{align}

According to the formal solution in Eq. (\ref{eq:FormalSolnG}) with
$\epsilon_f+V^\mathrm{H}=0$,
\begin{align}
    G_{ff,\sigma}(t>0,\beta)=-\frac{i}{2} \left[ 
    (1-\lambda)e^{-i\mathcal{C}t} + \lambda e^{-i(\mathcal{C}+\omega_1)t}
    \right]
\end{align}
with the corresponding spectral function
\begin{align}\label{eq:AwAnsatz}
    &A(\omega>0,\beta)=\frac{1-\lambda}{2\pi}
    \frac{|\mathrm{Im}\mathcal{C}|}
         {(\omega-\mathrm{Re}C)^2 + (\mathrm{Im}\mathcal{C})^2}
    \nonumber\\
    &+\frac{\lambda}{2\pi}
    \frac{|\mathrm{Im}(\mathcal{C}+\omega_1)|}
         {[\omega-\mathrm{Re}(C+\omega_1)]^2 
           + [\mathrm{Im}(\mathcal{C}+\omega_1)]^2}.
\end{align}
The particle-hole symmetry provides the relation
\begin{align}
    A(\omega>0,\beta)=A(-\omega,\beta).
\end{align}
The first term on the right-hand side of Eq. (\ref{eq:AwAnsatz})
can be recognised as the Hubbard side band located at
$\omega=\mathrm{Re}\mathcal{C}$ with a half width
$\Gamma_\mathrm{H}=|\mathrm{Im}C|$. The second term is the Kondo resonance
located at $\omega=\mathrm{Re}(\mathcal{C}+\omega_1)=0$ with a half width
$\Gamma_\mathrm{K}=|\mathrm{Im}(C+\omega_1)|$.

The parameters defining the ansatz can be determined from cluster calculations
with a bath consisting of fifty sites \cite{zhao2025}.
The calculated spectral functions are shown in Fig. \ref{fig:SIAM_Spectra}
and compared with the benchmark results obtained using
the numerical renormalisation group method.
The close agreement attests to the correctness of the ansatz.

\section{Summary and outlook}

The Vxc formalism is complementary to the traditional self-energy approach.
In contrast to the self-energy, which is usually expressed in
momentum and frequency domains,
Vxc is more naturally formulated in space and time
domains. In the Vxc formalism, the xc hole is a fundamental quantity that
generates Vxc as the Coulomb potential, and it is directly related to the
ground-state electron density. This is similar to the Slater exchange hole,
whose Coulomb potential is the Slater exchange potential.
Thus, in the Vxc formalism,
the many-electron problem associated with addition and
removal of an electron is reduced to constructing
for given $r$, $r'$, and $t$ an approximate
radial charge distribution (xc hole), which integrates to $-1$ (sum rule)
and whose value at the origin is equal to the negative
of the ground-state electron density (exact constraint).
It is worth noting that only the spherical average of the xc hole is needed
and only its first radial moment is relevant. This property suggests
that Vxc is not sensitive to the detailed structure of the xc hole, which
is known to partially explain the success of LDA in DFT. 
This opens the possibility of approximating Vxc as a functional of the
electron density, for example, within LDA.
The Vxc formalism can therefore be regarded as a density-functional approach to
calculating the Green function from which one-particle excitation spectra can
be extracted.

The Vxc formalism has been applied to a number of generic model Hamiltonians
representing strongly correlated systems.
These are the 1D half-filled Hubbard model, the 1D antiferromagnetic Heisenberg
model, and the half-filled single-impurity Anderson model.
By far, these applications are based on an extrapolation procedure in which
Vxc of a small cluster is calculated using an accurate numerical tool and
the resulting Vxc is extrapolated to the model Vxc by means of a parametrisation.
The results are in rather good agreement with benchmark results and suggest that
it may be feasible to parametrise Vxc as a functional of
the model parameters for a general filling.
Such a parametrised Vxc is highly advantageous from the numerical point
of view, since it would circumvent expensive computations in solving,
for example, the impurity problem in DMFT.

To apply the local-density approximation, the xc hole and
Vxc of the homogeneous electron gas
have been calculated within the random-phase approximation.
Due to the sum rule and the exact constraint,
it is found that the xc hole is rather robust in the sense that it
does not deviate drastically from the static xc hole. An encouraging
finding is that there is a large cancellation between exchange and correlation
potentials resulting in a generally smooth Vxc,
rendering a local-density approximation may well be reasonable.

The equation of motion
of the Green function in the Vxc formalism can be recast into a quasiparticle
equation of motion. Each quasiparticle interacts with the rest of the quasiparticles
through a (residual) xc field, thus providing a quantitative description of
a many-electron system as a set of interacting quasiparticles as in
Landau's phenomenological theory.
The coupling between the quasiparticles can be absorbed into an effective field.
An important finding is that this effective field, for systems that have
been investigated so far, has a simple form consisting of a static term
and a dynamic term. From the quasiparticle picture, the static term
corrects the reference mean-field quasiparticle energy, while the dynamic term
induces a coupling to the main collective excitation in the system,
for example, plasmon,
giving rise to an incoherent or a satellite feature.
This simple physical picture is very appealing, and in strong contrast
to the self-energy picture.

One of the main challenges of the Vxc formalism is to construct simple
but accurate approximations for Vxc or the effective field
$\Xi_q(r,t)$. The availability of
such approximations would avoid costly calculations of traditional
self-energy. For systems
with itinerant valence electrons as in alkali metals,
the effective field is modelled as a functional of electron density,
and the parameters
are determined from $GW$ calculations. A local-density
approximation, which reproduces the results
of the homogeneous electron gas in the limit of constant density,
is then proposed. For more general systems and materials,
machine learning and artificial intelligence
can potentially
play a crucial role in the search for accurate Vxc as a functional of
density or model parameters. There is a large amount of data available in
the literature on self-energy calculations based on the $GW$ approximation and
DMFT, which can be exploited to help construct the desired Vxc. This Vxc can
be used to study complex systems that
are too complicated to be treated using traditional self-energy methods.

The Vxc formalism originally developed for fermions
can be extended to systems of bosons. Model systems describing
a coupling between fermions and bosons can also be formulated
in the Vxc language. The Vxc formalism may offer an alternative
and simpler description of the many-body problem than traditional approaches.

\begin{acknowledgments}
Financial support from
the Swedish Research Council (Vetenskapsrådet, VR, Grant No. 2021\_04498)
is gratefully acknowledged.
\end{acknowledgments}







\bibliography{Refs} 

\end{document}